%% file: __main.tex
\documentclass[twocolumn]{aastex631}

\newcommand{\RN}[1]{%
  \textup{\uppercase\expandafter{\romannumeral#1}}%
}
\newcommand{\kms}{\,km\,s$^{-1}$}

\newcommand{\Msun}{\;M$_{\sun}$}
\newcommand{\s}[1]{\textsuperscript{#1}}
\defcitealias{Tanaka2020}{T20} 

\DeclareRobustCommand{\ion}[2]{%
\relax\ifmmode
\ifx\testbx\f@series
{\mathbf{#1\,\mathsc{#2}}}\else
{\mathrm{#1\,\mathsc{#2}}}\fi
\else\textup{#1\,{\mdseries\textsc{#2}}}%
\fi}

\shorttitle{MMT/Binospec: Spectroscopic Survey of Two $z$\,$\sim$\,0.8 Galaxy Clusters}
\shortauthors{Di et al.}

\begin{document}

\title{MMT/Binospec Spectroscopic Survey of Two \mbox{\boldmath $z$\,$\sim$\,0.8} Galaxy Clusters in the Eye of Horus Field}

\correspondingauthor{Eiichi Egami} 
\email{egami@arizona.edu}

\author{Jiyun Di}
\affiliation{Steward Observatory/Department of Astronomy, University of Arizona, Tucson, AZ 85721, USA}
\affiliation{Department of Physics and Astronomy, Stony Brook University, Stony Brook, NY 11794, USA}

\author{Eiichi Egami}
\affiliation{Steward Observatory/Department of Astronomy, University of Arizona, Tucson, AZ 85721, USA}

\author{Kenneth C. Wong}
\affiliation{Research Center for the Early Universe, Graduate School of Science, The University of Tokyo, 7-3-1 Hongo, Bunkyo-ku, Tokyo 113-0033, Japan}
\affiliation{National Astronomical Observatory of Japan, Mitaka, 181-8588 Tokyo, Japan}

\author{Chien-Hsiu Lee}
\affiliation{W. M. Keck Observatory, 65-1120 Mamalahoa Hwy, Kamuela, HI 96743, USA}

\author{Yuanhang Ning}
\affiliation{Department of Astronomy, Tsinghua University, Beijing 100084, China}

\author{Naomi Ota}
\affiliation{Department of Physics, Nara Women’s University, Kitauoyanishi-machi, Nara 630-8506, Japan}

\author{Masayuki Tanaka}
\affiliation{National Astronomical Observatory of Japan, Mitaka, 181-8588 Tokyo, Japan}

\begin{abstract}
The discovery of the Eye of Horus (EoH), a rare double source-plane lens system ($z_{\rm lens}$\,$=$\,0.795; $z_{\rm src}$\,$=$\, 1.302 and 1.988), has also led to the identification of two high-redshift ($z_{\rm phot}$\,$\sim$\,0.8) galaxy clusters in the same field based on the subsequent analysis of the Subaru/Hyper Suprime-Cam (HSC) optical and \textit{XMM-Newton} X-ray data.  The two brightest cluster galaxies (BCGs), one of which is the lensing galaxy of the EoH, are separated by only $\sim$100\arcsec\,($=$ 0.75\,Mpc\,$<$\,$r_{200}$) on the sky, raising the possibility that these two clusters may be physically associated.  Here, we present a follow-up optical spectroscopic survey of this EoH field, obtaining 218 secure redshifts using MMT/Binospec. We have confirmed that there indeed exist two massive ($M_{\rm dyn}$\,$>$\,$10^{14}$\,\Msun) clusters of galaxies at $z$\,$=$\,0.795 (the main cluster) and at $z=0.769$ (the NE cluster).  However, these clusters have a velocity offset of $\sim$4300 km s$^{-1}$, suggesting that this two-cluster system is likely a line-of-sight projection rather than a physically-related association (e.g., a cluster merger). In terms of the properties of the cluster-member galaxies, these two $z\sim0.8$ clusters appear well-developed, each harboring an old (age\,$=$\,3.6--6.0 Gyr) and massive ($M_\mathrm{*}$\,$=$\,4.2--9.5\,$\times$\,$10^{11}$\,\Msun) BCG and exhibiting a well-established red sequence (RS). This study underscores the importance of conducting a spectroscopic follow-up for high-redshift cluster candidates because RS-based cluster selections are susceptible to such a projection effect in general.
\end{abstract}


\section{Introduction} \label{sec:intro}

Clusters of galaxies are important laboratories for cosmological studies.  Observations of galaxy clusters at high redshifts (with an accurate mass calibration) produce valuable cosmological constraints \citep[][]{Mantz2010, Kravtsov2012}, allowing us to test cosmological models such as the now standard $\Lambda$CDM model.  We can also investigate physical processes driving the evolution of clusters and their member galaxies as well as probing the nature of dark energy and dark matter \citep[][]{Allen2011, Kravtsov2012}. 

Clusters of galaxies can be identified at a variety of wavelengths. In the X-ray, they show up as bright extended sources because hot plasma in the intracluster medium (ICM) emits strong bremsstrahlung radiation. In the optical and near-infrared, clusters can be identified as over-densities of evolved red galaxies through the detection of the so-called cluster red sequence (RS), and a number of such cluster-finding algorithms exist, such as the CAMIRA algorithm \citep[][]{Oguri2014, Oguri2018} applied to the HSC Subaru Strategic Program \citep[][]{Aihara2018}. At millimeter wavelengths, the spectral distortion of the cosmic microwave background due to the inverse-Compton scattering by the ICM provides a nearly redshift-independent signal for a given cluster mass, i.e. Sunyaev-Zel'dovich effect \citep[SZ effect;][]{SZ1972}, and this effect has been exploited by the Atacama Cosmology Telescope \citep[ACT;][]{Hasselfield2013} and South Pole Telescope SZ surveys \citep[][]{Reichardt2013, Bleem2015}.

The ``Eye of Horus'' (HSC J142449-005322; $\text{RA}=14^\mathrm{h}24^\mathrm{m}49^\mathrm{s}.0$ and $\text{Dec}=-0^{\circ}53'21''.7$, J2000; hereafter EoH) is a strong gravitational lens system discovered through the HSC Subaru Strategic Program \citep{Tanaka2016}. In this system, one foreground galaxy at $z_\mathrm{lens}=0.795$ is gravitationally lensing two distant background galaxies at $z_\mathrm{src}=1.988$ and $z_\mathrm{src}=1.302$, and thus it is called a double source plane (DSP) system. Such a DSP system is rare because it requires two background galaxies at different redshifts and a lensing galaxy perfectly aligned along the light of sight.  As a result, only a handful of DSP systems have been discovered so far \citep[e.g.,][]{Gavazzi2008, Tu2009}. 

\begin{figure*}[!b]
    \centering
	\includegraphics[height=0.86\textheight, keepaspectratio]{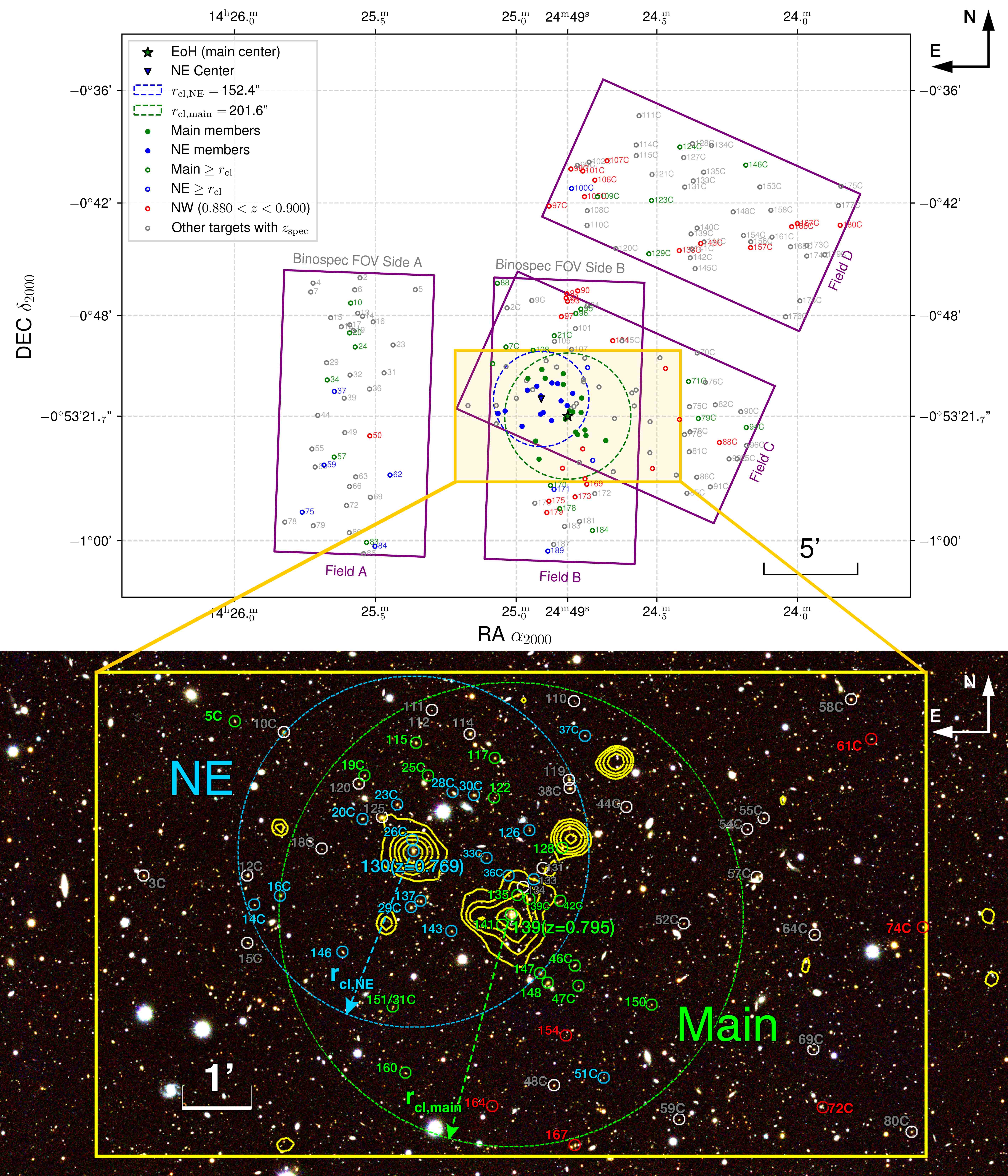}
    \caption{Positions of the main and NE cluster members. \textit{Upper panel} --- The MMT/Binospec field-of-views (FoVs) are denoted as Fields A--D, consisting of two mask designs of Side A+B for Run \#1 (Fields A and B; PA\,$=$\,$-2^{\circ}$)  and \#2 (Fields C and D; PA\,$=$\,66$^{\circ}$), respectively. 
    The member galaxies of the NE and main clusters are indicated with blue and green circles, respectively, while the blue triangle and green star indicate corresponding cluster BCGs.
    The green and blue dashed circles represent the virialized core radii $r_\mathrm{cl}$ ($r_\mathrm{cl, main}=201.6''$ and $r_\mathrm{cl, NE}=152.4''$). \textit{Lower panel} --- Spatial distribution of the MMT/Binospec-observed galaxies with slit numbers. The background is the three-color Subaru/HSC image with the \textit{r}, \textit{i}, \textit{z} bands, and the yellow contours show the \textit{XMM-Newton} X-ray map. The color definition is the same as the upper panel.}
    \label{fig: main}
\end{figure*}

To better understand the surrounding structure that could potentially affect the lensing potential of this system, the environment of EoH was examined for the existence of any mass concentration.  Recent studies based on the CAMIRA cluster catalog \citep[][]{Oguri2018} reported that two major clusters are located in the field of EoH with photometric redshifts of $z_\mathrm{phot}\sim0.8$, which was subsequently confirmed by the \textit{XMM-Newton} observation by \citet{Tanaka2020}, hereafter referred to as \citetalias{Tanaka2020}, detecting X-ray emission that peaks at the locations of the two brightest cluster galaxies (BCGs) determined by CAMIRA. One of the X-ray emission peaks is at the lensing galaxy of EoH while the other peak is at the location of the galaxy SDSS J142454.68-005226.3 ($\text{RA}=14^\mathrm{h}24^\mathrm{m}54^\mathrm{s}.7$, $\text{Dec}=-0^{\circ}52'26''.9$, J2000), which is positioned only $\sim100$\arcsec\ northeast (NE). The positions of the EoH lensing-galaxy BCG and second BCG are regarded as the centers of the corresponding galaxy clusters, denoted as the main and NE clusters, respectively, by \citetalias{Tanaka2020}. Their photometric redshifts suggest that the main and NE galaxy clusters are both at $z\sim0.8$.  Since their separation is only $\sim$\,100\arcsec\ on the sky, one question arises: are they physically related (e.g., a cluster merger) or the superposition of two structures along the line of sight? Spectroscopic information is needed to confirm the existence of the two $z_\mathrm{photo}\sim0.8$ clusters in this field and investigate their relationship. Although \citetalias{Tanaka2020} hinted that this system is probably not a cluster merger due to the absence of X-ray emission between the two cluster peaks, it is not possible to provide a definitive answer without having spectroscopic data.

In this work, we utilized the Binospec multi-object spectrograph on the MMT telescope to answer the question above by presenting the results from our spectroscopic survey of two $z\sim0.8$ clusters of galaxies conducted in the EoH field.  We organized the paper by introducing the observational method in Section~\ref{sec: obs}, presenting the results from our MMT/Binospec redshift survey in Section~\ref{sec: res}, and discussing cluster dynamical-mass estimates and the analysis of two BCG optical spectra in Section~\ref{sec: dis}. Our conclusions are placed in Section~\ref{sec: con}. This paper uses standard $\Lambda$CDM cosmological parameters: $H_0=100h=70\,\mathrm{km}\,\mathrm{s}^{-1}\mathrm{Mpc}^{-1}$, $\Omega_\mathrm{m}=0.3$, and $\Omega_\Lambda=0.7$.

\section{Observations and Data Reduction}
\label{sec: obs}

\subsection{MMT/Binospec}

We used Binospec \citep[][]{Fabricant2019}, a high-throughput imaging spectrograph with two $8'\times15'$ fields of view (Side A and B) at the f/5 focus of the 6.5m MMT telescope, to conduct our redshift survey. The main spectral line features we look for are the Ca H\&K ${\lambda}{\lambda}3933,3968$ absorption lines. For star-forming galaxies, we should also detect the [\ion{O}{II}] ${\lambda}{\lambda}3727,3729$ emission lines. Continuum features like the Balmer and 4000 \AA~ breaks may also be detectable in the spectra. To achieve the best sensitivity at $\lambda\sim7000-7600$~\AA, where all these features appear at $z\sim0.8$, we used the 600 lines/mm grating and set the central wavelength to 7300 \AA, which results in a dispersion of 0.61 \AA/pixel.  Each slit has a $1''$ width corresponding to 3.32 pixels, producing a spectral resolution of $\Delta\lambda\sim2$~\AA. The resultant spectral resolving power is $R=3650$ at 7300 \AA, high enough to clearly separate the Ca H\&K lines.

The observations were executed on two dates: UT 2019 July 8 (Run \#1; Field A and B) and UT 2022 January 31 (Run \#2; Field C and D).  For each run, we took 4 exposures of 1200 seconds each with a total integration time of 4800 seconds.  The seeing varied between $1.2''$ and $1.3''$. We placed one of the $8'\times15'$ field-of-views on top of the EoH field (Field B and C as shown in Figure~\ref{fig: main}). Binospec's one field-of-view is large enough to contain the two targeted $z\sim0.8$ galaxy clusters. With the minimum slit length of $\sim8''$, we were able to target a few tens of cluster-member candidates with one slitmask.
All the data have been processed by the standard Binospec data processing pipeline\footnote{ \url{http://mingus.mmto.arizona.edu/~bjw/mmt/binospec_info.html\#\#dataanalysis}}. 

\subsection{Target Selection}
\label{sec: target-select}

{\bf Run \#1:} Using the CAMIRA cluster catalog (S18A-Wide-v4), we first selected galaxies that were identified as the members of the main cluster at $z_{\rm CAMIRA}$\,$=$\,0.79.  These galaxies constitute our primary targets.  Note, however, that the CAMIRA catalog only includes red galaxies that form the cluster RS.  To recover cluster-member galaxies with bluer colors (e.g., due to star formation), we also used the original HSC Wide S18A source catalog and selected galaxies with photometric redshifts of $z$\,$\sim$\,0.8$\pm$0.5 without applying any other color selection.  This constitutes our secondary targets.  In each target group, we assigned higher priorities to more massive (and therefore generally more luminous) galaxies based on the stellar masses provided by the CAMIRA catalog.  The two BCGs were specifically targeted.

{\bf Run \#2:} For the 2nd run, we used an updated CAMIRA catalog (S20A-Wide-v2). 
We used the same target selection procedure as described above.  The HSC Wide S18A source catalog was also used to select bluer cluster members.  For this run, we adjusted the Binospec position angle (PA) such that the parallel field (Field D) will contain another high-redshift CAMIRA cluster candidate at $z_{\rm CAMIRA}$\,$=$\,0.896.

In total, we targeted $\sim$400 galaxies in the EoH field containing two clusters of galaxies at $z\sim0.8$ as identified by \citetalias{Tanaka2020}. Among the CAMIRA cluster-member candidates at $z_\mathrm{CAMIRA}=0.790, 0.759$, and $0.896$ of 131, 105, and 32 galaxies, we observed 32, 39, and 13 galaxies in the two runs, which yields the Binospec-observed fraction of 24, 37, and 41\% respectively.

\begin{figure}[!h]
    \includegraphics[width=0.99\columnwidth]{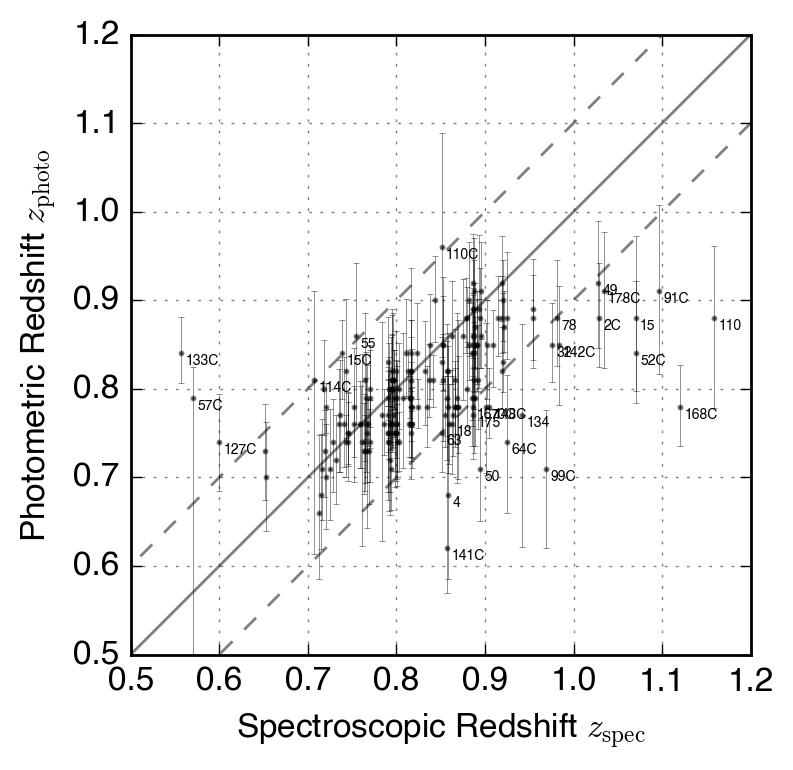}
    \caption{Spectroscopic vs. photometric redshifts. We used the photometric redshifts given by HSC/Mizuki \citep{Tanaka2015,Tanaka2018}. The majority (84\%) of measured redshifts fall within the $\Delta z \leq 0.1$ region, marked with the dashed lines centered with the identity line (in solid). Galaxies with $\Delta z > 0.1$ are additionally marked with their slit numbers. The scatter along the identity line is $\Delta (z_{\rm photo}-z_{\rm spec})/(1+z_{\rm spec})=0.07$}
    \label{fig: spec-photo}
\end{figure}

\subsection{Redshift Measurements}

\begin{figure*}[!t]
	\includegraphics[width=0.99\textwidth]{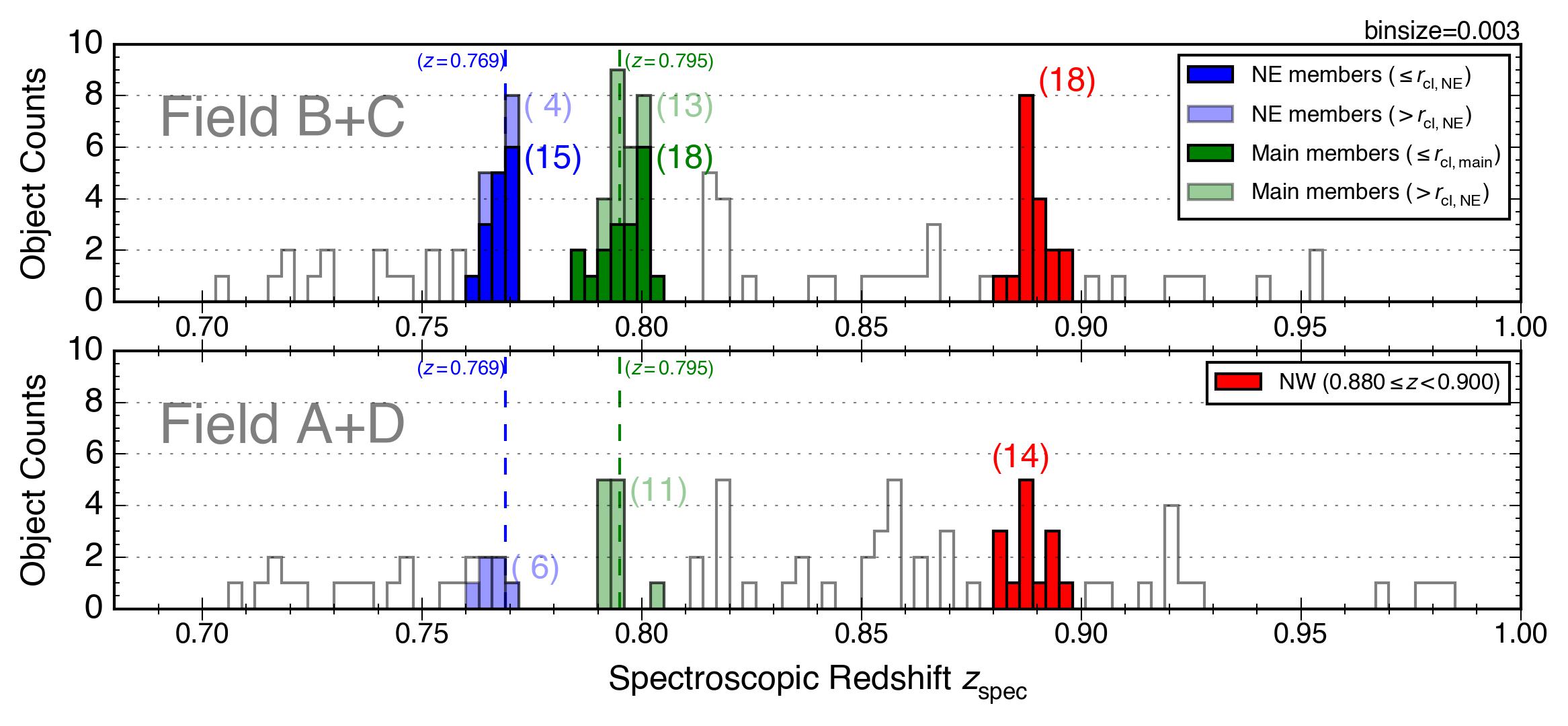}
    \caption{Redshift distributions of galaxies at $z$\,$\sim$\,0.7--1.0 in the B and C fields (\textit{Upper Panel}) and in the A and D fields (\textit{Lower Panel}). The main and NE cluster members inside the virialized core radii $r_\mathrm{cl}$ are marked with deep green and blue, respectively. The numbers of the main and NE cluster members within $r_\mathrm{cl}$ are shown in deeper colors. Galaxy members selected by the shifting gapper method but outside $r_\mathrm{cl}$ are marked in light green and blue. The galaxies at $0.880 \leq z < 0.900$ (the redshift-range of the NW-cluster candidate) are shown in red. The white bins show other galaxies with $z_\mathrm{spec}$. The vertical dashed lines indicate the redshifts of the cluster BCGs, which we also take as the redshifts of their host clusters (i.e. main: $z=0.795$; NE: $z=0.769$).}
    \label{fig: distr}
\end{figure*}

To measure redshifts, we used \textsc{SpecPro} \citep[][]{Masters2011b} software, which uses 19 galaxy templates with emission and absorption features. \textsc{SpecPro} has been adopted by the MMT observatory to analyze pipeline-processed Binospec 1D and 2D spectra, and for each manually-selected spectral template, it calculates multiple feasible redshift solutions.  By following the procedure given by \citet{Shah2020}, the most likely redshift solution was selected through visual inspection.

For each redshift determination, we assigned the ``confidence'' flag, indicating the reliability of the redshift measurement (3 = secure, 2 = likely, and 1 = no determination), which mainly depends on the signal-to-noise ratio of the spectrum being inspected. We record remarks and their issues in the ``notes'' column of Table~\ref{tab: RT01}. 

\section{Results}
\label{sec: res}

\subsection{Spectroscopic Redshifts}

All the measured spectroscopic redshifts are presented in Table~\ref{tab: RT01}. With the Run \#1 observation, we observed 190 sources, which produced the redshift confidence-flag distribution of 56, 1, and 43\% for 3, 2, and 1, respectively. With the Run \#2 observation, we observed 181 sources with a confidence-flag distribution of 61, 3, and 36\%. For our analyses, we only used spectra with a confidence flag of 3 or 2. In total, 221 secure redshifts were used in the following results and analysis. This includes 23, 28, and 6 secure redshifts from targets in the CAMIRA red candidate catalogs at $z_\mathrm{CAMIRA}=0.790, 0.759$, and $0.896$, with the success rate (secure redshifts versus targeted) of 72, 72, and 46\%.

Figure~\ref{fig: spec-photo} compares the measured spectroscopic redshifts with the HSC-derived photometric redshifts, which were used for the target selection. The agreement is good with with $\Delta (z_{\rm photo}-z_{\rm spec})/(1+z_{\rm spec})=0.07$. 

Figure~\ref{fig: distr} displays the redshift distribution for all the targets in both the Run \#1 and \#2 data. The Binospec spectral resolving power of $R = 3650$ yields a redshift measurement uncertainty of  $\Delta z = (1+z_{\rm spec})/R = 0.0005$. We set the binsize of the redshift distribution histogram to 0.003, which is $6\Delta z$. 

The spectroscopic redshift of the main-cluster BCG (Slit \#139) is determined as $z_\mathrm{main}=0.795$ and that of the NE-cluster BCG (Slit \#130) is $z_\mathrm{NE}=0.769$. We regard the locations of these two BCGs as the centers of the two clusters and treat these BCG redshifts as those of the respective clusters.

\input{Members-table}

\input{Centroids}

\subsection{Cluster Member Identification}%

\subsubsection{Shifting Gapper Method}
\label{sec: shifting}

The redshift boundaries of each galaxy cluster are often difficult to define based on a redshift histogram like Figure~\ref{fig: distr} alone. We therefore apply the ``shifting gapper method'', developed by \citet{Fadda1996} and implemented by \citet{Sifon2013, Sifon2016}, to assess the cluster membership of individual galaxies. More specifically, the following steps are adopted here: 

\begin{enumerate}
    \item 
        From the cluster center, we sort and bin galaxies by their radial distance, $r$. Each bin should have a width of at least 250 kpc.  If there are less than 10 galaxies within a bin, we will increase the bin width such that the resulting bin has at least 10 galaxies in it.
    \item 
        In each bin, we sort galaxies by the absolute light-of-sight velocity offsets ($|v|$) from the BCG's redshift, from low to high velocities. 
        Galaxies with $|v|$ larger than 4000\kms are not considered to be members of the cluster.
    \item 
        Inside each bin, we evaluate the difference (``gap'') in velocity $|v|$ between two galaxies adjacent in the order. 
        We keep including galaxies as cluster members as long as the $|v|$ difference is $< 500$\kms, and stop including when this 500\kms\ threshold is exceeded by a pair of galaxies adjacent in the $|v|$ order.  At this point, we record the maximum $|v|$ of each bin as $|v_\mathrm{max}|$.
    \item 
        In the next stage, we add galaxies outside the $|v_\mathrm{max}|$ boundary to the cluster if their $|v|$ values are $<1000$\kms\ from $|v_\mathrm{max}|$.  
\end{enumerate}

We iterated these steps until the membership became stable. Finally, we identified 42 member galaxies for the main cluster and 25 for the NE cluster.

\subsubsection{Boundary of the Virialized Core}

The virial radius of a galaxy cluster broadly defines the radius within which the motions of cluster-member galaxies can be considered dynamically relaxed (i.e., "virialized").  Theoretically, this virial radius is close to $r_{200}$, within which the average mass density of the galaxy cluster is equal to 200 times the critical mass density of the universe. Throughout this paper, we adopt the $r_{200}$ calculated by \citetalias{Tanaka2020} with the \textit{XMM-Newton} X-ray data: $r_{200,\text{main}}=168''$ and $r_{200,\text{NE}}=127''$. 

For the two clusters studied here, we define the boundary of the virialized core as $1.2 r_{200}$.  Though somewhat arbitrary, this boundary appears to demarcate nicely the higher-density cluster cores against the lower-density outskirts. We denote the virialized cores of the two clusters as $r_\mathrm{cl,main}$ and $r_\mathrm{cl,NE}$, which are 201.6\arcsec\,$=$\,1.510 Mpc and 152.4\arcsec\,$=$\,1.129 Mpc, respectively.

We list the members of the main and NE clusters in Table~\ref{tab: RT-new}, extending beyond the virialized core and out to the boundary defined by the cluster members identified in Section 3.2.1. \citetalias{Tanaka2020} points out that the BCG of the NE cluster identified by the CAMIRA cluster catalog \citep[][]{Oguri2018} is $\sim40$\arcsec\ displaced from the peak of the NE component of the X-ray emission, and that the real BCG is likely the bright galaxy close to the X-ray peak.  We consider the latter as the BCG marking the center of the NE cluster (Slit 130).

In Table~\ref{tab: centroids}, these BCG positions are compared with the unweighted centroids of the main and NE cluster members within $r_\mathrm{cl}$. The RA and Dec offsets and uncertainties ($\sigma_{\rm X}$) of X-ray emission peaks are given by \citetalias{Tanaka2020}. We calculate distance offset uncertainties by adding additional RA and Dec uncertainties of optical BCGs $\sigma_{\rm opt}=0.5$\arcsec, that is, $\sigma_{\rm total} = \sqrt{\sigma_{\rm X}^2 + \sigma_{\rm opt}^2}$. The main-cluster centroid is $13.1\pm13.3$\arcsec\ northeast to the main BCG while the NE cluster centroid is $10.8\pm9.9$\arcsec\ south of the NE BCG.

\begin{figure}[!b]
    \includegraphics[width=\columnwidth]{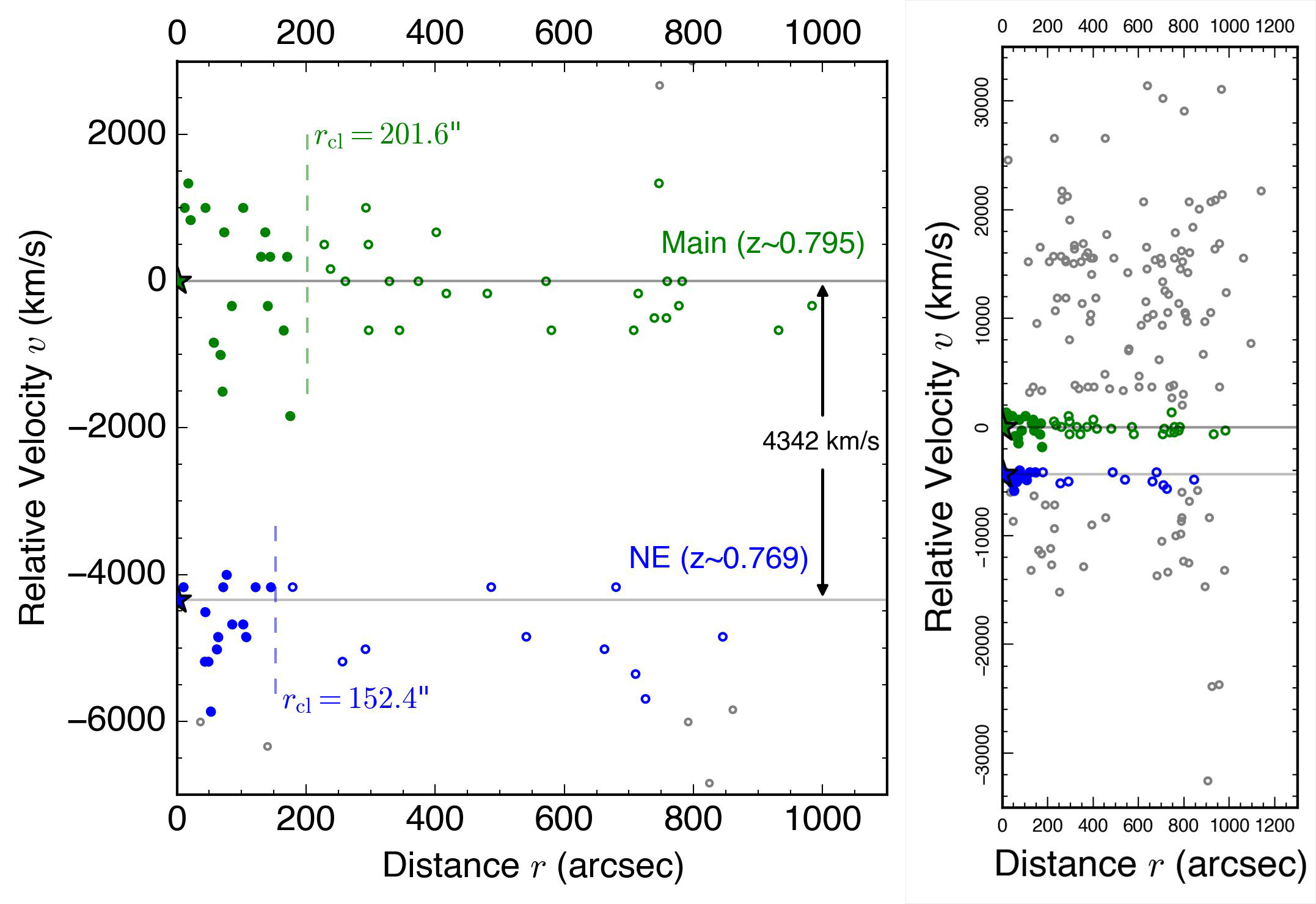}
    \caption{Plots of projected distance to cluster centre $r$ vs light-of-sight velocity $v$ for main and NE clusters. Green/Blue filled and open circles represent Main/NE cluster members inside and outside $r_\mathrm{cl}$ respectively. The grey open circles are non-cluster members. \textit{Left}: Galaxies near $0.76 \lesssim z_\mathrm{spec} \lesssim 0.81$ and \textit{right}: $0.6 \leq z_\mathrm{spec} \leq 1.0$.}
    \label{fig: rv}
\end{figure}

\subsubsection{Velocity Dispersion}
\label{sec: vel-dis}

Figure~\ref{fig: rv} plots the line-of-sight velocity distribution of galaxies, $v$, versus the projected distance from the cluster center, $r$. In this $r$-$v$ plot, the main cluster shows an expected trumpet shape with the velocity dispersion increasing toward the cluster center. For the NE cluster, the trumpet shape is not easy to identify, and many galaxies at larger $r$ show larger relative velocity offsets, making the cluster-member identification by the shifting gapper method critical.

The velocity dispersion of the NE and the main cluster members enclosed by $r_\mathrm{cl}$ are $\sigma_\mathrm{NE}=520\pm90$\kms and $\sigma_\mathrm{main}=930\pm120$\kms respectively, where the uncertainties of dispersion are derived from the Bootstrap resampling method\footnote{
\url{https://docs.scipy.org/doc/scipy/reference/generated/scipy.stats.bootstrap.html}}. 
These velocity dispersions evaluated inside virialized cores provide key inputs for estimating the virial masses. Moreover, the velocity dispersions of member galaxies within $r_{200}$ are $\sigma_\mathrm{200, NE}=520\pm100$ \kms\ and $\sigma_\mathrm{200, main}=850\pm110$ \kms from slightly smaller sample sizes (1 galaxy difference for NE and 2 for the main).

\subsection{Color-Magnitude Diagram}
\label{sec: c-m}

\begin{figure}[h]
    \includegraphics[width=\columnwidth]{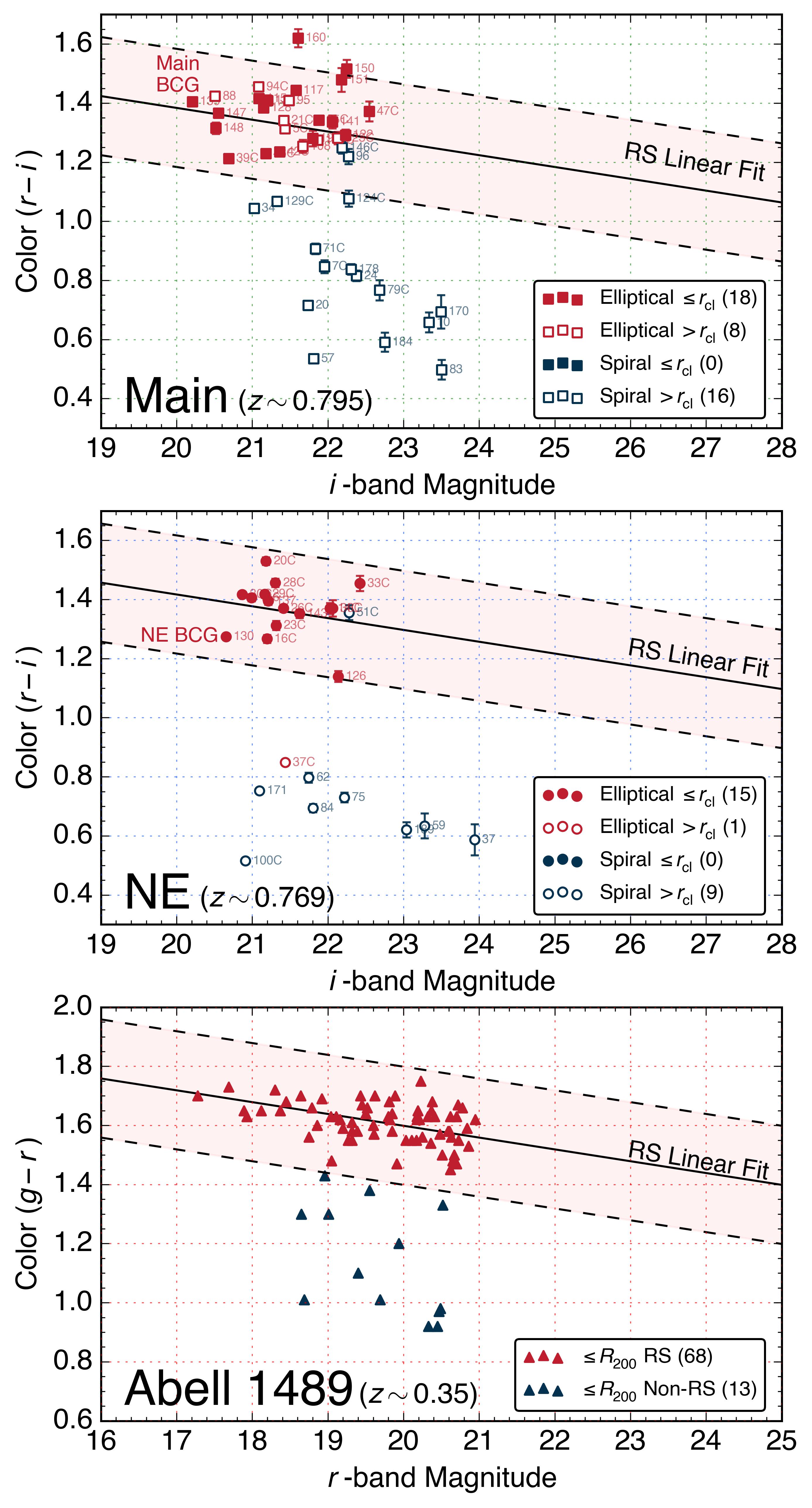}
    \caption{\textit{Top and Middle} --- Color-magnitude diagrams for the main and NE cluster-member galaxies with the HSC photometry. The red/blue color indicates the ``elliptical''/``spiral'' galaxies defined by the optical spectral features  (see Section~\ref{sec: c-m}).  
    The linear fit to the elliptical galaxies within $r_\mathrm{cl}$ (the solid lines) have the same fixed slope of $-0.04$. The dashed lines denote a range of $\Delta(r-i)=\pm0.2$ from the fits, enclosing the red-sequence (RS) galaxies. \textit{Bottom} --- Color-magnitude Diagram for Abell 1489 cluster-member galaxies within $R_{200}$ as identified by \citet{Rines2022}. }
    \label{fig: c-m}
\end{figure}

With the help of the HSC photometric catalog, we use confirmed cluster members to map out the color-magnitude diagrams. The HSC photometric catalog \citep[][]{Aihara2018} provides the galaxy's broad-band magnitudes in the $r$ band ($\lambda_\mathrm{eff}$\,$\sim$\,6175\,\AA, $\mathrm{FWHM}$\,$=$\,1503\,\AA) and $i$ band ($\lambda_\mathrm{eff}$\,$\sim$\,7711\,\AA, $\mathrm{FWHM}$\,$=$\,1574\,\AA), 
which measure the Balmer/4000 \AA\ breaks of $z\sim0.8$ galaxies. In our color-magnitude analysis, the galaxies are matched with the HSC photometry data. 
\citet{Huang2018} give the typical precision of the HSC photometry as $\sim0.023$ mag at $r=20$ mag, $\sim0.036$ mag at $r=24$ mag, $\sim0.015$ mag at $i=20$ mag, and $\sim0.030$ mag at $i=24$ mag. With this photometric precision and the standard deviation of $\sim0.004$ with the linear fit, the intercept uncertainty of the $(r-i)$ color is derived to be $0.03$.

We calculate the slope and intercept of the cluster RS as follows: First, we use spectral information to classify galaxy types: ``elliptical'' galaxies as those with strong Ca H\&K and G-band absorption lines and ``spiral'' galaxies as those with emission lines (e.g. [\ion{O}{II}] ${\lambda}$${\lambda}$3727, [\ion{O}{III}], H$\gamma$). 
Second, we iterate to refine the membership of RS galaxies by (1) selecting the ``elliptical'' galaxies among galaxy cluster members within $r_\mathrm{cl}$, (2) fitting them with the provided slope of $-0.04$ (plotted with solid lines), and (3) filtering out those that have more than 0.2 color difference (indicated with dashed lines). Typically, the RS membership achieves stability after the second iteration. 

In the main and NE cluster plots (Figure~\ref{fig: c-m}), their BCGs sit at the upper-left of the RS, confirming that BCGs are the brightest RS galaxies. The ``elliptical'' galaxies are located along the fitted RS line. The ``spiral'' galaxies look scattered downward from the RS line. The main cluster RS fit shows a $\Delta_{(r-i)}=0.11\pm0.02$ mag scatter and an intercept $(r-i)=1.47\pm0.03$ at $i=19$ mag. The scatter uncertainties are derived by the Bootstrap resampling (see Section~\ref{sec: vel-dis}). For the NE cluster, this gives a $\Delta_{(r-i)}=0.09\pm0.02$ mag and an intercept of $(r-i)=1.46\pm0.03$ as $i=19$ mag.

\section{Discussion}
\label{sec: dis}

\subsection{No Association Between the Two Clusters}

There is no significant overdensity of galaxies in the redshift/velocity space connecting the main and NE galaxy clusters at $0.77 \lesssim z \lesssim 0.80$ in Figure~\ref{fig: distr} and ~\ref{fig: rv}. We note, however, that a merging cluster system may have high relative velocities resulting in a large separation in redshift/velocity space, even if they are physically close to each other. To determine whether or not the two galaxy clusters are physically associated, we can use the velocity difference ($4300$\kms, see Figure~\ref{fig: rv}) and compare it with the possible maximum velocity offset of merging systems from simulations. 

\citet{Lee2010} used $N$-body simulations of high-velocity merging clusters with the physical parameters similar to the galaxy cluster 1E0657-56 (i.e. ``the Bullet Cluster'') to assess the likelihood of a satellite galaxy cluster merging with the main galaxy cluster. They showed that relative velocities of up to $\sim$\,2000--3000\kms, measured at $(2-3)r_{200}$, could be produced by a merging event, but that a relative velocity of $>3000$\kms\ is rare for a merging system with a probability of only $3.6\times10^{-9}$. Since the velocity offset between the main and NE clusters is $\sim4300$\kms, we conclude that the main and NE clusters are not physically associated with each other, and that their proximity is simply due to a projection along the line of sight. 

\input{Masses}

\subsection{Cluster Dynamical Masses}
\label{sec: masses}

\subsubsection{Virial Masses}

We derived the virial masses for the two clusters
by using the member galaxies within $r_\mathrm{cl}$ ($=1.2$\,$r_{200}$).  Below we follow the notation of \citet{Girardi1998}. With the line-of-sight velocity of each galaxy, $V_i$, and the projected distance between the members $i$ and $j$ in the cluster, $R_{ij}$, the cluster virial mass can be expressed as, 
\begin{equation}
    M_\mathrm{V}=\frac{3\pi}{2}\frac{\sigma_\mathrm{P}^2R_\mathrm{PV}}{G},
    \label{virial_mass}
\end{equation}
where $G$ is the gravitational constant, $\sigma_\mathrm{P}$ is the projected velocity dispersion, and $R_\mathrm{PV}$ is the projected virial radius, which are calculated as
\begin{equation}
    \sigma_\mathrm{P}=\sqrt{\frac{\textstyle \sum_{i}V_i^2}{N-1}},\;R_\mathrm{PV}=\frac{N(N-1)}{\textstyle \sum_{i>j}R_{ij}^{-1}},
\end{equation}
where $N$ is the total number of galaxies in the cluster.

By following Girardi's method, we derive the virial masses of the NE and main cluster as listed in Table~\ref{tab: masses}. Since the observed member galaxies might not represent the entire population of the cluster system, \citet{Girardi1998} warn that the calculated mass overestimates the true mass with a median systematic offset of 19 to 39\% because of different types of velocity anisotropy assumed for clusters (the so-called "surface term"). Here, we assume a systematic offset of 30\% and list the corrected masses $M_\mathrm{V}$ in Table~\ref{tab: masses}.

\subsubsection{Masses from the $\sigma_{200}$--M$_{200}$ Scaling Relation}

$N$-body dark matter particle simulations suggest that there exists a scaling relation between 
the velocity dispersion of galaxies and cluster mass within $r_{200}$ \citep[][]{Evrard2008}.
We adopt the following scaling relation presented in \citet{Sifon2013, Sifon2016}:

\begin{equation}
    \sigma_{200}=A_\mathrm{1D} \left( \frac{h E(z) M_{200}}{10^{15}\,\mathrm{M_{\sun}}} \right)^\alpha.
    \label{sig-m200}
\end{equation}
Here, $\sigma_{200}$ is the line-of-sight projected velocity dispersion of galaxies within $r_{200}$, and $M_{200}$ is the cluster mass within $r_{200}$.  $\alpha$ was assumed to be $0.364\pm0.002$. $A_\mathrm{1D}$ and $E(z)$ are expressed as follows: $A_\mathrm{1D}=1177\pm4.2$\kms, and $E(z)=\left( \Omega_\Lambda+(1+z)^3\Omega_\mathrm{m} \right)^{1/2}$. Table~\ref{tab: masses} lists the cluster masses derived from this method.
\medskip

\subsubsection{Comparison of Various Cluster-Mass Estimates}

There is a systematic offset between the cluster dynamical masses derived with the virial theorem (Equation~\ref{virial_mass}) and that with the $\sigma_{200}$--$M_{200}$ scaling relation (Equation~\ref{sig-m200}).  For example, if we assume a line-of-sight velocity dispersion of $\sigma_{\rm P}$\,$=$\,1000 km s$^{-1}$ and a virial radius of $R_{\rm PV}$\,$=$\,1 Mpc, the resultant viral mass $M_{\rm V}$ is $\approx$\,$10^{15}$\,\Msun, and with a 30\% surface-term correction, $\approx$\,$7\times10^{14}$\,\Msun.  On the other hand, if we assume $R_{200}$\,$\approx$\,$R_{\rm PV}$\,$=$\,1~Mpc and $\sigma_{200}$\,$=$\,$\sigma_{\rm P}$\,$=$\,1000 km s$^{-1}$, the resultant $M_{200}$ is $4\times10^{14}$\,\Msun\ at $z=0.8$, which is $\approx$\,60\% of $M_{\rm V}$.  As Table~\ref{tab: masses} shows, the observables themselves are similar ($R_{\rm PV}$\,$\approx$\,$R_{200}$ and $\sigma_{\rm P}$\,$\approx$\,$\sigma_{200}$), so the factor of a few difference between $M_{\rm V}$ and $M_{200}$ mainly comes from this systematic offset between the two methods. 

Based on the \textit{XMM-Newton} observations of this field, \citetalias{Tanaka2020} reported X-ray-derived cluster masses of $5.6\times10^{14}$\Msun\ for the EoH main cluster and $2.2\times10^{14}$\Msun\ for the NE cluster, respectively.  These X-ray-derived masses are between $M_{\rm V}$ and $M_{200}$.  In the case of the main cluster, the X-ray mass is closer to $M_{200}$ while in the case of NE, it is in the middle of the two dynamical-mass estimates.  Assuming that the X-ray-derived masses are closer to the true values, this indicates that both types of dynamical-mass estimates are equally good for the purpose of making a rough estimate (i.e., within a factor of $\sim$\,2 from the X-ray-derived masses) with a limited sampling of cluster members.  $M_{\rm V}$ and $M_{200}$ are seen to result in an overestimate/underestimate, but the sample of two clusters here is simply too small to draw any general conclusion. 
The overall agreement of the dynamical and X-ray cluster masses also suggests that the cores of these two $z\sim0.8$ clusters are already dynamically relaxed (i.e., virialized).

\subsection{The NW Cluster Candidate}
\label{sec: nw}

\begin{figure}
    \includegraphics[width=\columnwidth]{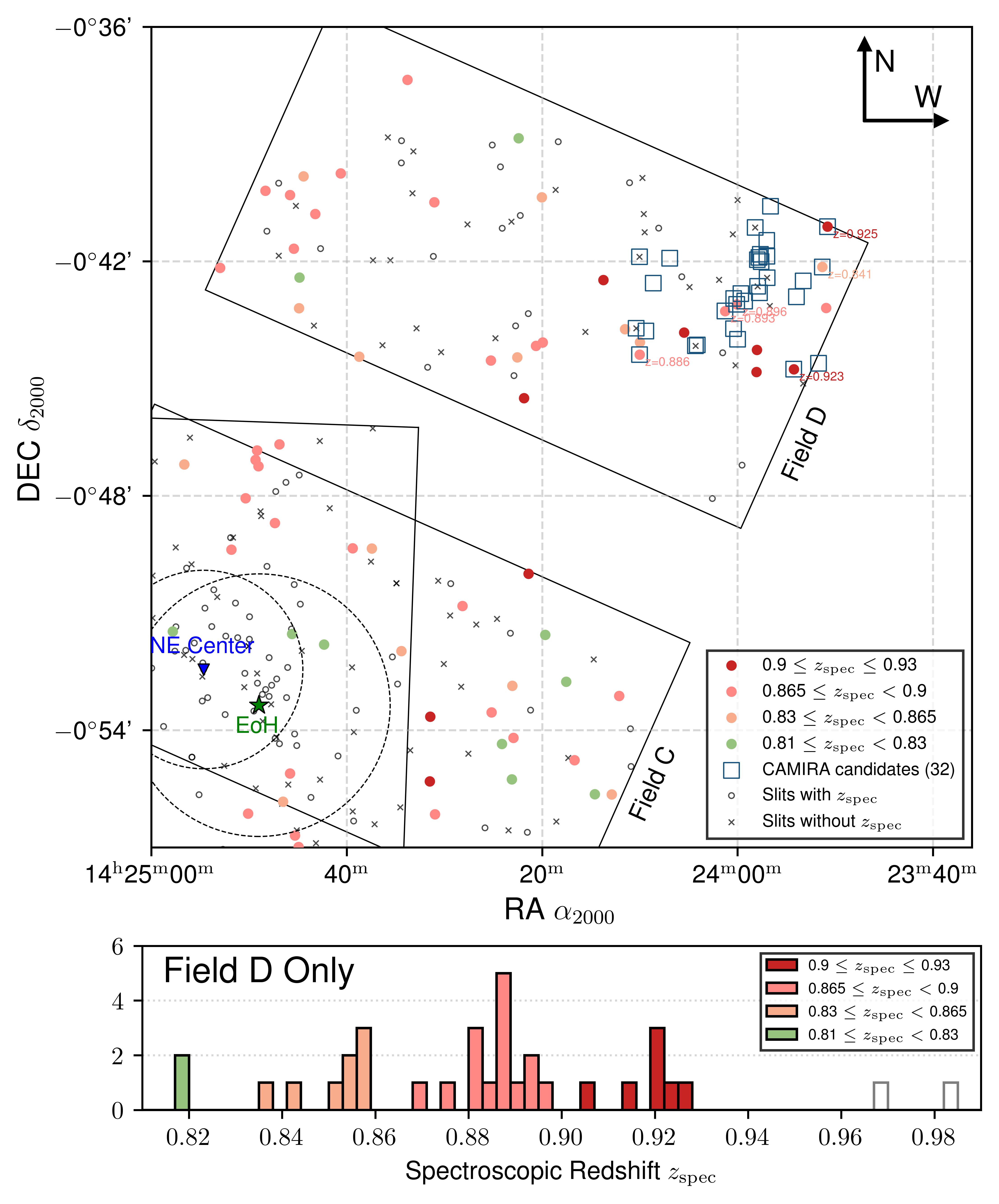}
    \caption{CAMIRA $z=0.896$ NW cluster candidate. \textit{Top} --- The sky positions of the Binospec slits are color-coded with redshifts.  The locations of the CAMIRA-identified cluster-member galaxies are marked with the orange circles. \textit{Bottom} --- The redshift distribution is shown for galaxies in Field D only.}
    \label{fig: nw}
\end{figure}

As described in Section~\ref{sec: target-select}, the slit mask in Run \#2 was designed to cover the CAMIRA cluster candidate with a photometric redshift of $z_\mathrm{CAMIRA}=0.896$ located at the north-west of EoH (Figure~\ref{fig: nw}, the blue squares in the top panel).  Figure~\ref{fig: distr} also shows a redshift peak at $z \sim 0.89$, which matches the redshift of this cluster candidate.  

Three of the CAMIRA-identified cluster members are found to have spectroscopic redshifts of 0.886, 0.893, and 0.896 (Figure~\ref{fig: nw}, the top panel), hinting the existence of a cluster.  However, three other cluster member candidates have spectroscopic redshifts outside the range of this putative cluster (which we define as $z$\,$\approx$\,0.88--0.90).  Although this is not surprising given the uncertainties of the photometric redshifts ($\sim$\,0.05 by \citealt{Tanaka2018}), it also prevents us from drawing a definitive conclusion about the reality of this CAMIRA-identified cluster. The Ca H\&K lines at $z\sim0.9$ are difficult to detect due to the B-band atmosphere absorption, which makes the redshift measurements difficult and only yields 6 galaxies with secure redshifts out of 13 galaxy candidates observed in the two observing runs. 

Figure~\ref{fig: nw} shows that $z\sim0.9$ galaxies in Field D can be grouped into three redshift ranges: $0.830 \leq z < 0.865$, $0.865 \leq z < 0.900$, and $0.900 \leq z < 0.930$.  However, none of these groupings exhibits any clear spatial clustering.  In particular, galaxies in the redshift range of the NW cluster ($z$\,$\approx$\,0.88--0.90) are distributed over Field D and even over Fields B and C, possibly suggesting a sheet-like structure (Figure~\ref{fig: main} and \ref{fig: distr}).  

\begin{figure*}[!t]
    \includegraphics[width=\textwidth]{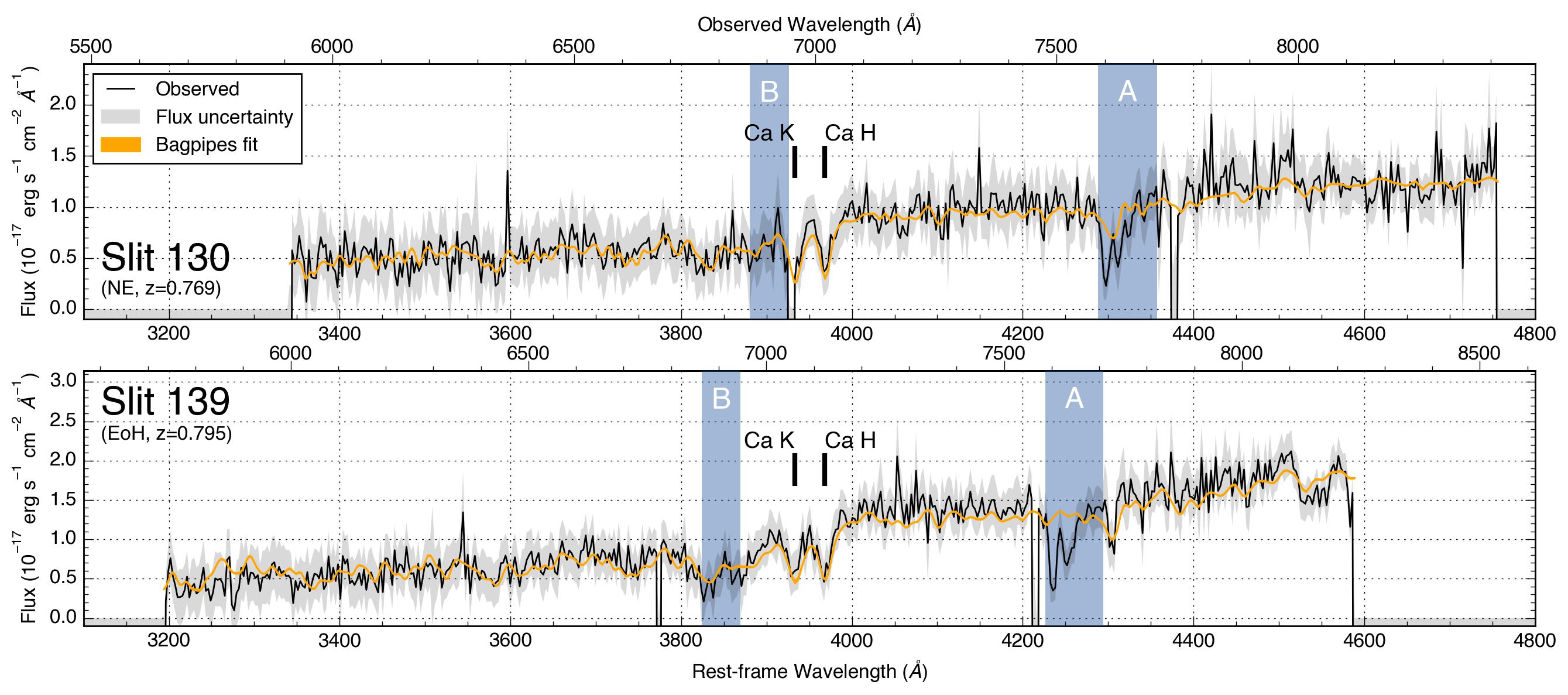}
    \includegraphics[width=\textwidth]{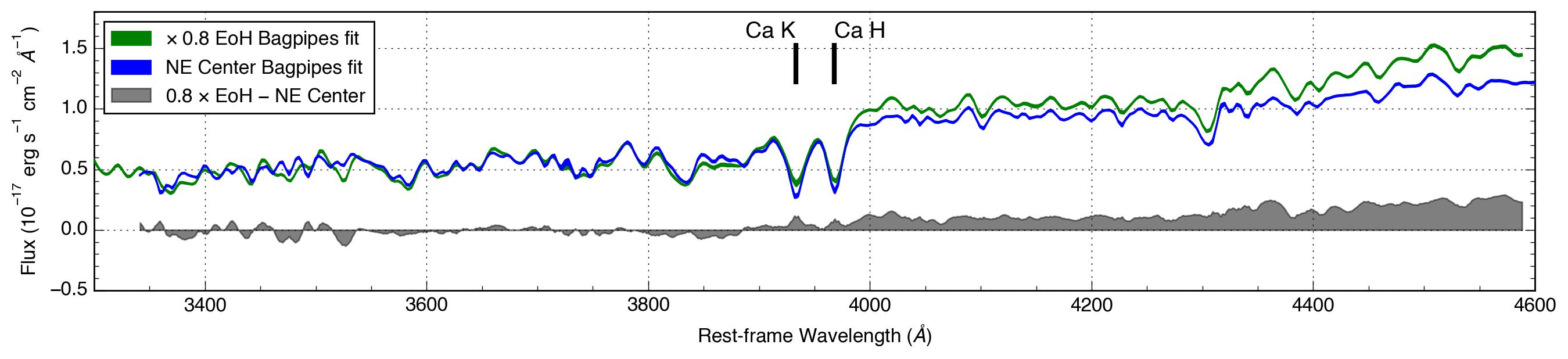}
    
    \caption{\textsc{Bagpipes} model fits to the spectra of the BCGs in the NE (\textit{Upper panel}) and main (\textit{Middle panel}) clusters. The orange model fits are generated by the \textsc{Bagpipes} posterior distribution with 1$\sigma$ error. The vertical blue bands at $\lambda\sim7586-7708$~\AA~ and $\lambda\sim6864-6945$~\AA~ are the A-band and B-band of the atmosphere absorption. The Ca H\&K ${\lambda}{\lambda}3933,3968$ absorption lines are also indicated. The best-fit model spectra for the two BCGs are compared in the bottom panel. The best-fit spectrum of the main-cluster BCG has been re-scaled by 0.8 to match the continuum level shortward of the Ca K line.}
    \label{fig: spec-analy}
\end{figure*}

\subsection{Spectral Analysis of the two BCGs}

Because of their brightnesses, the BCGs of the main cluster (EoH; Slit \#139) and the NE cluster (Slit \#130) produced high signal-to-noise Binospec spectra.  We therefore performed spectral model fitting for them by using the program Bayesian Analysis of Galaxies for Physical Inference and Parameter EStimation \citep[\textsc{Bagpipes};][]{Carnall2018}. More specifically, we used the exponentially-declining star-formation history (SFH) with the following priors:
the mass-weighted age $a/\mathrm{Gyr} \in (0.1, t_\mathrm{universe}(z_\mathrm{obs}))$, 
the exponential-decay parameter $\tau/\mathrm{Gyr} \in (0.3, 10.0)$, 
the formed stellar mass $M_\mathrm{formed}/\mathrm{M}_{\sun} \in (10^{1}, 10^{15})$, 
the metalicity $Z/\mathrm{Z}_{\sun} \in (0.1, 10.0)$, 
the absolute attenuation in V-band magnitudes $A_\mathrm{V}/\mathrm{mag} \in (0, 2)$, and 
the velocity dispersion of the stellar component in the galaxy $\sigma_\mathrm{*, vel}/\mathrm{km} ~ \mathrm{s}^{-1} \in (10, 1000)$. 
From derived quantities above after the fitting, \textsc{Bagpipes} will calculate the additional results, such as the current stellar mass $M_\mathrm{*}$. 

\input{Bagpipes}

Figure~\ref{fig: spec-analy} compares the observed spectra of the two BCGs with the best-fit spectral models from \textsc{Bagpipes}. The model fits are seen to match the observed spectra quite well, reproducing the overall continuum shapes as well as those of the Ca H\&K absorption lines.  Table~\ref{tab: spec-analy} lists the resultant best-fit parameters. The stellar velocity dispersion of the EoH BCG is measured to be $\sigma_\mathrm{*, EoH} = 340^{+20}_{-20}$\kms. Note that the error bars listed in the table only reflect the statistical uncertainties associated with the \textsc{Bagpies} fitting.  For example, the uncertainties associated with the stellar mass $M_\mathrm{*}$ are $<$\,0.1 dex, consistent with those of other studies using \textsc{Bagpipes} \citep[$\sim 0.06$ dex by][]{Man2021}, but the choice of the initial mass function alone could introduce a systematic uncertainty of $\sim 0.3$ dex \citep[e.g.,][]{Mobasher2015,Carnall2019a}.

As expected, these two BCGs are old (age\,$=$\,3.6--6.0 Gyr) and massive ($M_\mathrm{*}$\,$=$\,4.2--9.5\,$\times$\,$10^{11}$\,\Msun).  These spectroscopically derived stellar masses agree well with the photometrically derived ones from the HSC/Mizuki catalog as listed in Table~\ref{tab: spec-analy} \citep{Tanaka2015,Tanaka2018}.  Note, however, that these Mizuki stellar masses were derived with the Mizuki photometric redshifts for the BCGs, which are slightly different from our spectroscopic ones. The formation redshifts ($z_{\rm form}$) calculated from the Bagpipes-derived ages are $z\sim2$ for the main-cluster BCG and $z\sim7$ for the NE-cluster BCG, which are consistent with the range of $z_{\rm form}$ derived for quiescent galaxies in the SPT clusters in the same redshift range \citep{Khullar2022}.

Figure~\ref{fig: spec-analy} overplots the two best-fit model spectra after scaling, showing that these two BCGs have very similar spectra.  This reflects the fact that galaxy spectra do not change substantially at such old ages. Note that the 4000 \AA\ break amplitude ($D_{4000}$) is larger for the EoH BCG although its derived stellar age is younger. This is likely due to the effect of this BCG having a larger metallicity (Table~\ref{tab: spec-analy}), reflecting the well-known age-metallicity degeneracy with old stellar populations \citep[e.g.,][]{Worthey1994}.

\subsection{Red-Sequence Scatter}

High-redshift clusters are expected to show an increased scatter with its RS because member galaxies are younger and exhibit a larger color variation.  
Observationally, however, it is not clear if this is really the case with $z\sim1$ clusters.  For example, some previous studies seem to indicate that the RS is already well-established in $z\sim1$ galaxies, showing an RS color scatter that is as tight as those observed for lower-redshift clusters \citep[e.g.,][]{Menci2008, Nishizawa2018}. 

In the bottom panel of Figure~\ref{fig: c-m}, we plot the color-magnitude diagram for the $z=0.351$ cluster Abell~1489 in comparison with those of the two $z\sim0.8$ clusters presented in Section~\ref{sec: c-m}. Starting with the spectroscopically-confirmed cluster members within $R_{200}$ reported by \citet{Rines2022} and using the photometric magnitudes from the Dark Energy Camera Legacy Survey \citep[][]{Dey2019}, we have identified 68 galaxies in the RS with a scatter of $\Delta_{(g-r)}=0.062\pm0.005$ mag in the $(g-r)$ color (Figure~\ref{fig: c-m}, the bottom penel). This RS color scatter is significantly smaller (1.5--1.8\,$\times$) than $0.11\pm0.02$ with the main cluster and $0.09\pm0.02$ with the NE cluster, suggesting the possibility that the RS scatter does increase with $z\sim1$ clusters.

\section{Conclusions}
\label{sec: con}

In this paper, we used MMT/Binospec to conduct a spectroscopic survey to study two $z\sim0.8$ galaxy clusters in the ``Eye of Horus'' lens system field. Our findings are as follows:

\begin{enumerate}
    \item We have spectroscopically confirmed the existence of two galaxy clusters in the EoH field, the main cluster at $z=0.795$ (42 $z_{\rm spec}$ members) and the NE cluster (25 $z_{\rm spec}$ members) at $0.769$, corresponding to the two X-ray sources detected by {\it XMM-Newton}. 
    The two clusters have a small separation on the sky ($\sim$\,100\arcsec\,$=$\,0.75\,Mpc\,$<r_{200}$), but a velocity offset of $\sim$4300 km s$^{-1}$ suggests that this two-cluster system is likely a line-of-sight projection rather than a physically-related association (e.g., a cluster merger).
    This underscores the importance of conducting a spectroscopic follow-up for high-redshift cluster candidates because RS-based cluster selections are susceptible to such a projection effect in general.

    \item The masses of the main and NE clusters are estimated dynamically to be 8.6 and 3.3\,$\times$\,$10^{14}$\,\Msun\ with the virial theorem and 3.8 and 1.0\,$\times$\,$10^{14}$\,\Msun\ with the $\sigma_{200}$\,--\,$M_\mathrm{200}$ scaling relation. Broadly speaking, these dynamical masses agree with the X-ray-derived ones within a factor of two; However, the virial masses are systematically larger than the $\sigma_{200}$\,--\,$M_\mathrm{200}$ scaling masses by a factor of a few.

    \item The spectral analysis of the BCGs in the main and NE clusters shows that these galaxies are already old (Age\,$=$\,3.6--6.0 Gyr) and massive ($M_\mathrm{*}$\,$=$\,4.2--9.5\,$\times$\,$10^{11}$\,\Msun), similar to the BCGs in lower-redshift clusters.

    \item The color-magnitude diagrams of the main and NE clusters show an RS scatter of $\sim$0.1 mag, which appears significantly larger (1.5--1.8\,$\times$) than that of a $z=0.35$ cluster in comparison.  A more extensive comparison of spectrscopically-confirmed members in low- and high-redshift clusters is needed to make a more robust assessment.

\end{enumerate}

\medskip

\section*{acknowledgments}
    Observations reported in this work were obtained at the MMT Observatory, a joint facility of the University of Arizona and the Smithsonian Institution. We thank the MMT telescope operators and staff for their services during the observation nights, as well as our thanks to the programmers who contribute to MMT data processing pipelines and convert Binospec data to \textsc{SpecPro} readable files.

    The Hyper Suprime-Cam (HSC) collaboration includes the astronomical communities of Japan and Taiwan, and, Princeton University. The HSC instrumentation and software were developed by the National Astronomical Observatory of Japan (NAOJ), the Kavli Institute for the Physics and Mathematics of the Universe (Kavli IPMU), the University of Tokyo, the High Energy Accelerator Research Organization (KEK), the Academia Sinica Institute for Astronomy and Astrophysics in Taiwan (ASIAA), and Princeton University. Funding was contributed by the FIRST program from the Japanese Cabinet Office, the Ministry of Education, Culture, Sports, Science and Technology (MEXT), the Japan Society for the Promotion of Science (JSPS), Japan Science and Technology Agency (JST), the Toray Science Foundation, NAOJ, Kavli IPMU, KEK, ASIAA, and Princeton University. 

    This paper makes use of software developed for the Vera C. Rubin Observatory. We thank the Rubin Observatory for making their code available as free software at http://pipelines.lsst.io/.

    The Figure~\ref{fig: main} background optical image uses the data collected at the Subaru Telescope and this paper is based on data retrieved from the HSC data archive system, which is operated by the Subaru Telescope and Astronomy Data Center (ADC) at the National Astronomical Observatory of Japan (NAOJ). Data analysis was in part carried out with the cooperation of the Center for Computational Astrophysics (CfCA), NAOJ. We are honored and grateful for the opportunity to observe the Universe from Maunakea, which has cultural, historical, and natural significance in Hawaii.

    J.D. appreciates Masamune Oguri (Chiba University) for the timely and constructive comments and suggestions, Keigo Tanaka (Kanazawa University) for providing the \textit{Newton-XMM} X-ray map, Anja von der Linden (Stony Brook University) for \textsc{Bagpipes} installation advice, and Tianyu Jiao (Nio Inc.) for providing hardware and MacOS support. 
    
\facility{MMT:6.5m (BINOSPEC)}

\software{
        \textsc{IPython} \citep[][]{Perez2007}{}{}, \textsc{NumPy} \citep[][]{Harris2020}{}{}, \textsc{SciPy} \citep[][]{Virtanen2020}{}{}, \textsc{Astropy} \citep[][]{Robitaille2013}{}{},   \textsc{Matplotlib} \citep[][]{Hunter2007}{}{}, \textsc{SciencePlot} \citep[][]{SciencePlots}.
        }


\appendix

\section{Redshift Table}
\label{sec:appendix-a}

We provide the full redshift catalog including 371 sources in Table~\ref{tab: RT01}, which combines the 2019 and 2022 MMT/Binospec observations over the four fields shown in Figure~\ref{fig: main}: Slit~\#1--86 (Field A) and Slit~\#87--190 (Field B) in 2019; Slit~\#1C--96C (Field C) and Slit~\#97C--181C (Field D) in 2022.
In the ``Notes'' column, the reasons are provided for those slits that did not produce secure redshifts.

\input{RT-Appendix-1}

\section{Galaxy Spectra}
\label{sec:appendix-b}

This appendix displays the 1D spectra and 2D spectral images.  The $x$-axis is the observed wavelength in \AA, and
the $y$-axis is flux in counts.
The vertical purple stripes on the plots indicate the A-band and B-band of the atmospheric absorption at $\lambda\sim7586-7708$~\AA and $\lambda\sim6864-6945$~\AA. On the left edge of the 2D spectra, we use dark blue arrows to show the locations of extraction/trace lines. In the 1D spectral plots, the [\ion{O}{II}] ${\lambda\lambda}$3727 emission line is indicated by the green dashed line and Ca H\&K absorption lines by the orange solid lines. Additionally, [\ion{O}{III}] ${\lambda\lambda}$5007,4959, \ion{H}{$\alpha$} ${\lambda}$6563, \ion{H}{$\beta$} ${\lambda}$4861, \ion{H}{$\gamma$} ${\lambda}$4340 are shown in some cases (e.g., lower-redshift interlopers).  
The complete figure set (47 images) will also be provided in the online ApJ journal.

\input{show_2dspec_to_arXiv}



\bibliography{__abbrlibrary}{}
\bibliographystyle{aasjournal}



\end{document}

%% file: Members-table.tex
\begin{deluxetable*}{rlccrrrclccrrr}[!t]
    \label{tab: RT-new}
    \tabletypesize{\footnotesize}
    \tablecaption{Member Galaxies of the Main and NE Clusters}
    \tablehead{\colhead{} & \multicolumn6c{NE BCG: Slit \#130, ($\alpha$, $\delta$)\,$=$\,(216.22772$^{\circ}$, -0.87414$^{\circ}$)} & & \multicolumn6c{Main BCG: Slit \#139, ($\alpha$, $\delta$)\,$=$\, (216.20415$^{\circ}$,\,$-0.88935^{\circ}$) } \\  
    \cline{2-7} \cline{9-14}
    \colhead{} & \colhead{Slit \#} &  \colhead{$z$} & \colhead{$|v|$}      & \colhead{$\Delta$RA} &  \colhead{$\Delta$DEC} &  \colhead{$r$} & & \colhead{Slit \#} &  \colhead{$z$} & \colhead{$|v|$}      & \colhead{$\Delta$RA} &  \colhead{$\Delta$DEC} &  \colhead{$r$}  \\
     \colhead{} & \colhead{} &  \colhead{} & \colhead{(\kms)}      & \colhead{($''$)} &  \colhead{($''$)} &  \colhead{($''$)} & & \colhead{} &  \colhead{} & \colhead{(\kms)}      & \colhead{($''$)} &  \colhead{($''$)} &  \colhead{($''$)}}
    \startdata
             1 & 130 $^\bullet$ & $0.769$ & $   0$ & $   0$ & $   0$ & $   0$ & & 139 $^\bullet$ & $0.795$ & $   0$ & $   0$ & $   0$ & $   0$\\
             2 & 26C $^\bullet$ & $0.770$ & $ 170$ & $   1$ & $  10$ & $  10$ & & 141 $^\bullet$ & $0.801$ & $1003$ & $   8$ & $  -9$ & $  12$\\
             3 & 23C $^\bullet$ & $0.764$ & $ 848$ & $  14$ & $  41$ & $  43$ & & 135 $^\bullet$ & $0.803$ & $1337$ & $  -5$ & $  17$ & $  17$\\
             4 & 137 $^\bullet$ & $0.768$ & $ 170$ & $  -6$ & $ -43$ & $  44$ & & 39C $^\bullet$ & $0.800$ & $ 836$ & $ -16$ & $  13$ & $  21$\\
             5 & 29C $^\bullet$ & $0.764$ & $ 848$ & $   2$ & $ -48$ & $  48$ & & 42C $^\bullet$ & $0.801$ & $1003$ & $ -43$ & $  12$ & $  44$\\
             6 & 20C $^\bullet$ & $0.760$ & $1526$ & $  44$ & $  28$ & $  52$ & & 147 $^\bullet$ & $0.790$ & $ 836$ & $ -25$ & $ -51$ & $  57$\\
             7 & 28C $^\bullet$ & $0.765$ & $ 678$ & $ -34$ & $  51$ & $  62$ & & 148 $^\bullet$ & $0.789$ & $1003$ & $ -32$ & $ -59$ & $  67$\\
             8 & 33C $^\bullet$ & $0.766$ & $ 509$ & $ -63$ & $  -6$ & $  64$ & & 46C $^\bullet$ & $0.786$ & $1504$ & $ -55$ & $ -44$ & $  71$\\
             9 & 30C $^\bullet$ & $0.770$ & $ 170$ & $ -52$ & $  49$ & $  72$ & & 128 $^\bullet$ & $0.799$ & $ 669$ & $ -44$ & $  58$ & $  73$\\
            10 & 143 $^\bullet$ & $0.771$ & $ 339$ & $ -33$ & $ -69$ & $  77$ & & 47C $^\bullet$ & $0.793$ & $ 334$ & $ -58$ & $ -62$ & $  85$\\
            11 & 36C $^\bullet$ & $0.767$ & $ 339$ & $ -83$ & $ -21$ & $  85$ & & 122 $^\bullet$ & $0.801$ & $1003$ & $  15$ & $ 101$ & $ 102$\\
            12 & 126 $^\bullet$ & $0.767$ & $ 339$ & $-101$ & $  18$ & $ 102$ & & 151 $^\bullet$ & $0.797$ & $ 334$ & $ 103$ & $ -80$ & $ 130$\\
            13 & 146 $^\bullet$ & $0.766$ & $ 509$ & $  61$ & $ -87$ & $ 107$ & & 117 $^\bullet$ & $0.799$ & $ 669$ & $  14$ & $ 136$ & $ 137$\\
            14 & 16C $^\bullet$ & $0.770$ & $ 170$ & $ 116$ & $ -38$ & $ 122$ & & 25C $^\bullet$ & $0.793$ & $ 334$ & $  72$ & $ 121$ & $ 141$\\
            15 & 14C $^\circ$   & $0.770$ & $ 170$ & $ 138$ & $ -46$ & $ 146$ & & 150 $^\bullet$ & $0.797$ & $ 334$ & $-122$ & $ -78$ & $ 144$\\ \cline{2-7}
            16 & 37C            & $0.770$ & $ 170$ & $-148$ & $ 100$ & $ 179$ & & 160 $^\bullet$ & $0.791$ & $ 669$ & $  92$ & $-138$ & $ 165$\\
            17 & 51C            & $0.764$ & $ 848$ & $-165$ & $-196$ & $ 256$ & & 115 $^\circ$   & $0.797$ & $ 334$ & $  83$ & $ 149$ & $ 170$\\
            18 & 171            & $0.765$ & $ 678$ & $ -41$ & $-289$ & $ 292$ & & 19C $^\circ$   & $0.784$ & $1838$ & $ 127$ & $ 120$ & $ 175$\\ \cline{9-14}
            19 & 189            & $0.770$ & $ 170$ & $ -21$ & $-486$ & $ 487$ & & 170            & $0.798$ & $ 501$ & $  55$ & $-221$ & $ 228$\\
            20 & 62             & $0.766$ & $ 509$ & $ 483$ & $-243$ & $ 541$ & & 108            & $0.796$ & $ 167$ & $ 111$ & $ 210$ & $ 238$\\
            21 & 37             & $0.765$ & $ 678$ & $ 662$ & $  24$ & $ 662$ & & 21C            & $0.795$ & $   0$ & $  43$ & $ 257$ & $ 261$\\
            22 & 100C           & $0.770$ & $ 170$ & $ -97$ & $ 673$ & $ 680$ & & 5C             & $0.801$ & $1003$ & $ 240$ & $ 168$ & $ 293$\\
            23 & 84             & $0.763$ & $1018$ & $ 532$ & $-471$ & $ 710$ & & 7C             & $0.798$ & $ 501$ & $ 198$ & $ 221$ & $ 297$\\
            24 & 59             & $0.761$ & $1357$ & $ 694$ & $-211$ & $ 726$ & & 178            & $0.791$ & $ 669$ & $  25$ & $-296$ & $ 297$\\
            25 & 75             & $0.766$ & $ 509$ & $ 764$ & $-362$ & $ 845$ & & 96             & $0.795$ & $   0$ & $ -26$ & $ 328$ & $ 329$\\
            26 & -              &    -    &     -  &     -  &     -  &     -  & & 95             & $0.791$ & $ 669$ & $ -42$ & $ 342$ & $ 345$\\
            27 & -              &    -    &     -  &     -  &     -  &     -  & & 184            & $0.795$ & $   0$ & $ -79$ & $-366$ & $ 374$\\
            28 & -              &    -    &     -  &     -  &     -  &     -  & & 71C            & $0.799$ & $ 669$ & $-386$ & $ 110$ & $ 402$\\
            29 & -              &    -    &     -  &     -  &     -  &     -  & & 79C            & $0.794$ & $ 167$ & $-417$ & $  -8$ & $ 417$\\
            30 & -              &    -    &     -  &     -  &     -  &     -  & & 88             & $0.794$ & $ 167$ & $ 225$ & $ 425$ & $ 481$\\
            31 & -              &    -    &     -  &     -  &     -  &     -  & & 94C            & $0.795$ & $   0$ & $-570$ & $ -36$ & $ 572$\\
            32 & -              &    -    &     -  &     -  &     -  &     -  & & 129C           & $0.791$ & $ 669$ & $-259$ & $ 519$ & $ 580$\\
            33 & -              &    -    &     -  &     -  &     -  &     -  & & 109C           & $0.791$ & $ 669$ & $ -95$ & $ 701$ & $ 707$\\
            34 & -              &    -    &     -  &     -  &     -  &     -  & & 24             & $0.794$ & $ 167$ & $ 679$ & $ 221$ & $ 714$\\
            35 & -              &    -    &     -  &     -  &     -  &     -  & & 123C           & $0.792$ & $ 501$ & $-268$ & $ 689$ & $ 739$\\
            36 & -              &    -    &     -  &     -  &     -  &     -  & & 20             & $0.803$ & $1337$ & $ 698$ & $ 266$ & $ 747$\\
            37 & -              &    -    &     -  &     -  &     -  &     -  & & 57             & $0.792$ & $ 501$ & $ 747$ & $-131$ & $ 758$\\
            38 & -              &    -    &     -  &     -  &     -  &     -  & & 83             & $0.795$ & $   0$ & $ 643$ & $-404$ & $ 759$\\
            39 & -              &    -    &     -  &     -  &     -  &     -  & & 34             & $0.793$ & $ 334$ & $ 769$ & $ 116$ & $ 777$\\
            40 & -              &    -    &     -  &     -  &     -  &     -  & & 10             & $0.795$ & $   0$ & $ 694$ & $ 361$ & $ 782$\\
            41 & -              &    -    &     -  &     -  &     -  &     -  & & 124C           & $0.791$ & $ 669$ & $-358$ & $ 860$ & $ 932$\\
            42 & -              &    -    &     -  &     -  &     -  &     -  & & 146C           & $0.793$ & $ 334$ & $-569$ & $ 802$ & $ 984$
    \enddata
    \tablecomments{
            $\bullet$ $r_{200}$--enclosed cluster members ($r \leq r_{200}$): $r_{200,\text{NE}}=127''=940$ kpc, $r_{200,\text{main}}=168''=1259$ kpc;
           $\circ$ Members outside $r_{200}$ but enclosed by the virialized core radii: $r_{200} < r \leq r_\mathrm{cl} (= 1.2 \times r_{200})$. }
\end{deluxetable*}

%% file: Centroids.tex
\begin{deluxetable*}{rccccccc}
        \label{tab: centroids}
        \tablewidth{0pt}
        \tablecaption{Offsets between the BCGs and Cluster-Member/X-ray Centroids}
	\tablehead{\colhead{} & \multicolumn3c{From Main BCG: ($\alpha$,$\delta$)$=$(216.20415$^{\circ}$,$-$0.88935$^{\circ}$)}  & & \multicolumn3c{From NE BCG: ($\alpha$,$\delta$)$=$(216.22772$^{\circ}$,$-$0.87414$^{\circ}$)}\\
    \cline{2-4} \cline{6-8}
              \colhead{} & \colhead{$\Delta$RA} & \colhead{$\Delta$Dec} & \colhead{$\Delta r$} & &\colhead{$\Delta$RA} & \colhead{$\Delta$Dec} & \colhead{$\Delta r$}\\
       \colhead{} & \colhead{(\arcsec)} & \colhead{(\arcsec)} & \colhead{(\arcsec)} & &\colhead{(\arcsec)} & \colhead{(\arcsec)} & \colhead{(\arcsec)}}
    \startdata
            \hline
            Member-galaxy centroids                     
            & $+6.4\pm15.1$  
            & $+11.4\pm19.6$ 
            & $13.1\pm13.3$   &
            & $+0.3\pm16.9$ 
            & $-10.8\pm10.9$ 
            & $10.8\pm9.9$     \\
            X-ray peaks (\citetalias{Tanaka2020})
            & $-1.0\pm{1.0}$  
            & $-3.6\pm{1.0}$ 
            & $3.7\pm{1.5}$  &       
            & $+2.3\pm{0.4}$ 
            & $-2.7\pm{0.5}$ 
            & $3.5\pm{0.8}$    \\
    \enddata
    \tablecomments{The main/NE cluster centroids are at ($\alpha$, $\delta$)\,$=$\,(216.20592$^{\circ}$, -0.88618$^{\circ}$) and ($\alpha$, $\delta$)\,$=$\,$(216.22780^{\circ}, -0.87715^{\circ})$.}
\end{deluxetable*}

%% file: Masses.tex
\begin{deluxetable*}{rccccccccccc}[!t]
	\label{tab: masses}
	\tablecaption{The Main and NE Cluster Dynamical Masses}
	\tablehead{\colhead{} & \multicolumn5c{Virial mass ($<r_\mathrm{cl}$ $=1.2 \times r_{200}$)} &  &\multicolumn4c{Scaling-relation mass ($<r_{200}$)} & \colhead{X-ray (\citetalias{Tanaka2020})} \\
         \cline{2-6} \cline{8-11}
         \colhead{} & \colhead{$N_\mathrm{mem}$} & \colhead{$R_\mathrm{cl}$} & \colhead{$R_\mathrm{PV}$} & \colhead{$\sigma_\mathrm{P}$} & \colhead{$M_\mathrm{V}$} & & \colhead{$N_\mathrm{mem}$} & \colhead{$R_{200}$} & \colhead{$\sigma_{200}$} & \colhead{$M_\mathrm{200}$}       & \colhead{$M_\mathrm{200}$} \\
         \colhead{} & \colhead{} & \colhead{(Mpc)}           & \colhead{(Mpc)}  & \colhead{(\kms)} & \colhead{$(10^{14}$ M$_{\sun})$} & & \colhead{} & \colhead{(Mpc)} & \colhead{(\kms)} & \colhead{$(10^{14}$ M$_{\sun})$} & \colhead{$(10^{14}$ M$_{\sun})$} }
		\startdata
        Main    & 18 & 1.5 & 1.2 & $930\pm120$  & $8.6\pm1.5$ & & 16 & 1.3 & $850\pm110$  & $3.8\pm1.4$ & $5.6^{+1.3}_{-0.8}$ \\
        NE      & 15 & 1.1 & 1.0 & $520\pm~~90$ & $3.3\pm0.1$ & & 14 & 0.9& $520\pm100$ & $1.0\pm0.5$ & $2.2^{+0.5}_{-0.3}$\\
		\enddata
\end{deluxetable*}

%% file: Bagpipes.tex
\begin{deluxetable*}{rcccccccc}
	\label{tab: spec-analy}
	\tablecaption{BCG Physical Properties from the \textsc{Bagpipes} Spectral-Model Fitting.}
	\tablehead{\colhead{} & \multicolumn7c{\textsc{Bagpipes}} & \colhead{HSC/Mizuki} \\
         \cline{2-8}
              \colhead{}       
              & \colhead{Age $a/\mathrm{Gyr}$ }
              & \colhead{$\tau/\mathrm{Gyr}$}
              & \colhead{$\log{\left(M_\mathrm{formed}/\mathrm{M}_{\sun}\right)}$}
              & \colhead{$\log{\left(Z/\mathrm{Z}_{\sun}\right)}$}
              & \colhead{$A_\mathrm{V}/\mathrm{mag}$}
              & \colhead{$\sigma_\mathrm{*,vel}/$\kms}
              & \colhead{$\log{\left(M_*/\mathrm{M}_{\sun}\right)}$} 
              & \colhead{$\log{\left(M_\mathrm{*, Mizuki}/\mathrm{M}_{\sun}\right)}$} }   
\startdata
            EoH       
            & $3.6^{+0.4}_{-0.4}$ 
            & $0.6^{+0.1}_{-0.1}$ 
            & $11.98^{+0.04}_{-0.04}$
            & $0.8^{+0.1}_{-0.1}$
            & $0.08^{+0.07}_{-0.05}$
            & $340^{+20}_{-20}$ 
            & $11.61^{+0.04}_{-0.05}$   
            & $11.53^{+0.04}_{-0.05}$ ($z_{\rm mizuki}$\,$=$\,0.73) \\
            NE 
            & $6.0^{+0.6}_{-0.8}$ 
            & $0.5^{+0.2}_{-0.1}$ 
            & $11.62^{+0.04}_{-0.04}$
            & $0.4^{+0.1}_{-0.1}$
            & $0.07^{+0.09}_{-0.05}$
            & $310^{+60}_{-60}$ 
            & $11.30^{+0.01}_{-0.02}$   
            & $11.29^{+0.04}_{-0.12}$ ($z_{\rm mizuki}$\,$=$\,0.75)        \\
            \enddata
\end{deluxetable*}

%% file: RT-Appendix-1.tex
\startlongtable
\begin{deluxetable*}{rcccccl}
\tablecaption{Redshift Table for the 2019 and 2022 MMT/Binospec Observations}
\label{tab: RT01}
\tablehead{
\colhead{Slit} & \colhead{Redshift} & \colhead{RA} & \colhead{DEC} & \colhead{Lines} & \colhead{Confidence} & \colhead{Notes} 
}
\startdata
      1 & -     & 216.38562 & -0.76420 & -                               & 1 & \\
      2 & 0.813 & 216.38863 & -0.76642 & Ca H\&K                         & 3 & \\
      3 & -     & 216.44077 & -0.76689 & -                               & 1 & noise\\
      4 & 0.858 & 216.43033 & -0.77128 & [\ion{O}{II}] ${\lambda}$3727 & 3 & \\
      5 & 0.864 & 216.33951 & -0.77705 & Ca H\&K                         & 3 & \\
      6 & 0.759 & 216.39340 & -0.77723 & [\ion{O}{II}] ${\lambda}$3727 & 3 & \\
      7 & 0.745 & 216.43243 & -0.77912 & [\ion{O}{II}] ${\lambda}$3727 & 3 & \\
      8 & -     & 216.39641 & -0.78259 & -                               & 1 & continue (no spectrum feature)\\
      9 & -     & 216.39623 & -0.78598 & -                               & 1 & faint continue (no spectrum feature)\\
     10 & 0.795 & 216.39694 & -0.78897 & [\ion{O}{II}] ${\lambda}$3727 & 3 & \\
     11 & -     & 216.39259 & -0.79164 & -                               & 1 & noise\\
     12 & -     & 216.39078 & -0.79438 & -                               & 1 & point source, [\ion{O}{II}] ${\lambda}$3727-overlapped\\
     13 & 0.811 & 216.39064 & -0.79803 & [\ion{O}{II}] ${\lambda}$3727 & 3 & \\
     14 & 0.858 & 216.38635 & -0.80063 & Ca H\&K                         & 3 & \\
     15 & 1.070 & 216.41468 & -0.80203 & [\ion{O}{II}] ${\lambda}$3727 & 3 & \\
     16 & 0.832 & 216.37693 & -0.80564 & [\ion{O}{II}] ${\lambda}$3727 & 3 & \\
     17 & 0.818 & 216.39775 & -0.80834 & [\ion{O}{II}] ${\lambda}$3727 & 3 & \\
     18 & 0.863 & 216.40519 & -0.81001 & [\ion{O}{II}] ${\lambda}$3727 & 3 & \\
     19 & 0.817 & 216.39466 & -0.81343 & [\ion{O}{II}] ${\lambda}$3727 & 3 & \\
     20 & 0.803 & 216.39791 & -0.81548 & [\ion{O}{II}] ${\lambda}$3727 & 3 & \\
     21 & -     & 216.39976 & -0.81736 & -                               & 1 & noise\\
     22 & -     & 216.40978 & -0.82074 & -                               & 1 & faint continue (no spectrum feature)\\
     23 & 0.817 & 216.35927 & -0.82625 & Ca H\&K                         & 3 & \\
     24 & 0.794 & 216.39286 & -0.82796 & [\ion{O}{II}] ${\lambda}$3727 & 3 & \\
     25 & -     & 216.35929 & -0.83292 & -                               & 3 & M6 star - monitor atm\\
     26 & -     & 216.41304 & -0.83380 & -                               & 1 & point source, [\ion{O}{II}] ${\lambda}$3727-overlapped\\
     27 & -     & 216.37556 & -0.83763 & -                               & 1 & noise, as faint as 048\\
     28 & -     & 216.36190 & -0.84013 & -                               & 1 & [\ion{O}{II}] ${\lambda}$3727-overlapped\\
     29 & 0.743 & 216.41831 & -0.84203 & Ca H\&K                         & 3 & \\
     30 & 0.796 & 216.39969 & -0.84611 & [\ion{O}{II}] ${\lambda}$3727 & 1 & \\
     31 & 0.823 & 216.36726 & -0.85097 & [\ion{O}{II}] ${\lambda}$3727 & 3 & \\
     32 & 0.976 & 216.39745 & -0.85292 & [\ion{O}{II}] ${\lambda}$3727 & 3 & \\
     33 & -     & 216.38826 & -0.85619 & -                               & 1 & [\ion{O}{II}] ${\lambda}$3727-overlapped\\
     34 & 0.793 & 216.41769 & -0.85725 & [\ion{O}{II}] ${\lambda}$3727 & 3 & \\
     35 & -     & 216.41113 & -0.86031 & -                               & 1 & noise\\
     36 & 0.855 & 216.38017 & -0.86529 & [\ion{O}{II}] ${\lambda}$3727 & 3 & \\
     37 & 0.765 & 216.41153 & -0.86742 & [\ion{O}{II}] ${\lambda}$3727 & 3 & \\
     38 & 0.942 & 216.38731 & -0.87103 & [\ion{O}{II}] ${\lambda}$3727 & 1 & \\
     39 & 0.870 & 216.40269 & -0.87345 & [\ion{O}{II}] ${\lambda}$3727 & 3 & two galaxies, redshift for lower one\\
     40 & 0.848 & 216.35439 & -0.87967 & Ca H\&K                         & 1 & first type 0.858, then adjust\\
     41 & -     & 216.42898 & -0.88117 & -                               & 1 & noise\\
     42 & -     & 216.42185 & -0.88452 & -                               & 1 & noise\\
     43 & -     & 216.43421 & -0.88606 & -                               & 1 & noise\\
     44 & 0.721 & 216.42609 & -0.88873 & [\ion{O}{II}] ${\lambda}$3727 & 3 & \\
     45 & -     & 216.36506 & -0.89389 & -                               & 1 & noise\\
     46 & -     & 216.42594 & -0.89454 & -                               & 1 & noise\\
     47 & -     & 216.40032 & -0.89771 & -                               & 1 & continue (no spectrum feature)\\
     48 & -     & 216.40650 & -0.90146 & -                               & 1 & noise, as faint as 027\\
     49 & 1.028 & 216.40205 & -0.90398 & [\ion{O}{II}] ${\lambda}$3727 & 3 & \\
     50 & 0.894 & 216.37999 & -0.90689 & [\ion{O}{II}] ${\lambda}$3727 & 3 & \\
     51 & 1.030 & 216.37827 & -0.90895 & [\ion{O}{II}] ${\lambda}$3727 & 1 & \\
     52 & -     & 216.40155 & -0.91211 & -                               & 1 & noise\\
     53 & -     & 216.40548 & -0.91505 & -                               & 1 & faint continue (no spectrum feature)\\
     54 & -     & 216.43441 & -0.91627 & -                               & 1 & faint continue (no spectrum feature)\\
     55 & 0.754 & 216.43136 & -0.91855 & [\ion{O}{II}] ${\lambda}$3727 & 3 & \\
     56 & -     & 216.40117 & -0.92251 & -                               & 1 & noise\\
     57 & 0.792 & 216.41158 & -0.92571 & [\ion{O}{II}] ${\lambda}$3727 & 3 & \\
     58 & -     & 216.33630 & -0.93211 & -                               & 1 & noise\\
     59 & 0.761 & 216.42056 & -0.93288 & [\ion{O}{II}] ${\lambda}$3727 & 3 & \\
     60 & 0.919 & 216.42847 & -0.93454 & [\ion{O}{II}] ${\lambda}$3727 & 3 & \\
     61 & 0.710 & 216.39817 & -0.93757 & Ca H\&K                         & 1 & \\
     62 & 0.766 & 216.36198 & -0.94169 & [\ion{O}{II}] ${\lambda}$3727 & 3 & \\
     63 & 0.851 & 216.39249 & -0.94331 & [\ion{O}{II}] ${\lambda}$3727 & 3 & \\
     64 & -     & 216.37143 & -0.94660 & -                               & 1 & faint continue (no spectrum feature)\\
     65 & -     & 216.43833 & -0.94805 & -                               & 1 & faint continue (no spectrum feature)\\
     66 & 0.868 & 216.39797 & -0.95193 & [\ion{O}{II}] ${\lambda}$3727 & 3 & \\
     67 & -     & 216.37452 & -0.95666 & -                               & 1 & noise\\
     68 & -     & 216.41021 & -0.95747 & -                               & 1 & faint continue (no spectrum feature)\\
     69 & 0.713 & 216.37940 & -0.96138 & [\ion{O}{II}] ${\lambda}$3727 & 3 & two galaxies, redshift for middle\\
     70 & -     & 216.40184 & -0.96304 & -                               & 1 & faint continue (no spectrum feature)\\
     71 & -     & 216.40042 & -0.96664 & -                               & 1 & faint continue (no spectrum feature)\\
     72 & 0.902 & 216.40058 & -0.96880 & [\ion{O}{II}] ${\lambda}$3727 & 3 & \\
     73 & -     & 216.42254 & -0.97060 & -                               & 1 & $z=$0.894 if [\ion{O}{II}] ${\lambda}$3727 overlapped\\
     74 & -     & 216.37381 & -0.97441 & -                               & 1 & $z=$0.986 if CaK,CaH overlapped\\
     75 & 0.766 & 216.43996 & -0.97465 & [\ion{O}{II}] ${\lambda}$3727 & 3 & \\
     76 & -     & 216.39145 & -0.97886 & -                               & 1 & $z=$0.904 if [\ion{O}{II}] ${\lambda}$3727 overlapped\\
     77 & -     & 216.43429 & -0.98117 & -                               & 1 & $z=$0.690 if [\ion{O}{II}] ${\lambda}$3727 overlapped\\
     78 & 0.981 & 216.45543 & -0.98347 & [\ion{O}{II}] ${\lambda}$3727 & 3 & \\
     79 & 0.835 & 216.43017 & -0.98669 & [\ion{O}{II}] ${\lambda}$3727 & 3 & \\
     80 & 0.745 & 216.39819 & -0.99283 & [\ion{O}{II}] ${\lambda}$3727 & 3 & \\
     81 & -     & 216.40618 & -0.99636 & -                               & 1 & noise\\
     82 & -     & 216.39960 & -0.99873 & -                               & 1 & $z=$0.389 if Mg-Na\\
     83 & 0.795 & 216.38280 & -1.00149 & [\ion{O}{II}] ${\lambda}$3727 & 3 & \\
     84 & 0.763 & 216.37536 & -1.00503 & [\ion{O}{II}] ${\lambda}$3727 & 3 & \\
     85 & -     & 216.38575 & -1.00944 & -                               & 1 & cannot tell\\
     86 & 0.736 & 216.38540 & -1.01161 & [\ion{O}{II}] ${\lambda}$3727 & 3 & \\
     \hline
      87 & -     & 216.15565 & -0.77128 & - & 1 & cannot tell\\
      88 & 0.794 & 216.26665 & -0.77135 & Ca H\&K & 3 & \\
      89 & -     & 216.17817 & -0.77663 & - & 1 & noise\\
      90 & 0.888 & 216.19541 & -0.77813 & Ca H\&K,G & 3 & \\
      91 & 0.888 & 216.20494 & -0.78072 & [\ion{O}{II}] ${\lambda}$3727 & 3 & \\
      92 & 0.891 & 216.20558 & -0.78467 & Ca H\&K & 3 & \\
      93 & 0.889 & 216.20434 & -0.78738 & [\ion{O}{II}] ${\lambda}$3727 & 3 & \\
      94 & 0.718 & 216.18694 & -0.79120 & [\ion{O}{II}] ${\lambda}$3727 & 3 & \\
      95 & 0.791 & 216.19260 & -0.79432 & Ca H\&K & 3 & \\
      96 & 0.795 & 216.19694 & -0.79822 & [\ion{O}{II}] ${\lambda}$3727 & 3 & \\
      97 & 0.893 & 216.20984 & -0.80107 & [\ion{O}{II}] ${\lambda}$3727 & 3 & \\
      98 & -     & 216.17398 & -0.80534 & - & 1 & noise\\
      99 & -     & 216.20353 & -0.80667 & - & 1 & noise\\
     100 & 0.823 & 216.20321 & -0.80868 & Ca H\&K & 1 & suspicious spot above continuum\\
     101 & 0.866 & 216.19738 & -0.81166 & [\ion{O}{II}] ${\lambda}$3727 & 3 & \\
     102 & -     & 216.26411 & -0.81330 & - & 1 & noise\\
     103 & -     & 216.21523 & -0.81777 & - & 1 & noise\\
     104 & 0.886 & 216.16421 & -0.82239 & [\ion{O}{II}] ${\lambda}$3727 & 3 & \\
     105 & 0.866 & 216.21594 & -0.82295 & [\ion{O}{II}] ${\lambda}$3727 & 3 & \\
     106 & -     & 216.15638 & -0.82810 & - & 1 & $z=$0.885?, cannot tell\\
     107 & 0.728 & 216.20069 & -0.83015 & [\ion{O}{II}] ${\lambda}$3727 & 3 & \\
     108 & 0.796 & 216.23497 & -0.83090 & Ca H\&K & 3 & \\
     109 & -     & 216.14556 & -0.83733 & - & 1 & noise\\
     110 & 1.159 & 216.18907 & -0.83791 & [\ion{O}{II}] ${\lambda}$3727 & 3 & \\
     111 & 0.752 & 216.22337 & -0.84014 & [\ion{O}{II}] ${\lambda}$3727 & 3 & \\
     112 & 0.922 & 216.22195 & -0.84322 & Ca H\&K,G & 1 & \\
     113 & -     & 216.22195 & -0.84322 & - & 1 & faint continue (no spectrum feature)\\
     114 & 0.727 & 216.21412 & -0.84580 & Ca H\&K & 3 & \\
     115 & 0.797 & 216.22715 & -0.84796 & Ca H\&K & 3 & \\
     116 & 0.796 & 216.22715 & -0.84796 & Ca H\&K & 3 & \\
     117 & 0.799 & 216.20813 & -0.85164 & Ca H\&K & 3 & \\
     118 & -     & 216.24958 & -0.85213 & - & 1 & faint continue (no spectrum feature)\\
     119 & 0.716 & 216.19024 & -0.85687 & Ca H\&K & 3 & \\
     120 & 0.815 & 216.24089 & -0.85787 & Ca H\&K & 3 & \\
     121 & 0.814 & 216.24089 & -0.85787 & Ca H\&K & 3 & \\
     122 & 0.801 & 216.20820 & -0.86123 & Ca H\&K & 3 & \\
     123 & 0.801 & 216.20820 & -0.86123 & Ca H\&K & 3 & \\
     124 & -     & 216.20860 & -0.86411 & - & 1 & \\
     125 & 0.757 & 216.23533 & -0.86596 & Ca H\&K,G & 3 & \\
     126 & 0.767 & 216.19977 & -0.86907 & Ca H\&K,G & 3 & \\
     127 & 0.767 & 216.19977 & -0.86907 & Ca H\&K,G & 3 & \\
     128 & 0.799 & 216.19187 & -0.87327 & Ca H\&K,G & 3 & \\
     129 & 0.799 & 216.19187 & -0.87327 & Ca H\&K,G & 3 & \\
     130 & 0.769 & 216.22772 & -0.87414 & Ca H\&K & 3 & \\
     131 & 0.743 & 216.19654 & -0.87817 & G & 3 & \\
     132 & 0.758 & 216.19654 & -0.87817 & - & 1 & \\
     133 & 0.759 & 216.19869 & -0.88085 & Ca H\&K & 3 & \\
     134 & 0.942 & 216.20126 & -0.88268 & [\ion{O}{II}] ${\lambda}$3727 & 2 & \\
     135 & 0.803 & 216.20265 & -0.88476 & Ca H\&K,G & 3 & \\
     136 & 0.802 & 216.20265 & -0.88476 & Ca H\&K,G & 3 & \\
     137 & 0.768 & 216.22598 & -0.88614 & Ca H\&K & 3 & \\
     138 & 0.768 & 216.22598 & -0.88614 & Ca H\&K & 3 & \\
     139 & 0.795 & 216.20415 & -0.88935 & Ca H\&K & 3 & \\
     140 & 0.795 & 216.20415 & -0.88935 & Ca H\&K & 3 & \\
     141 & 0.801 & 216.20628 & -0.89187 & Ca H\&K & 3 & \\
     142 & 0.801 & 216.20628 & -0.89187 & Ca H\&K & 3 & \\
     143 & 0.771 & 216.21853 & -0.89336 & Ca H\&K,G & 3 & \\
     144 & -     & 216.20257 & -0.89612 & - & 1 & faint continue (no spectrum feature)\\
     145 & -     & 216.20257 & -0.89612 & - & 1 & two galaxies\\
     146 & 0.766 & 216.24479 & -0.89843 & Ca H\&K & 3 & \\
     147 & 0.790 & 216.19721 & -0.90350 & Ca H\&K & 3 & \\
     148 & 0.789 & 216.19537 & -0.90580 & Ca H\&K,G & 3 & \\
     149 & -     & 216.27081 & -0.90572 & - & 1 & noise\\
     150 & 0.797 & 216.17039 & -0.91103 & Ca H\&K,G & 3 & \\
     151 & 0.797 & 216.23269 & -0.91152 & Ca H\&K,G & 3 & \\
     152 & -     & 216.23269 & -0.91152 & - & 1 & noise\\
     153 & -     & 216.21902 & -0.91517 & - & 1 & noise\\
     154 & 0.886 & 216.19092 & -0.91847 & [\ion{O}{II}] ${\lambda}$3727 & 3 & \\
     155 & -     & 216.17814 & -0.92103 & - & 1 & noise\\
     156 & -     & 216.17814 & -0.92103 & - & 1 & noise\\
     157 & -     & 216.19165 & -0.92317 & - & 1 & noise\\
     158 & -     & 216.19165 & -0.92317 & - & 1 & noise\\
     159 & -     & 216.20553 & -0.92485 & - & 1 & noise\\
     160 & 0.791 & 216.22966 & -0.92755 & Ca H\&K,G & 3 & \\
     161 & 0.791 & 216.22966 & -0.92755 & Ca H\&K & 3 & \\
     162 & -     & 216.19424 & -0.93132 & - & 1 & noise\\
     163 & -     & 216.16414 & -0.93462 & - & 1 & noise\\
     164 & 0.894 & 216.20880 & -0.93560 & [\ion{O}{II}] ${\lambda}$3727 & 3 & \\
     165 & -     & 216.15486 & -0.94014 & - & 1 & noise\\
     166 & -     & 216.18801 & -0.94266 & - & 1 & noise\\
     167 & 0.886 & 216.18882 & -0.94494 & [\ion{O}{II}] ${\lambda}$3727 & 3 & \\
     168 & 0.752 & 216.17984 & -0.94817 & G & 1 & \\
     169 & 0.889 & 216.18723 & -0.94986 & Ca H\&K,G & 3 & \\
     170 & 0.798 & 216.21948 & -0.95084 & [\ion{O}{II}] ${\lambda}$3727 & 3 & \\
     171 & 0.765 & 216.21633 & -0.95445 & [\ion{O}{II}] ${\lambda}$3727 & 3 & \\
     172 & 0.920 & 216.18011 & -0.95817 & - & 2 & several absorption lines with continue\\
     173 & 0.889 & 216.19767 & -0.96103 & [\ion{O}{II}] ${\lambda}$3727 & 3 & \\
     174 & -     & 216.17484 & -0.96385 & - & 1 & noise\\
     175 & 0.887 & 216.22087 & -0.96487 & [\ion{O}{II}] ${\lambda}$3727 & 3 & \\
     176 & 0.909 & 216.23340 & -0.96672 & [\ion{O}{II}] ${\lambda}$3727 & 3 & \\
     177 & -     & 216.23521 & -0.96883 & - & 1 & noise\\
     178 & 0.791 & 216.21121 & -0.97154 & [\ion{O}{II}] ${\lambda}$3727 & 3 & \\
     179 & 0.885 & 216.22268 & -0.97496 & [\ion{O}{II}] ${\lambda}$3727 & 3 & \\
     180 & -     & 216.21390 & -0.97788 & - & 1 & noise\\
     181 & 0.816 & 216.19365 & -0.98274 & [\ion{O}{II}] ${\lambda}$3727 & 3 & \\
     182 & -     & 216.19471 & -0.98489 & - & 1 & noise\\
     183 & 0.863 & 216.20702 & -0.98722 & [\ion{O}{II}] ${\lambda}$3727 & 3 & \\
     184 & 0.795 & 216.18229 & -0.99090 & [\ion{O}{II}] ${\lambda}$3727 & 3 & \\
     185 & -     & 216.24355 & -0.99683 & - & 1 & noise\\
     186 & -     & 216.23331 & -0.99970 & - & 1 & faint continue (no spectrum feature)\\
     187 & 0.866 & 216.21658 & -1.00330 & [\ion{O}{II}] ${\lambda}$3727 & 3 & \\
     188 & -     & 216.20184 & -1.00635 & - & 1 & faint continue (no spectrum feature)\\
     189 & 0.770 & 216.22189 & -1.00922 & [\ion{O}{II}] ${\lambda}$3727 & 3 & \\
     190 & -     & 216.18153 & -1.01652 & - & 1 & continue (no spectrum feature)\\
     \hline
      1C & -     & 216.26140 & -0.79481 & - & 1 & noise\\
      2C & 1.029 & 216.25806 & -0.79324 & [\ion{O}{II}] ${\lambda}$3727 & 3 & \\
      3C & 0.818 & 216.29276 & -0.87998 & Ca H\&K & 3 & \\
      4C & -     & 216.24861 & -0.78536 & - & 1 & faint continue (no spectrum feature)\\
      5C & 0.801 & 216.27079 & -0.84277 & Ca H\&K & 3 & \\
      6C & -     & 216.26300 & -0.82987 & - & 1 & noise\\
      7C & 0.798 & 216.25903 & -0.82793 & [\ion{O}{II}] ${\lambda}$3727 & 3 & \\
      8C & -     & 216.23356 & -0.77519 & - & 1 & emission line $\sim6400$~\AA\\
      9C & 0.853 & 216.23602 & -0.78668 & [\ion{O}{II}] ${\lambda}$3727 & 3 & \\
     10C & 0.704 & 216.25897 & -0.84544 & Ca H\&K & 3 & \\
     11C & -     & 216.24960 & -0.83405 & - & 1 & noise\\
     12C & 0.752 & 216.26768 & -0.87993 & [\ion{O}{II}] ${\lambda}$3727 & 2 & CaHK behind B-Band\\
     13C & -     & 216.23961 & -0.82203 & - & 1 & noise\\
     14C & 0.770 & 216.26605 & -0.88703 & Ca H\&K & 3 & \\
     15C & 0.739 & 216.26768 & -0.89618 & Ca H\&K & 2 & CaHK behind B-Band\\
     16C & 0.770 & 216.25982 & -0.88481 & Ca H\&K & 3 & \\
     17C & -     & 216.23293 & -0.82923 & - & 1 & faint continue (no spectrum feature)\\
     18C & 0.725 & 216.24985 & -0.87355 & [\ion{O}{II}] ${\lambda}$3727 & 3 & \\
     19C & 0.784 & 216.23953 & -0.85591 & Ca H\&K & 3 & \\
     20C & 0.760 & 216.23997 & -0.86629 & Ca H\&K & 3 & \\
     21C & 0.795 & 216.21621 & -0.81791 & Ca H\&K & 3 & \\
     22C & -     & 216.23613 & -0.86787 & - & 1 & no signal\\
     23C & 0.764 & 216.23169 & -0.86283 & Ca H\&K & 3 & bad sky subtraction\\
     24C & -     & 216.23255 & -0.86959 & - & 1 & \\
     25C & 0.793 & 216.22423 & -0.85582 & Ca H\&K & 3 & bad sky subtraction\\
     26C & 0.770 & 216.22806 & -0.87133 & Ca H\&K,G & 3 & \\
     27C & -     & 216.22826 & -0.87711 & - & 1 & noise\\
     28C & 0.765 & 216.21826 & -0.85989 & Ca H\&K,G & 3 & two continues, take lower one\\
     29C & 0.764 & 216.22830 & -0.88758 & Ca H\&K,G & 3 & wonderful features\\
     30C & 0.770 & 216.21318 & -0.86057 & Ca H\&K,G & 3 & \\
     31C & 0.797 & 216.23269 & -0.91152 & Ca H\&K & 2 & (=Slit~\#151/152) \\
     32C & -     & 216.20860 & -0.86412 & - & 1 & (=Slit~\#124) CaHK behind A-Band\\
     33C & 0.766 & 216.21012 & -0.87577 & Ca H\&K & 3 & faint continue (no spectrum feature)\\
     34C & -     & 216.21044 & -0.88164 & - & 1 & noise\\
     35C & -     & 216.20496 & -0.87575 & - & 1 & noise\\
     36C & 0.767 & 216.20475 & -0.87996 & Ca H\&K & 3 & bad sky subtraction\\
     37C & 0.770 & 216.18647 & -0.84626 & Ca H\&K,G & 3 & \\
     38C & 0.814 & 216.19002 & -0.85895 & Ca H\&K,G & 3 & \\
     39C & 0.800 & 216.19978 & -0.88561 & Ca H\&K,G & 3 & \\
     40C & -     & 216.19898 & -0.88894 & - & 1 & faint continue (no spectrum feature)\\
     41C & -     & 216.18590 & -0.86616 & - & 1 & noise\\
     42C & 0.801 & 216.19233 & -0.88613 & Ca H\&K,G & 3 & \\
     43C & -     & 216.19664 & -0.90295 & - & 1 & noise\\
     44C & 0.817 & 216.17641 & -0.86347 & Ca H\&K,G & 3 & \\
     45C & 0.843 & 216.15603 & -0.82251 & [\ion{O}{II}] ${\lambda}$3727 & 3 & \\
     46C & 0.786 & 216.18890 & -0.90164 & Ca H\&K,G & 3 & \\
     47C & 0.793 & 216.18799 & -0.90658 & Ca H\&K,G & 3 & \\
     48C & 0.852 & 216.19391 & -0.93053 & Ca H\&K & 3 & \\
     49C & -     & 216.16449 & -0.87315 & - & 1 & noise\\
     50C & -     & 216.14556 & -0.83733 & - & 1 & (=Slit~\#109) noise\\
     51C & 0.764 & 216.18193 & -0.92870 & [\ion{O}{II}] ${\lambda}$3727 & 3 & \\
     52C & 1.070 & 216.16261 & -0.89155 & [\ion{O}{II}] ${\lambda}$3727 & 3 & \\
     53C & -     & 216.16368 & -0.90076 & - & 1 & no signal\\
     54C & 0.719 & 216.14721 & -0.86878 & [\ion{O}{II}] ${\lambda}$3727 & 3 & \\
     55C & 0.859 & 216.14340 & -0.86622 & [\ion{O}{II}] ${\lambda}$3727 & 3 & \\
     56C & -     & 216.12775 & -0.83622 & - & 1 & faint continue (no spectrum feature)\\
     57C & 0.570 & 216.14499 & -0.88034 & Ca H\&K,G & 3 & \\
     58C & 0.083 & 216.12232 & -0.83752 & M6 & 3 & M6 star\\
     59C & 0.954 & 216.16373 & -0.93877 & [\ion{O}{II}] ${\lambda}$3727 & 3 & \\
     60C & -     & 216.12265 & -0.85105 & - & 1 & bad sky subtraction\\
     61C & 0.886 & 216.11735 & -0.84704 & [\ion{O}{II}] ${\lambda}$3727 & 3 & \\
     62C & -     & 216.12537 & -0.87170 & - & 1 & bad sky subtraction\\
     63C & -     & 216.13969 & -0.90858 & - & 1 & faint continue (no spectrum feature)\\
     64C & 0.925 & 216.13108 & -0.89423 & [\ion{O}{II}] ${\lambda}$3727 & 3 & \\
     65C & -     & 216.11152 & -0.85490 & - & 1 & bad sky subtraction\\
     66C & -     & 216.14387 & -0.93289 & - & 1 & faint continue (no spectrum feature)\\
     67C & -     & 216.11791 & -0.88143 & - & 1 & faint continue (no spectrum feature)\\
     68C & -     & 216.10263 & -0.85251 & - & 1 & no signal\\
     69C & 0.922 & 216.13127 & -0.92183 & [\ion{O}{II}] ${\lambda}$3727 & 3 & \\
     70C & 0.901 & 216.08915 & -0.83328 & [\ion{O}{II}] ${\lambda}$3727 & 3 & \\
     71C & 0.799 & 216.09688 & -0.85866 & [\ion{O}{II}] ${\lambda}$3727 & 3 & \\
     72C & 0.895 & 216.12903 & -0.93589 & [\ion{O}{II}] ${\lambda}$3727 & 3 & bottom: blue galaxy at $z=$0.102\\
     73C & -     & 216.11488 & -0.90999 & - & 1 & faint continue (no spectrum feature)\\
     74C & 0.896 & 216.10496 & -0.89238 & Ca H\&K,G & 3 & \\
     75C & 0.857 & 216.09606 & -0.88104 & [\ion{O}{II}] ${\lambda}$3727 & 3 & \\
     76C & 0.824 & 216.08203 & -0.85937 & Ca H\&K & 3 & faint with distinguishable feature\\
     77C & 0.817 & 216.10053 & -0.90584 & [\ion{O}{II}] ${\lambda}$3727 & 3 & \\
     78C & 0.879 & 216.09559 & -0.90328 & [\ion{O}{II}] ${\lambda}$3727 & 3 & \\
     79C & 0.794 & 216.08827 & -0.89147 & [\ion{O}{II}] ${\lambda}$3727 & 3 & \\
     80C & 0.741 & 216.10771 & -0.94168 & Ca H\&K & 3 & \\
     81C & 0.817 & 216.09630 & -0.92098 & Ca H\&K & 3 & \\
     82C & 0.816 & 216.07308 & -0.87937 & Ca H\&K & 3 & \\
     83C & -     & 216.09774 & -0.94265 & - & 1 & no signal\\
     84C & -     & 216.06135 & -0.86983 & - & 1 & noise\\
     85C & 0.954 & 216.09830 & -0.95793 & [\ion{O}{II}] ${\lambda}$3727 & 3 & \\
     86C & 0.745 & 216.08964 & -0.94358 & [\ion{O}{II}] ${\lambda}$3727 & 3 & \\
     87C & -     & 216.07222 & -0.91188 & - & 1 & noise\\
     88C & 0.888 & 216.06954 & -0.91276 & Ca H\&K & 3 & \\
     89C & -     & 216.07734 & -0.93897 & - & 1 & faint continue (no spectrum feature)\\
     90C & 0.880 & 216.05048 & -0.88543 & [\ion{O}{II}] ${\lambda}$3727 & 3 & \\
     91C & 1.097 & 216.07754 & -0.95224 & [\ion{O}{II}] ${\lambda}$3727 & 3 & \\
     92C & -     & 216.08556 & -0.97561 & - & 1 & faint continue (no spectrum feature)\\
     93C & 0.815 & 216.06094 & -0.92736 & [\ion{O}{II}] ${\lambda}$3727 & 3 & \\
     94C & 0.795 & 216.04568 & -0.89943 & Ca H\&K,G & 3 & wonderful features\\
     95C & 0.838 & 216.05368 & -0.92740 & [\ion{O}{II}] ${\lambda}$3727 & 3 & \\
     96C & 0.134 & 216.04542 & -0.91551 & \ion{H}{$\alpha$} ${\lambda}$6563 & 3 & green galaxy or quasar\\
     \hline
     97C & 0.887 & 216.22066 & -0.70284 & [\ion{O}{II}] ${\lambda}$3727 & 3 & \\
     98C & 0.892 & 216.20140 & -0.66991 & [\ion{O}{II}] ${\lambda}$3727 & 3 & \\
     99C & 0.969 & 216.19570 & -0.66670 & [\ion{O}{II}] ${\lambda}$3727 & 3 & \\
    100C & 0.770 & 216.20072 & -0.68718 & [\ion{O}{II}] ${\lambda}$3727 & 3 & wonderful features\\
    101C & 0.882 & 216.19088 & -0.67172 & [\ion{O}{II}] ${\lambda}$3727 & 3 & \\
    102C & 0.853 & 216.18517 & -0.66378 & [\ion{O}{II}] ${\lambda}$3727 & 3 & \\
    103C & -     & 216.18842 & -0.67635 & - & 1 & noise\\
    104C & -     & 216.19569 & -0.69759 & - & 1 & continue but no spectrum feature\\
    105C & 0.885 & 216.18930 & -0.69458 & [\ion{O}{II}] ${\lambda}$3727 & 2 & another emission at $7700$~\AA\\
    106C & 0.888 & 216.18011 & -0.67982 & Ca H\&K,G & 3 & faint with distinguishable feature\\
    107C & 0.891 & 216.16926 & -0.66255 & [\ion{O}{II}] ${\lambda}$3727 & 3 & \\
    108C & 0.817 & 216.18688 & -0.70697 & [\ion{O}{II}] ${\lambda}$3727 & 3 & \\
    109C & 0.791 & 216.17784 & -0.69461 & Ca H\&K & 3 & \\
    110C & 0.851 & 216.18703 & -0.72003 & [\ion{O}{II}] ${\lambda}$3727 & 3 & \\
    111C & 0.869 & 216.14085 & -0.62258 & Ca H\&K & 3 & \\
    112C & -     & 216.14917 & -0.64708 & - & 1 & noise\\
    113C & -     & 216.18067 & -0.72742 & - & 1 & noise\\
    114C & 0.707 & 216.14331 & -0.64860 & [\ion{O}{II}] ${\lambda}$3727 & 3 & \\
    115C & 0.760 & 216.14338 & -0.65805 & [\ion{O}{II}] ${\lambda}$3727 & 3 & \\
    116C & -     & 216.13829 & -0.65307 & - & 1 & continue but no spectrum feature\\
    117C & -     & 216.15551 & -0.69953 & - & 1 & noise\\
    118C & -     & 216.13871 & -0.67097 & - & 1 & continue but no spectrum feature\\
    119C & -     & 216.14815 & -0.69955 & - & 1 & continue but no spectrum feature\\
    120C & 0.837 & 216.16138 & -0.74069 & [\ion{O}{II}] ${\lambda}$3727 & 3 & \\
    121C & 0.880 & 216.12934 & -0.67482 & [\ion{O}{II}] ${\lambda}$3727 & 3 & wonderful features\\
    122C & -     & 216.15143 & -0.73580 & - & 1 & faint continue (no spectrum feature)\\
    123C & 0.792 & 216.12973 & -0.69792 & Ca H\&K & 3 & \\
    124C & 0.791 & 216.10461 & -0.65042 & [\ion{O}{II}] ${\lambda}$3727 & 3 & \\
    125C & -     & 216.13700 & -0.72860 & - & 1 & too bright background\\
    126C & -     & 216.11520 & -0.68423 & - & 1 & noise\\
    127C & 0.600 & 216.10112 & -0.65974 & [\ion{O}{II}] ${\lambda}$3727,\ion{H}{$\gamma$},\ion{H}{$\beta$},[\ion{O}{III}] & 3 & \\
    128C & 0.817 & 216.09341 & -0.64750 & [\ion{O}{II}] ${\lambda}$3727 & 3 & \\
    129C & 0.791 & 216.13217 & -0.74523 & [\ion{O}{II}] ${\lambda}$3727 & 3 & \\
    130C & -     & 216.12651 & -0.73876 & - & 1 & noise\\
    131C & 0.720 & 216.10037 & -0.68593 & [\ion{O}{II}] ${\lambda}$3727 & 3 & \\
    132C & -     & 216.09644 & -0.68313 & - & 1 & noise\\
    133C & 0.557 & 216.09266 & -0.68048 & [\ion{O}{II}] ${\lambda}$3727 & 3 & \\
    134C & 0.716 & 216.07647 & -0.64902 & [\ion{O}{II}] ${\lambda}$3727 & 3 & \\
    135C & 0.853 & 216.08352 & -0.67267 & [\ion{O}{II}] ${\lambda}$3727 & 3 & \\
    136C & -     & 216.07764 & -0.66950 & - & 1 & noise\\
    137C & -     & 216.10327 & -0.73276 & - & 1 & faint continue (no spectrum feature)\\
    138C & 0.882 & 216.10522 & -0.74232 & [\ion{O}{II}] ${\lambda}$3727 & 3 & \\
    139C & 0.732 & 216.09465 & -0.72757 & [\ion{O}{II}] ${\lambda}$3727 & 3 & \\
    140C & 0.715 & 216.08925 & -0.72234 & Ca H\&K,G & 3 & \\
    141C & 0.857 & 216.09401 & -0.74096 & [\ion{O}{II}] ${\lambda}$3727 & 3 & \\
    142C & 0.983 & 216.09535 & -0.74879 & [\ion{O}{II}] ${\lambda}$3727 & 3 & \\
    143C & 0.888 & 216.08603 & -0.73606 & [\ion{O}{II}] ${\lambda}$3727 & 3 & \\
    144C & 0.875 & 216.08314 & -0.73465 & [\ion{O}{II}] ${\lambda}$3727 & 3 & \\
    145C & 0.919 & 216.09109 & -0.75833 & [\ion{O}{II}] ${\lambda}$3727 & 3 & \\
    146C & 0.793 & 216.04606 & -0.66651 & [\ion{O}{II}] ${\lambda}$3727 & 3 & \\
    147C & -     & 216.04060 & -0.66446 & - & 1 & noise\\
    148C & 0.905 & 216.05727 & -0.70798 & [\ion{O}{II}] ${\lambda}$3727 & 3 & wonderful features\\
    149C & -     & 216.06491 & -0.73007 & - & 1 & bad sky subtraction\\
    150C & -     & 216.04005 & -0.67986 & - & 1 & continue but no spectrum feature\\
    151C & -     & 216.03961 & -0.68761 & - & 1 & noise\\
    152C & -     & 216.04183 & -0.69807 & - & 1 & noise\\
    153C & 0.653 & 216.03363 & -0.68587 & Ca H\&K,G & 3 & \\
    154C & 0.858 & 216.04804 & -0.72895 & [\ion{O}{II}] ${\lambda}$3727 & 3 & wonderful features\\
    155C & -     & 216.04325 & -0.72855 & - & 1 & faint continue (no spectrum feature)\\
    156C & 0.857 & 216.04163 & -0.73452 & [\ion{O}{II}] ${\lambda}$3727 & 3 & \\
    157C & 0.886 & 216.04181 & -0.73980 & Ca H\&K,G & 3 & \\
    158C & 0.652 & 216.02378 & -0.70661 & Ca H\&K & 3 & \\
    159C & -     & 216.02047 & -0.71095 & - & 1 & continue but no spectrum feature\\
    160C & -     & 215.99999 & -0.67382 & - & 1 & noise\\
    161C & 0.915 & 216.02282 & -0.73041 & [\ion{O}{II}] ${\lambda}$3727 & 3 & \\
    162C & -     & 216.00204 & -0.68845 & - & 1 & noise\\
    163C & -     & 216.00794 & -0.70785 & - & 1 & continue but no spectrum feature\\
    164C & -     & 216.01796 & -0.73603 & - & 1 & faint continue (no spectrum feature)\\
    165C & -     & 215.99251 & -0.68559 & - & 1 & faint continue (no spectrum feature)\\
    166C & 0.893 & 216.00536 & -0.72122 & Ca H\&K & 3 & faint continue (no spectrum feature)\\
    167C & 0.896 & 216.00037 & -0.71840 & [\ion{O}{II}] ${\lambda}$3727 & 3 & \\
    168C & 1.120 & 216.00634 & -0.73904 & [\ion{O}{II}] ${\lambda}$3727 & 3 & small peak redwards [\ion{O}{II}] ${\lambda}$3727\\
    169C & -     & 215.99143 & -0.71067 & - & 1 & bright continue but not like a science\\
    170C & -     & 215.98740 & -0.70703 & - & 1 & faint continue (no spectrum feature)\\
    171C & -     & 216.00090 & -0.74438 & - & 1 & bright continue but not like a science\\
    172C & -     & 215.98628 & -0.71906 & - & 1 & noise\\
    173C & 0.920 & 215.99179 & -0.73781 & [\ion{O}{II}] ${\lambda}$3727 & 3 & wonderful features\\
    174C & 0.919 & 215.99195 & -0.74722 & [\ion{O}{II}] ${\lambda}$3727 & 3 & wonderful features\\
    175C & 0.925 & 215.96163 & -0.68523 & Ca H\&K & 2 & CaHK behind B-Band\\
    176C & 0.735 & 216.01075 & -0.80120 & [\ion{O}{II}] ${\lambda}$3727 & 3 & \\
    177C & 0.841 & 215.96387 & -0.70244 & Ca H\&K & 3 & \\
    178C & 1.034 & 215.99808 & -0.78697 & [\ion{O}{II}] ${\lambda}$3727 & 3 & \\
    179C & 0.923 & 215.97603 & -0.74603 & Ca H\&K & 2 & CaHK behind B-Band\\
    180C & 0.888 & 215.96226 & -0.71991 & [\ion{O}{II}] ${\lambda}$3727 & 3 & \\
    181C & -     & 215.97205 & -0.75215 & - & 1 & noise\\
\enddata
\end{deluxetable*}

%% file: show_2dspec_to_arXiv.tex
\clearpage
\begin{figure*}
    \centering
    \includegraphics[width=\textwidth, height=0.96\textheight, keepaspectratio]{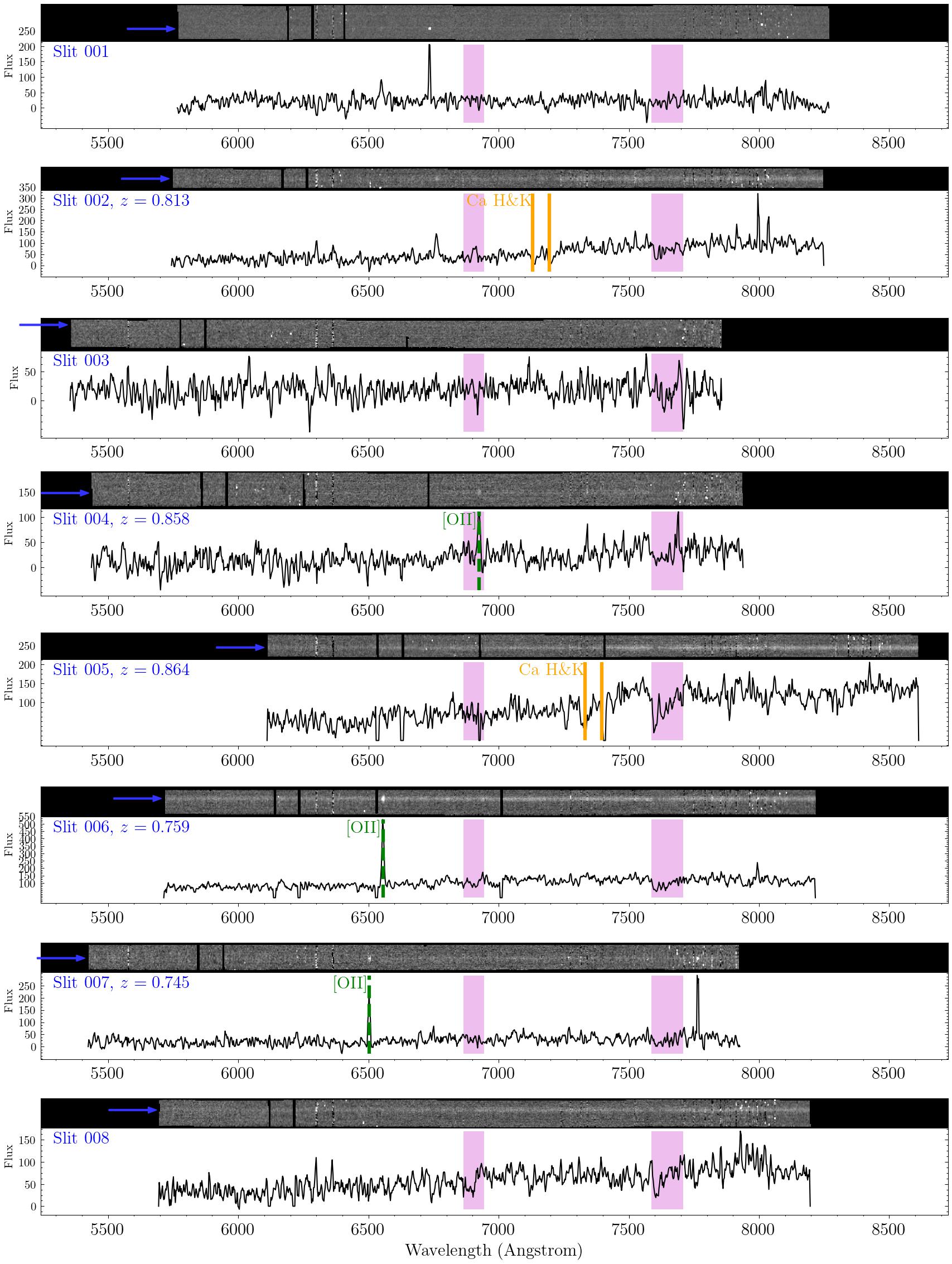}
    \caption{Galaxy Spectra of Slit~\#1--8.}
\end{figure*}

\clearpage
\begin{figure*}
    \centering     \includegraphics[width=\textwidth, height=0.96\textheight, keepaspectratio]{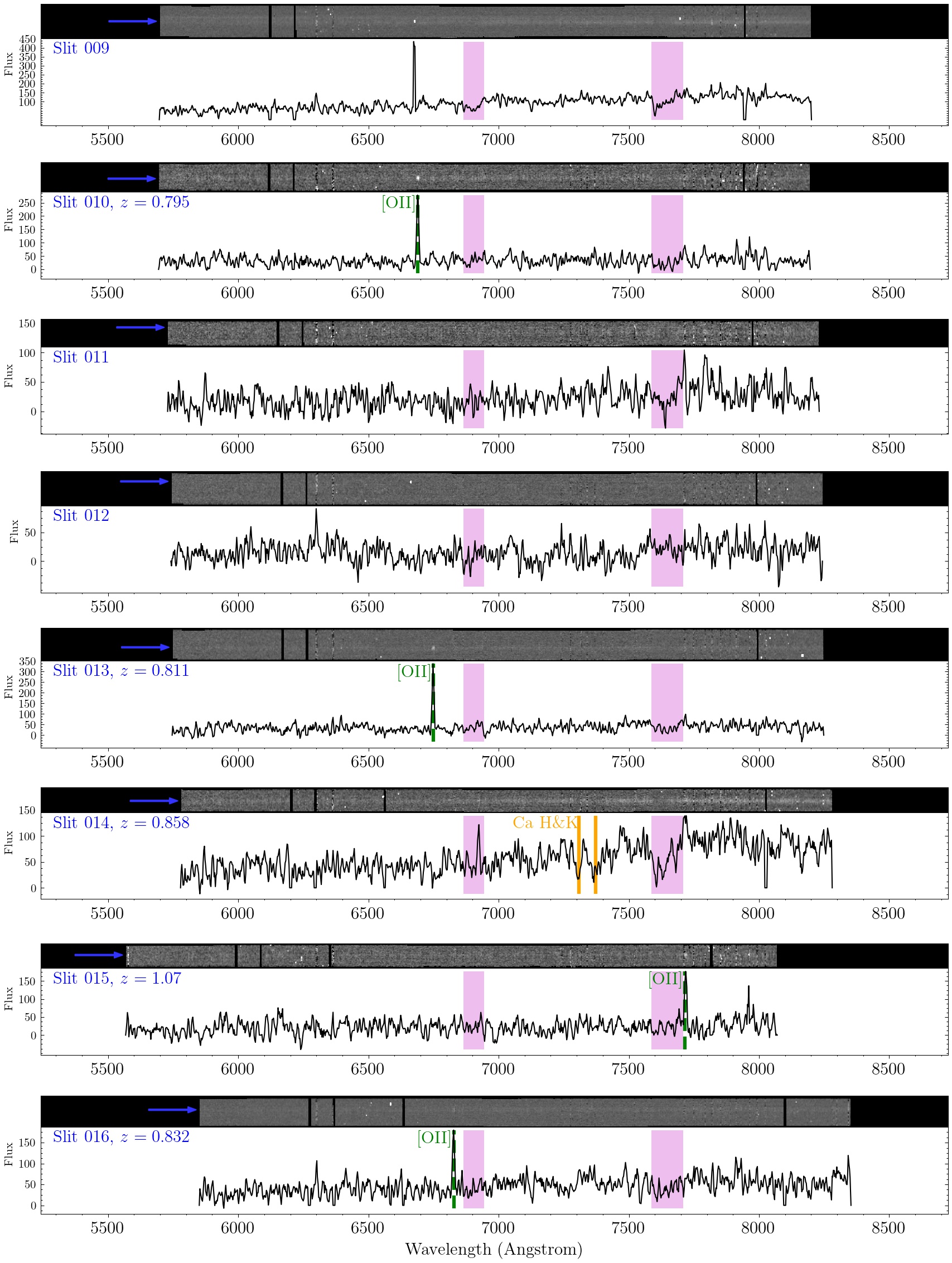}
    \caption{Galaxy Spectra of Slit~\#9--16.}
\end{figure*}

\clearpage
\begin{figure*}
    \centering     \includegraphics[width=\textwidth, height=0.96\textheight, keepaspectratio]{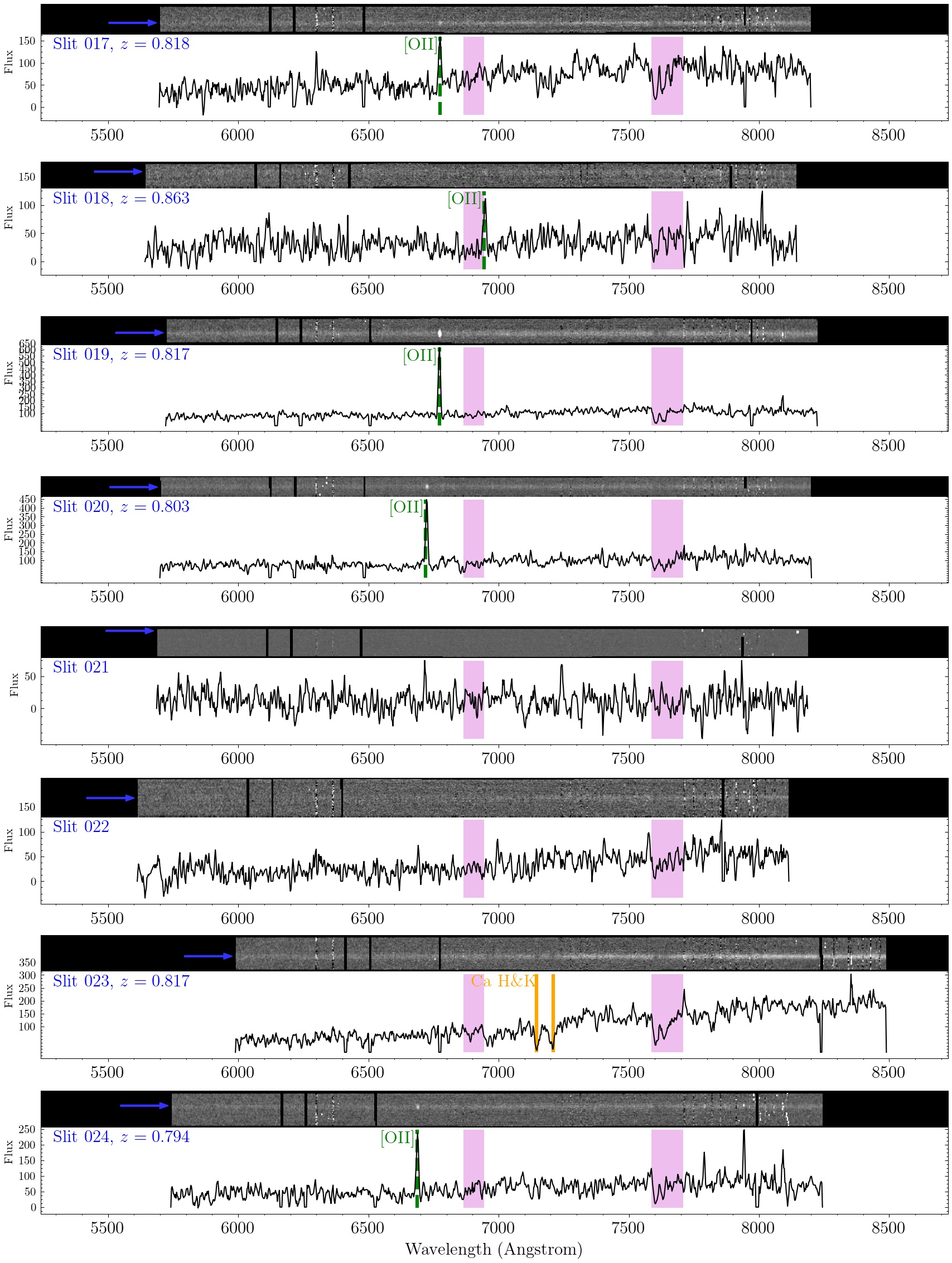}
    \caption{Galaxy Spectra of Slit~\#17--24.}
\end{figure*}

\clearpage
\begin{figure*}
    \centering     \includegraphics[width=\textwidth, height=0.96\textheight, keepaspectratio]{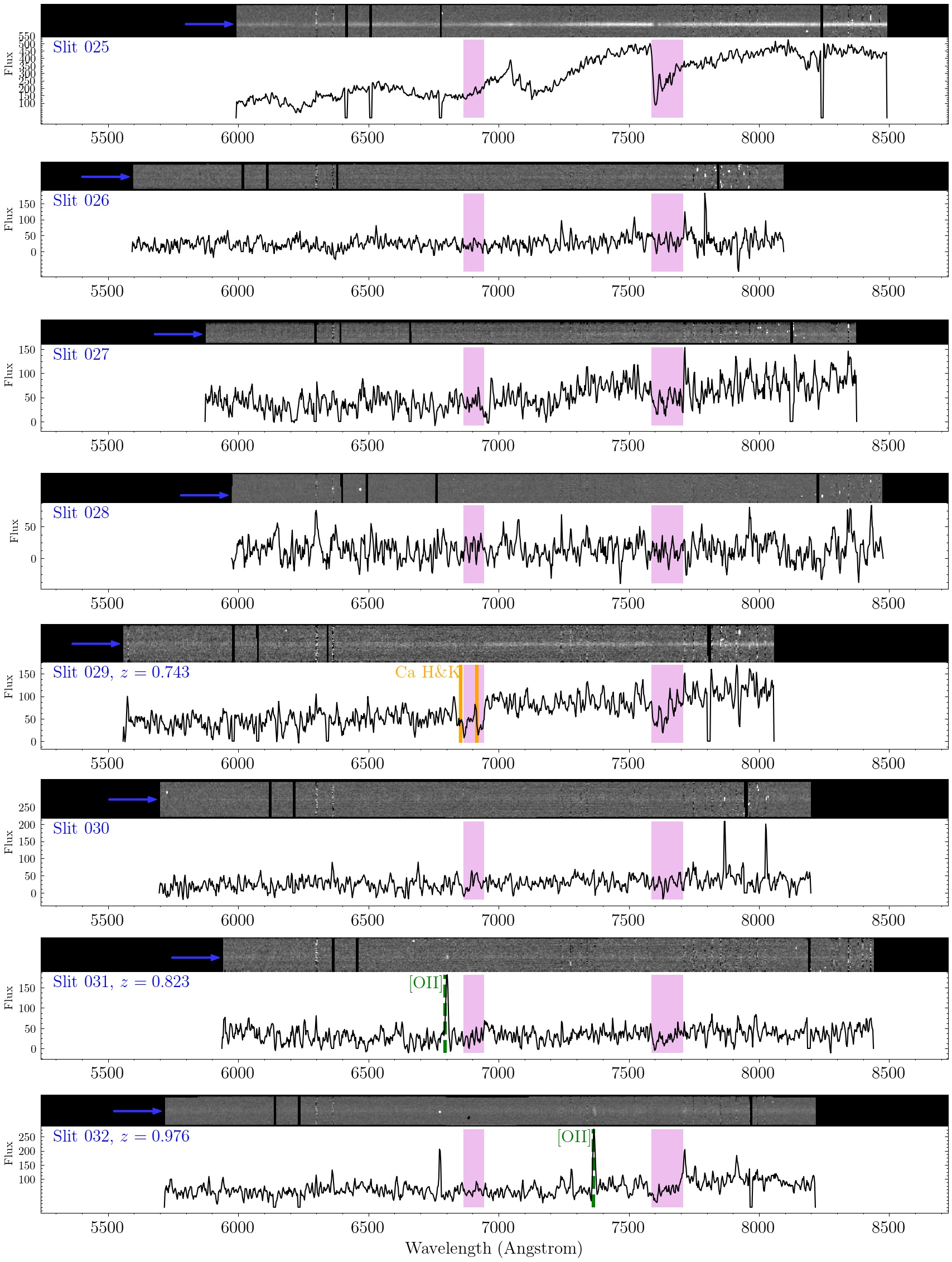}
    \caption{Galaxy Spectra of Slit~\#25--32.}
\end{figure*}

\clearpage
\begin{figure*}
    \centering     \includegraphics[width=\textwidth, height=0.96\textheight, keepaspectratio]{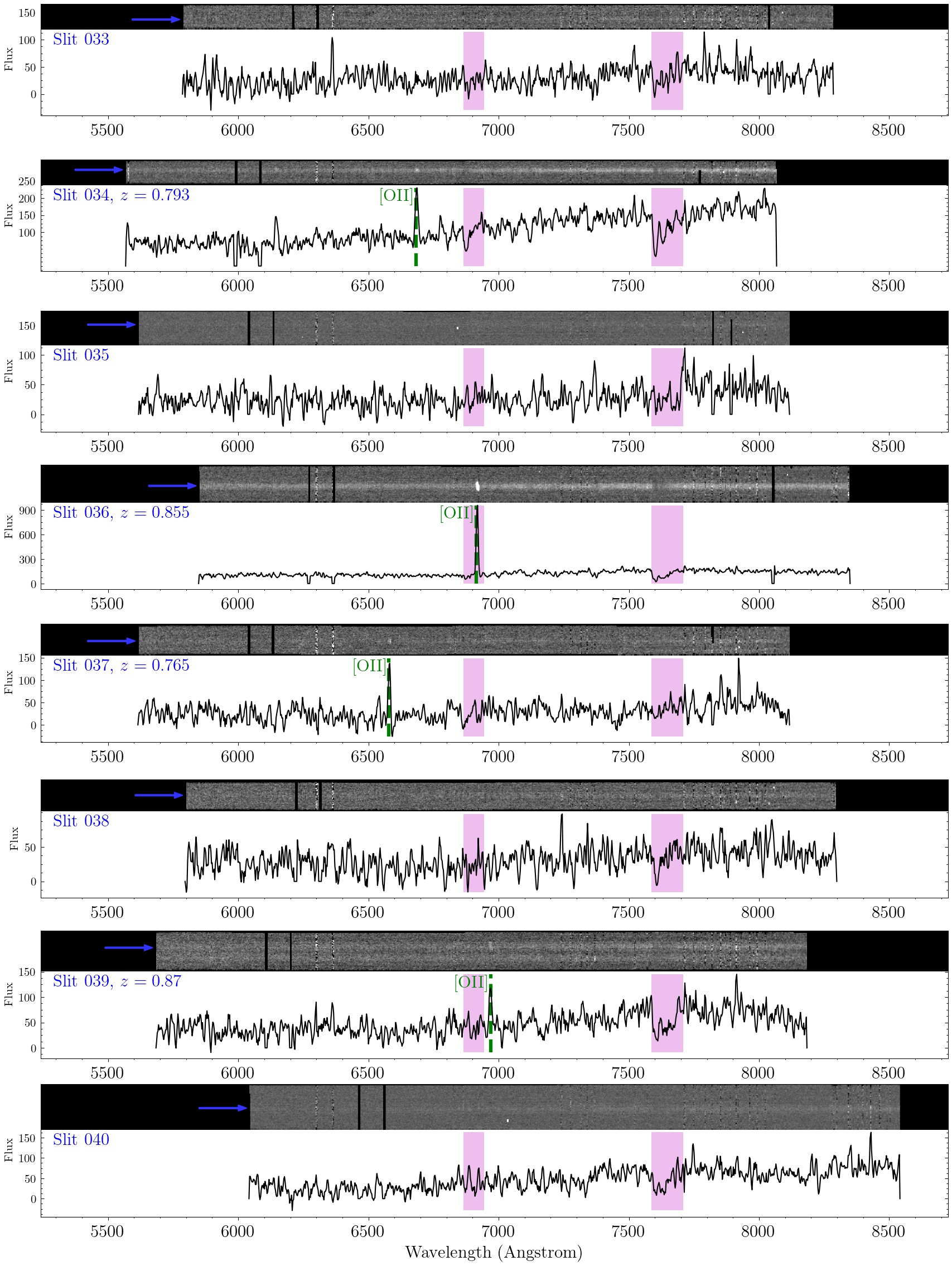}
    \caption{Galaxy Spectra of Slit~\#33--40.}
\end{figure*}

\clearpage
\begin{figure*}
    \centering     \includegraphics[width=\textwidth, height=0.96\textheight, keepaspectratio]{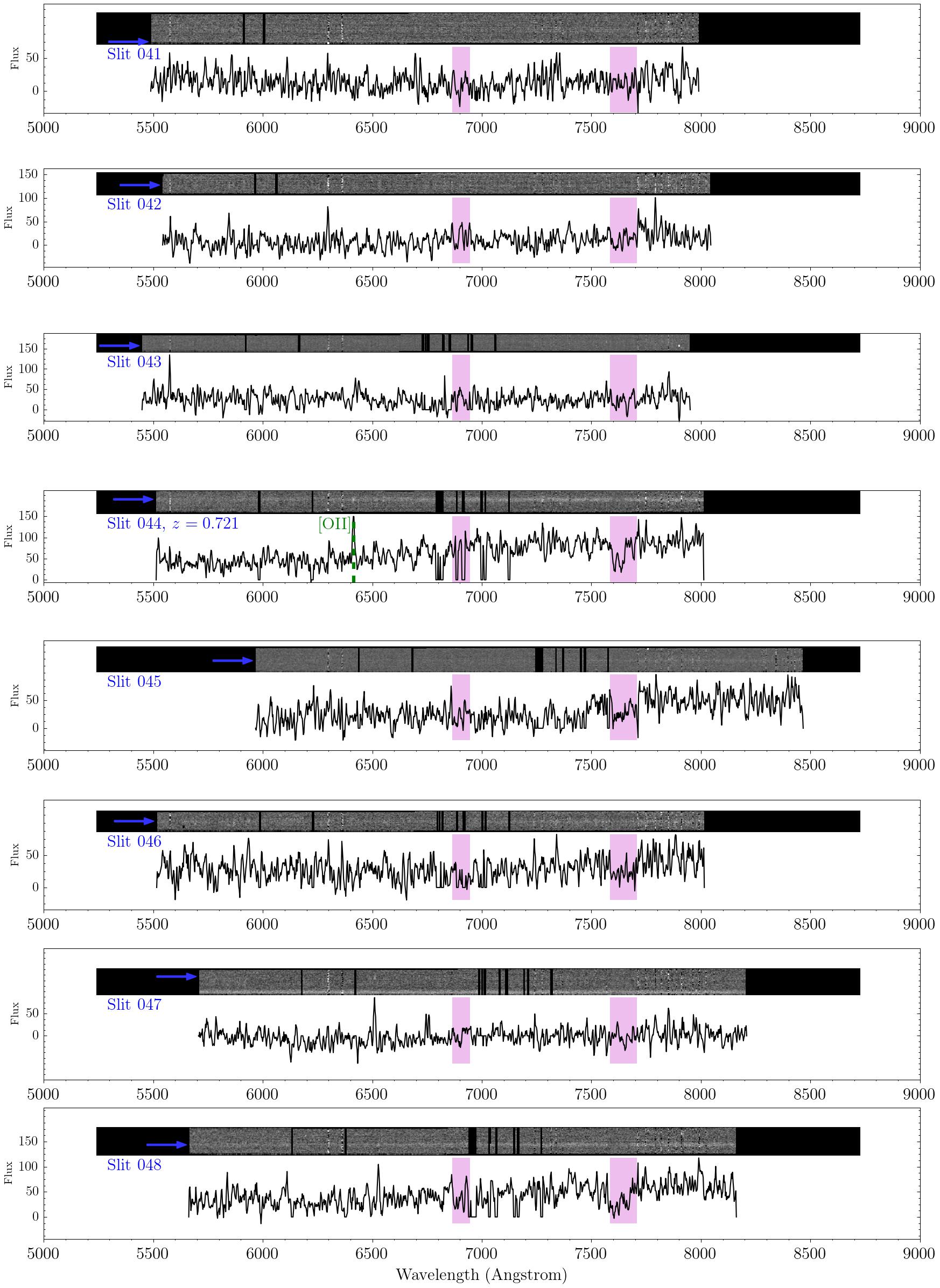}
    \caption{Galaxy Spectra of Slit~\#41--48.}
\end{figure*}

\clearpage
\begin{figure*}
    \centering     \includegraphics[width=\textwidth, height=0.96\textheight, keepaspectratio]{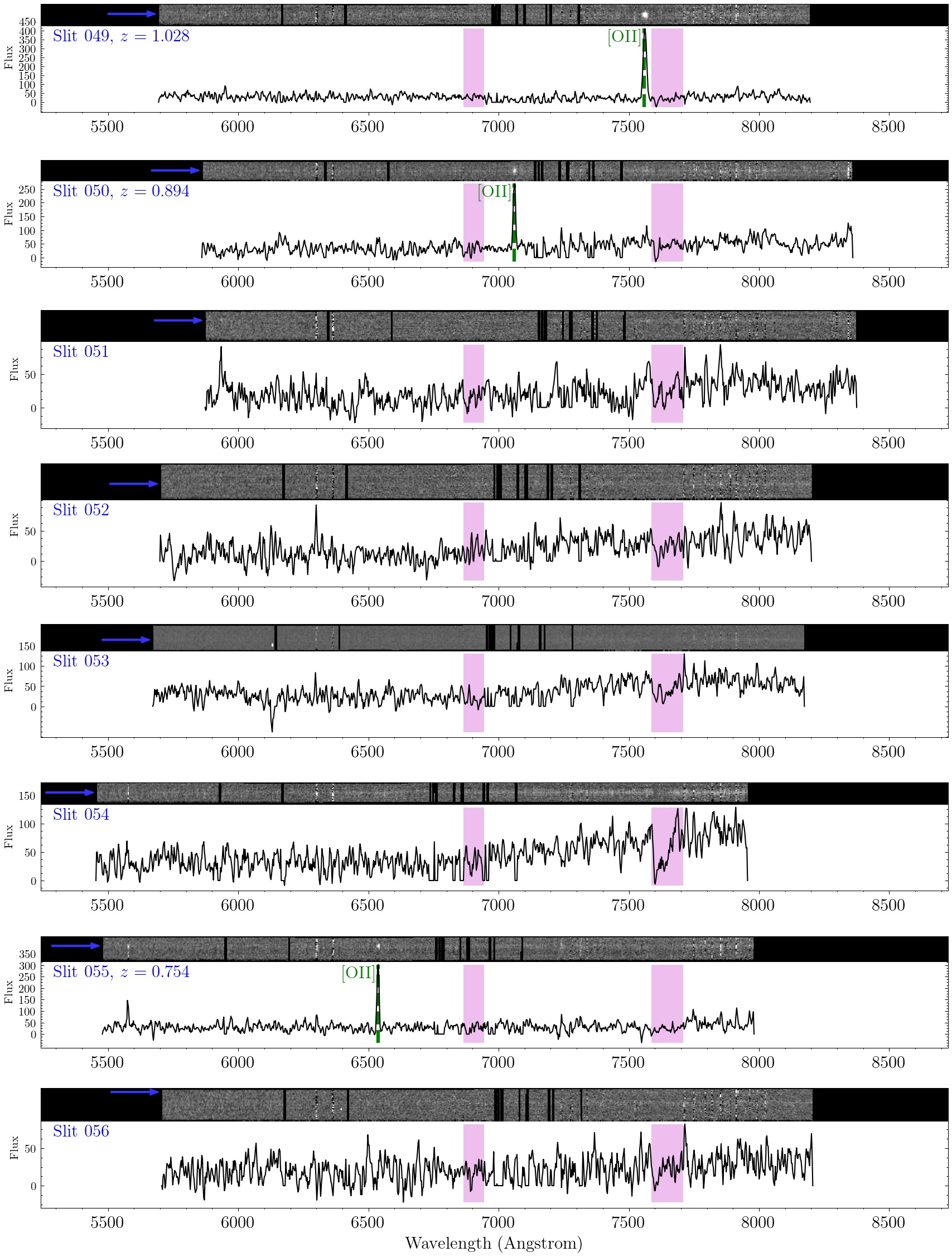}
    \caption{Galaxy Spectra of Slit~\#49--56.}
\end{figure*}

\clearpage
\begin{figure*}
    \centering     \includegraphics[width=\textwidth, height=0.96\textheight, keepaspectratio]{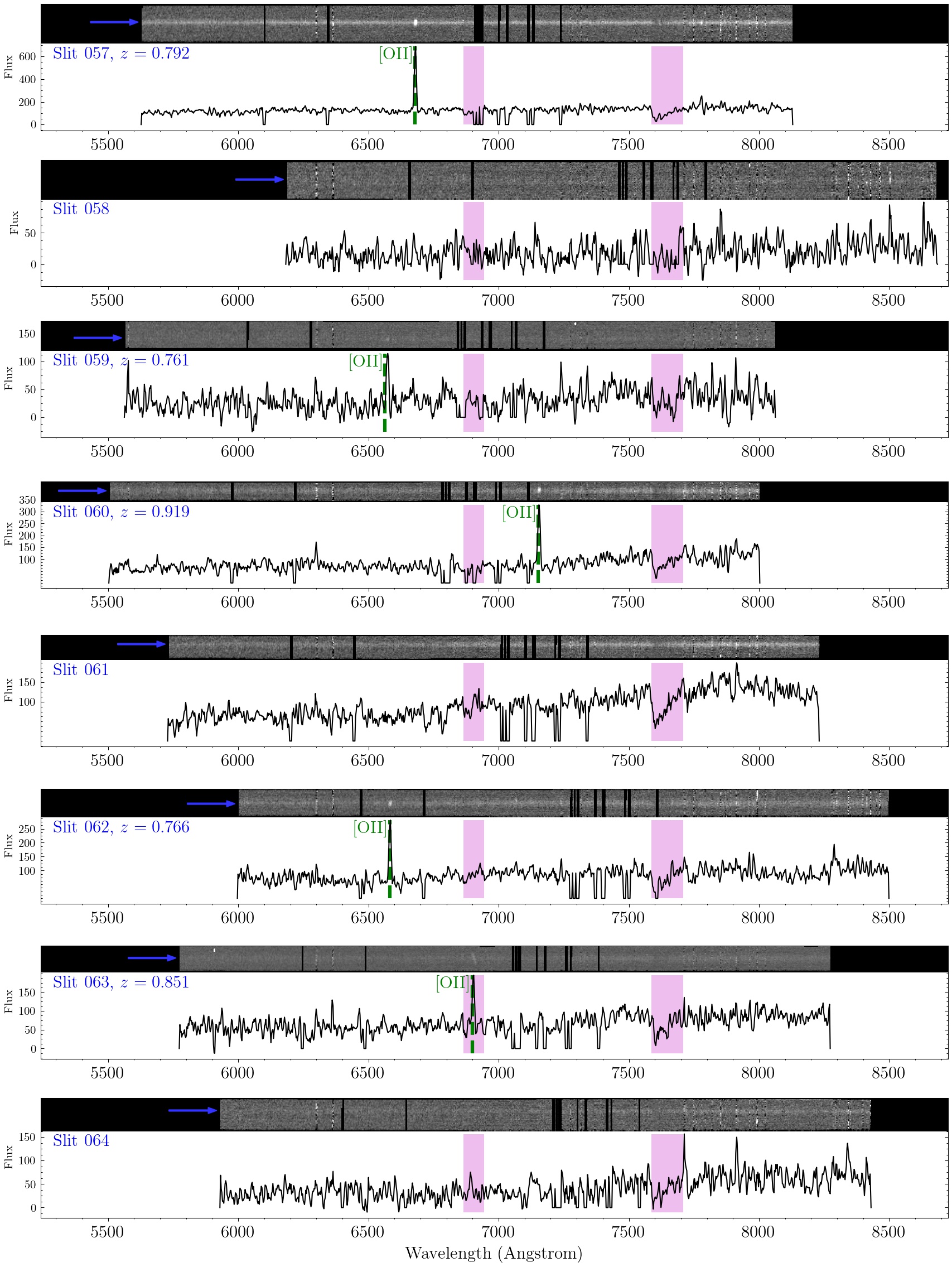}
    \caption{Galaxy Spectra of Slit~\#57--64.}
\end{figure*}

\clearpage
\begin{figure*}
    \centering     \includegraphics[width=\textwidth, height=0.96\textheight, keepaspectratio]{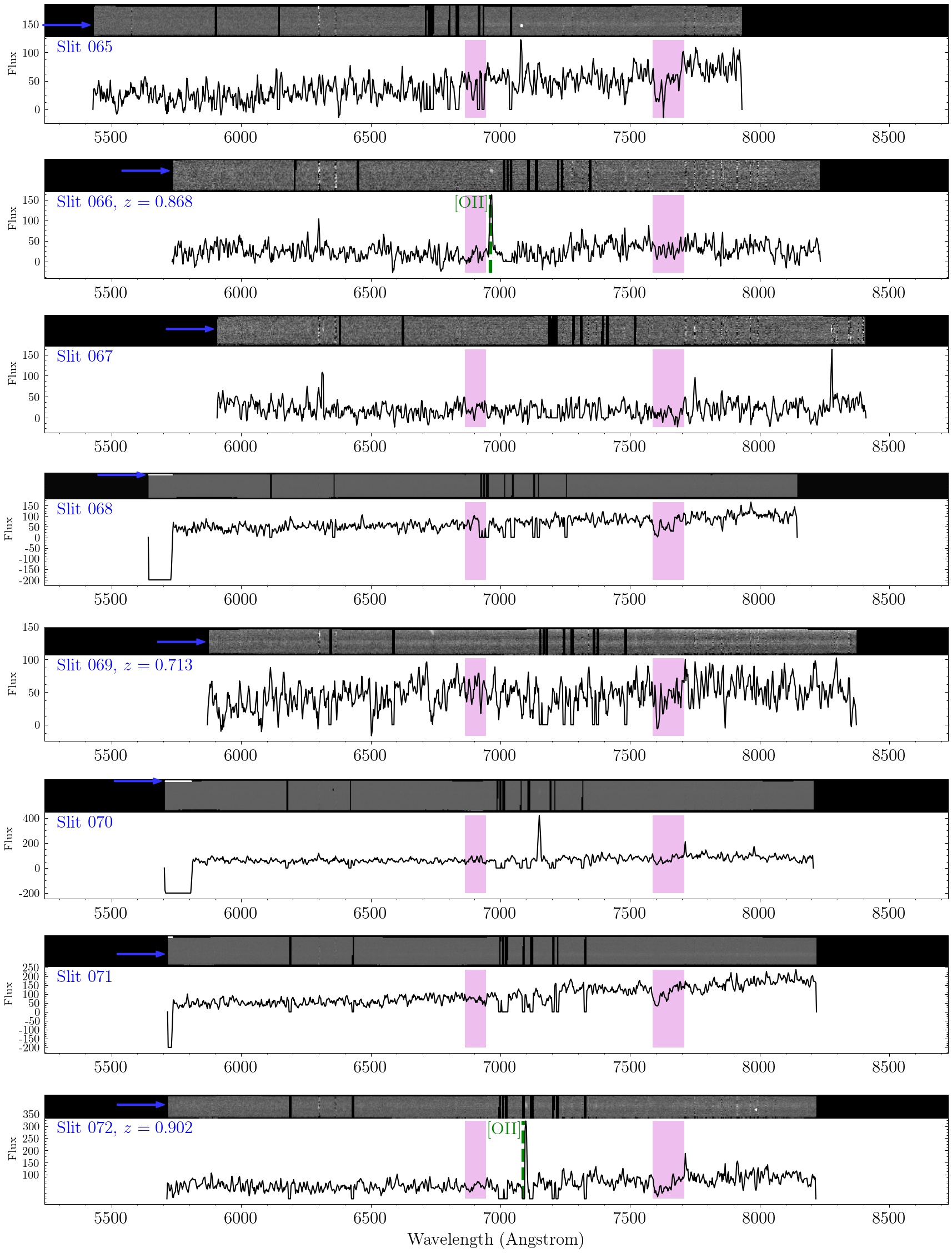}
    \caption{Galaxy Spectra of Slit~\#65--72.}
\end{figure*}

\clearpage
\begin{figure*}
    \centering     \includegraphics[width=\textwidth, height=0.96\textheight, keepaspectratio]{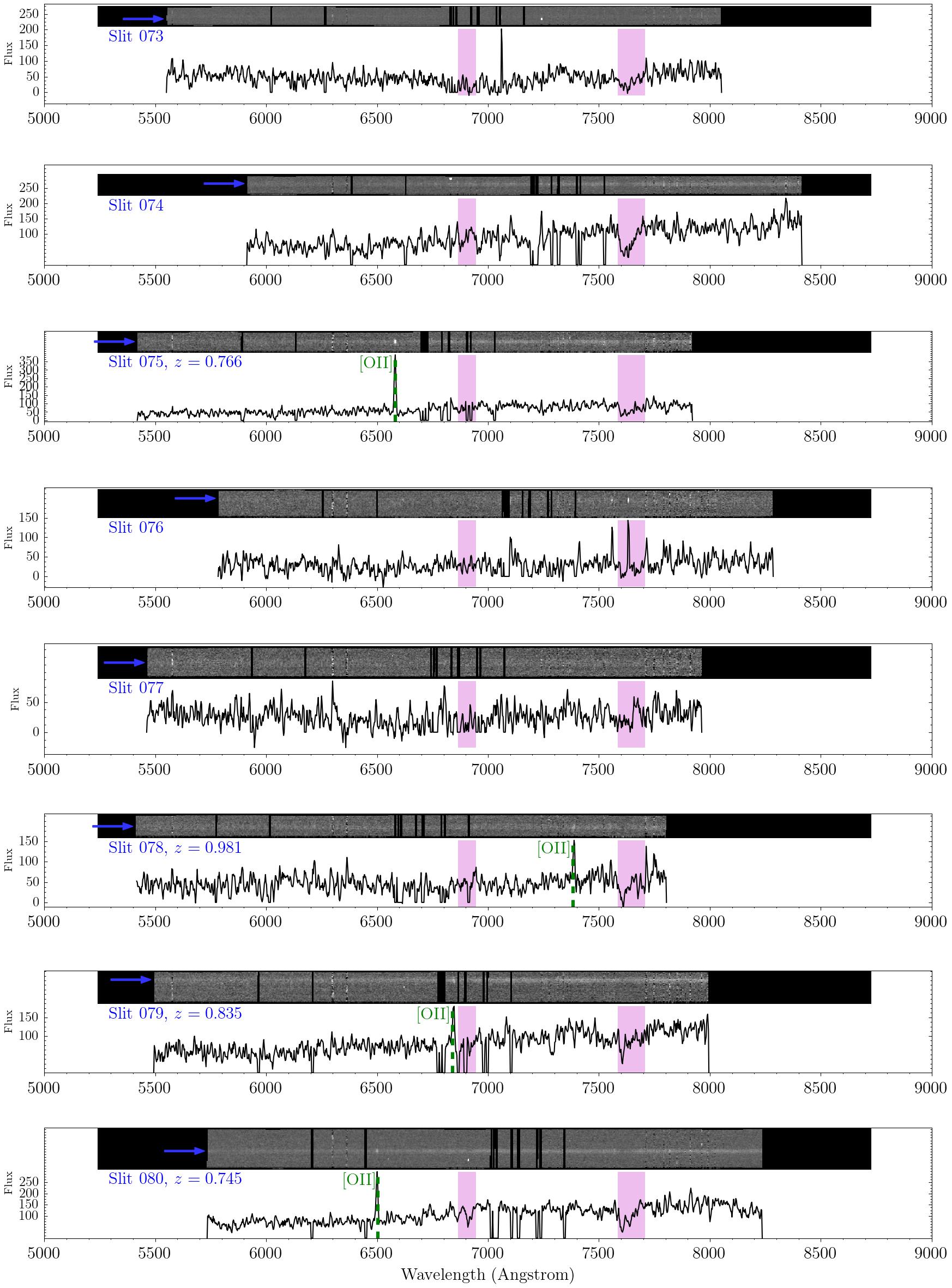}
    \caption{Galaxy Spectra of Slit~\#73--80.}
\end{figure*}

\clearpage
\begin{figure*}
    \centering     \includegraphics[width=\textwidth, height=0.96\textheight, keepaspectratio]{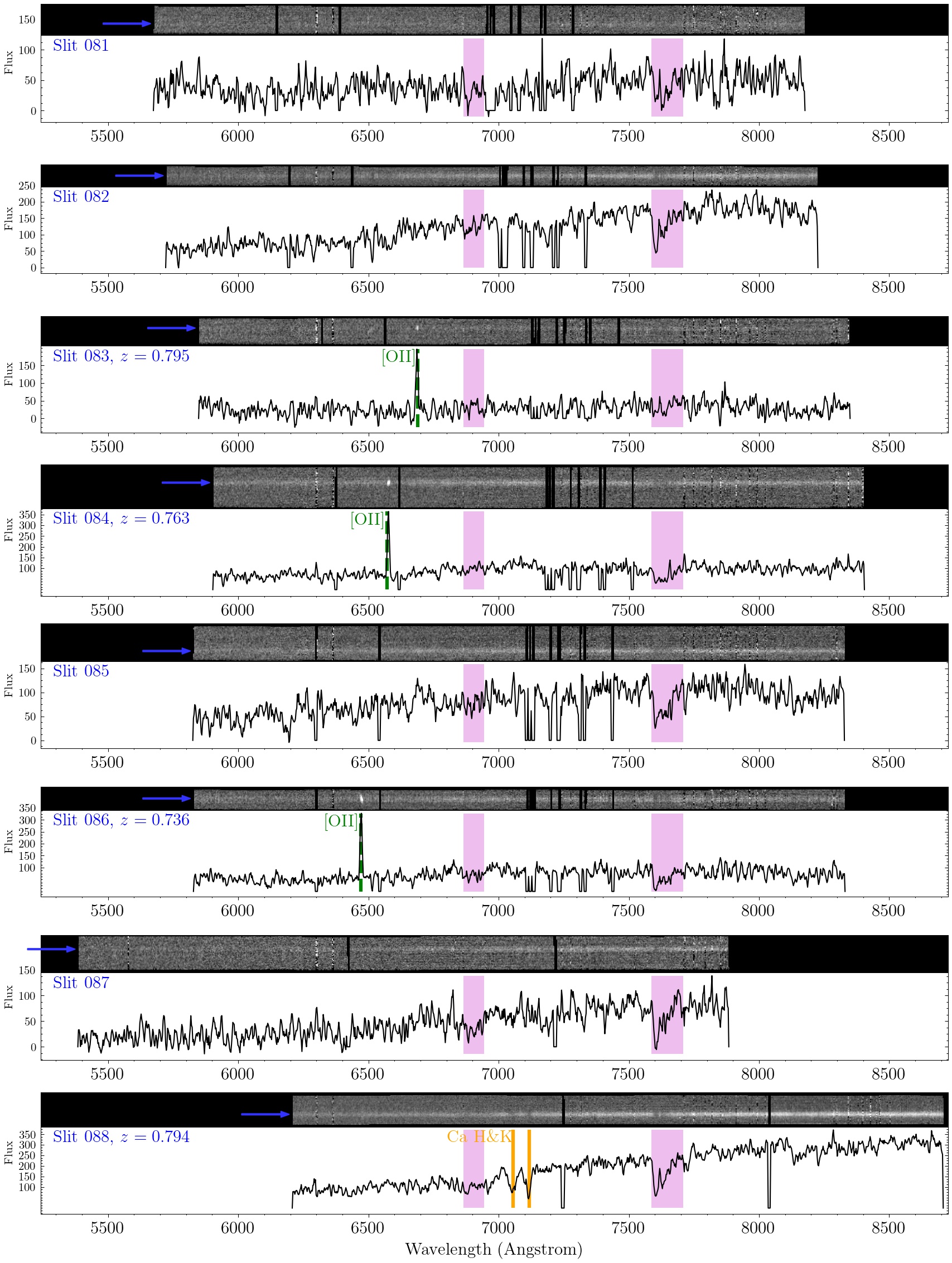}
    \caption{Galaxy Spectra of Slit~\#81--88.}
\end{figure*}

\clearpage
\begin{figure*}
    \centering     \includegraphics[width=\textwidth, height=0.96\textheight, keepaspectratio]{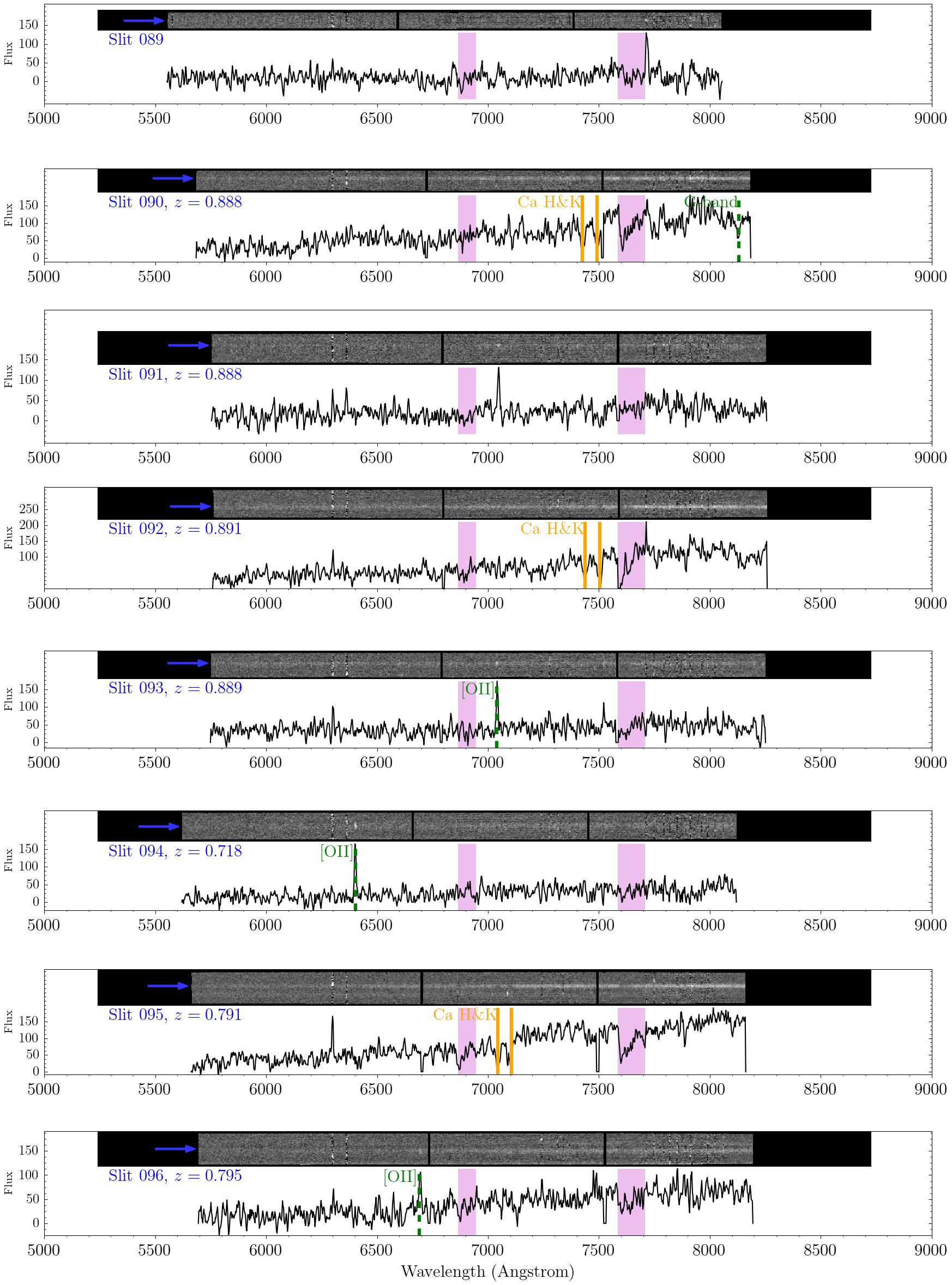}
    \caption{Galaxy Spectra of Slit~\#89--96.}
\end{figure*}

\clearpage
\begin{figure*}
    \centering     \includegraphics[width=\textwidth, height=0.96\textheight, keepaspectratio]{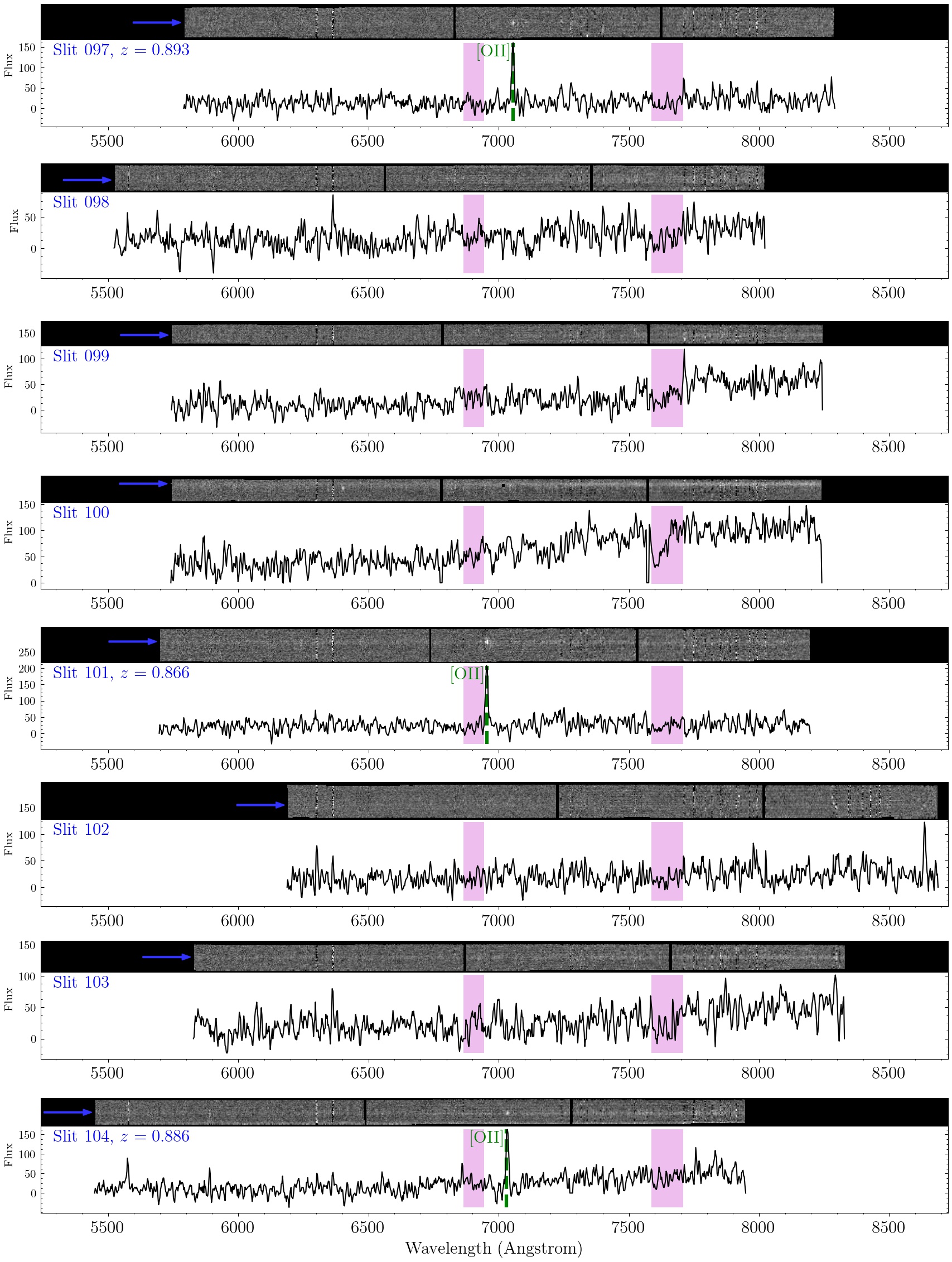}
    \caption{Galaxy Spectra of Slit~\#97--104.}
\end{figure*}

\clearpage
\begin{figure*}
    \centering     \includegraphics[width=\textwidth, height=0.96\textheight, keepaspectratio]{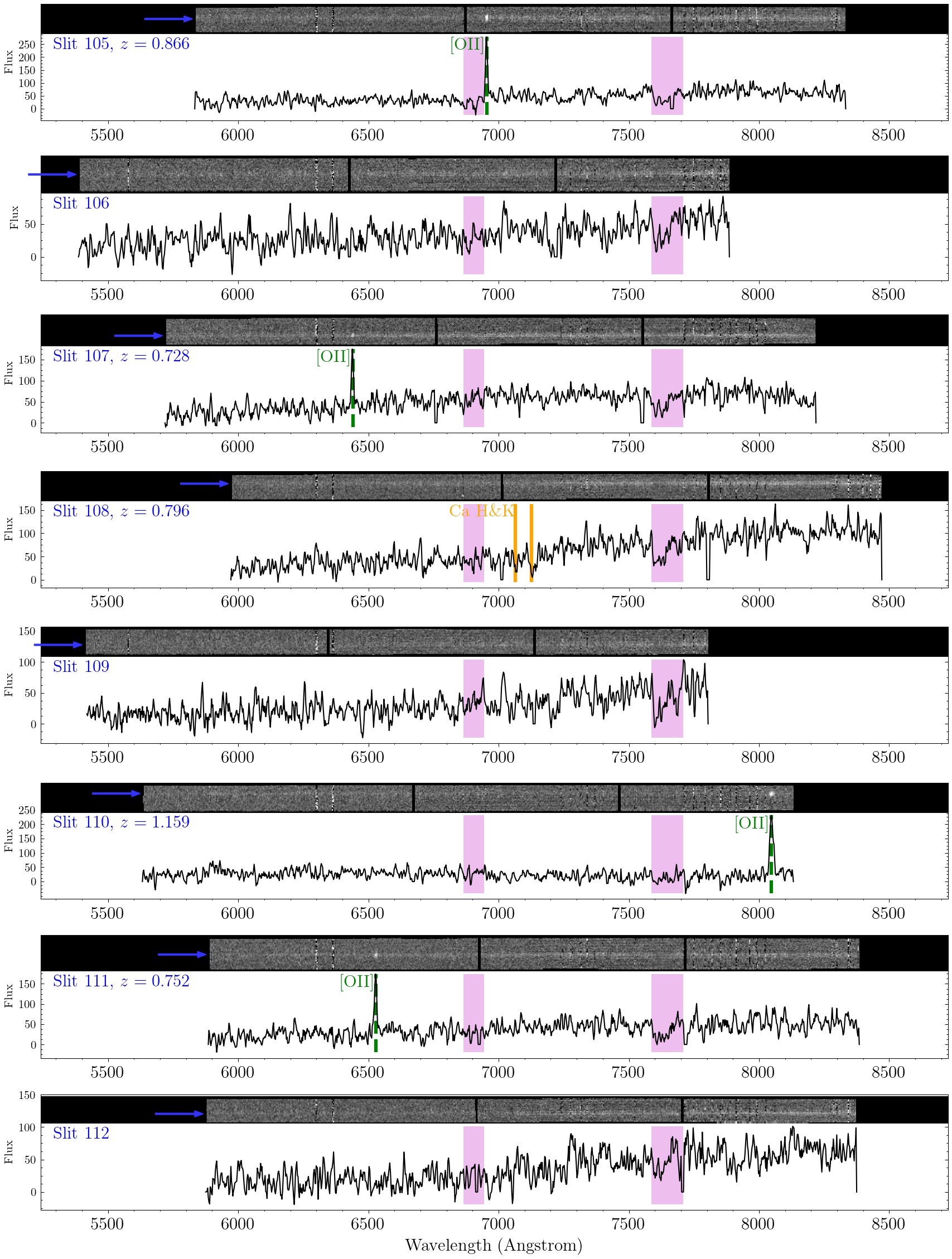}
    \caption{Galaxy Spectra of Slit~\#105--112.}
\end{figure*}

\clearpage
\begin{figure*}
    \centering     \includegraphics[width=\textwidth, height=0.96\textheight, keepaspectratio]{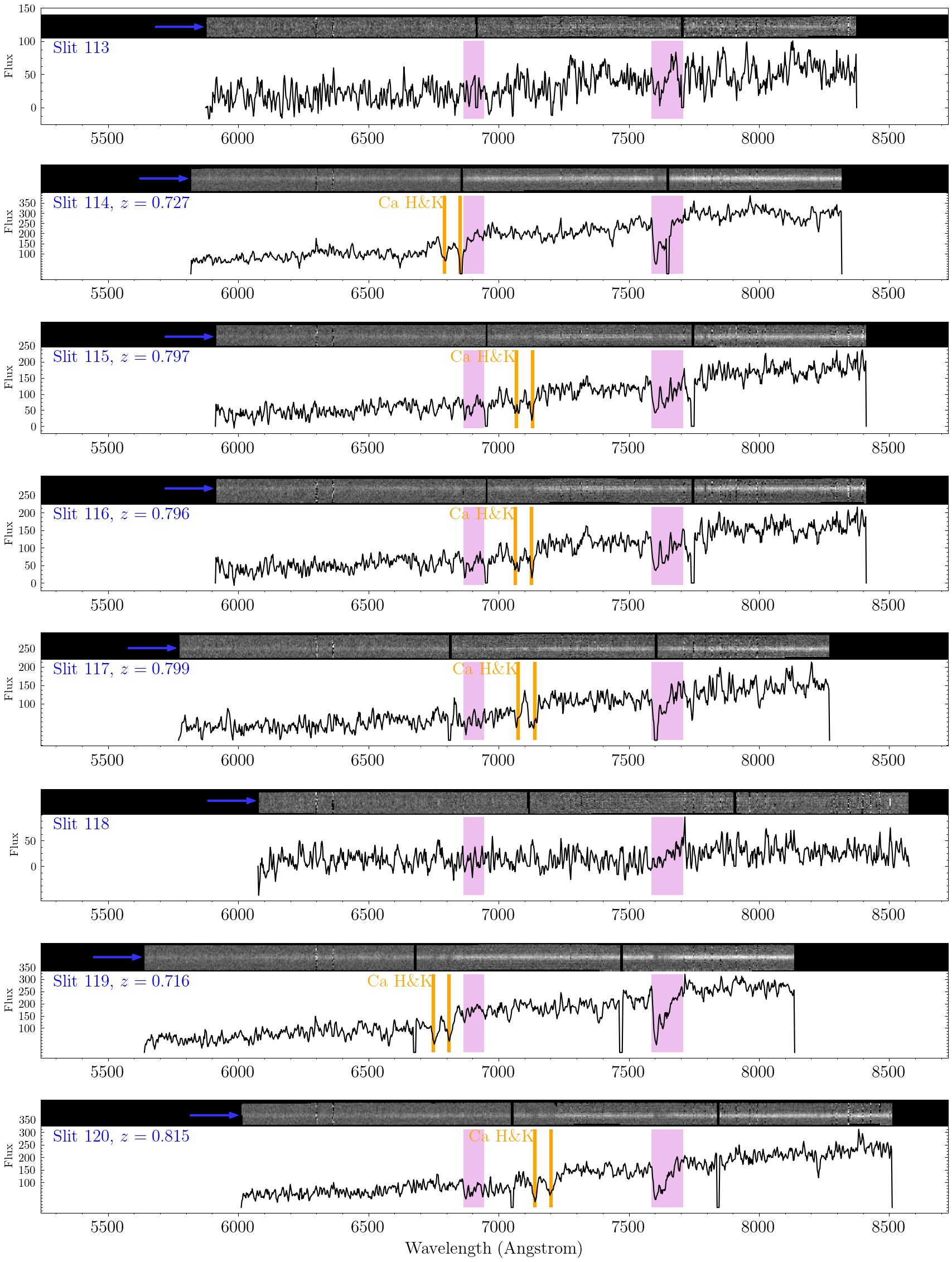}
    \caption{Galaxy Spectra of Slit~\#113--120.}
\end{figure*}

\clearpage
\begin{figure*}
    \centering     \includegraphics[width=\textwidth, height=0.96\textheight, keepaspectratio]{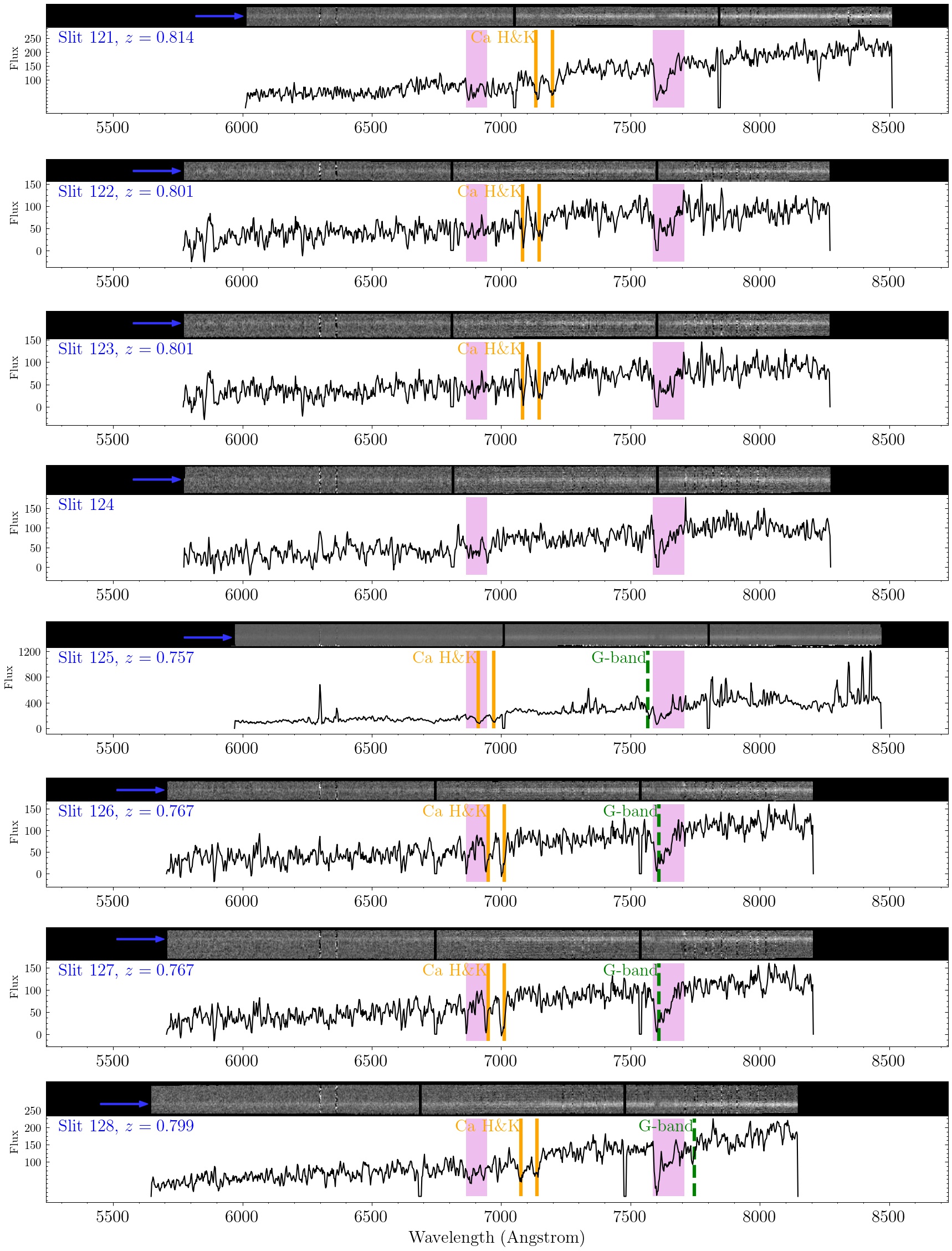}
    \caption{Galaxy Spectra of Slit~\#121--128.}
\end{figure*}

\clearpage
\begin{figure*}
    \centering     \includegraphics[width=\textwidth, height=0.96\textheight, keepaspectratio]{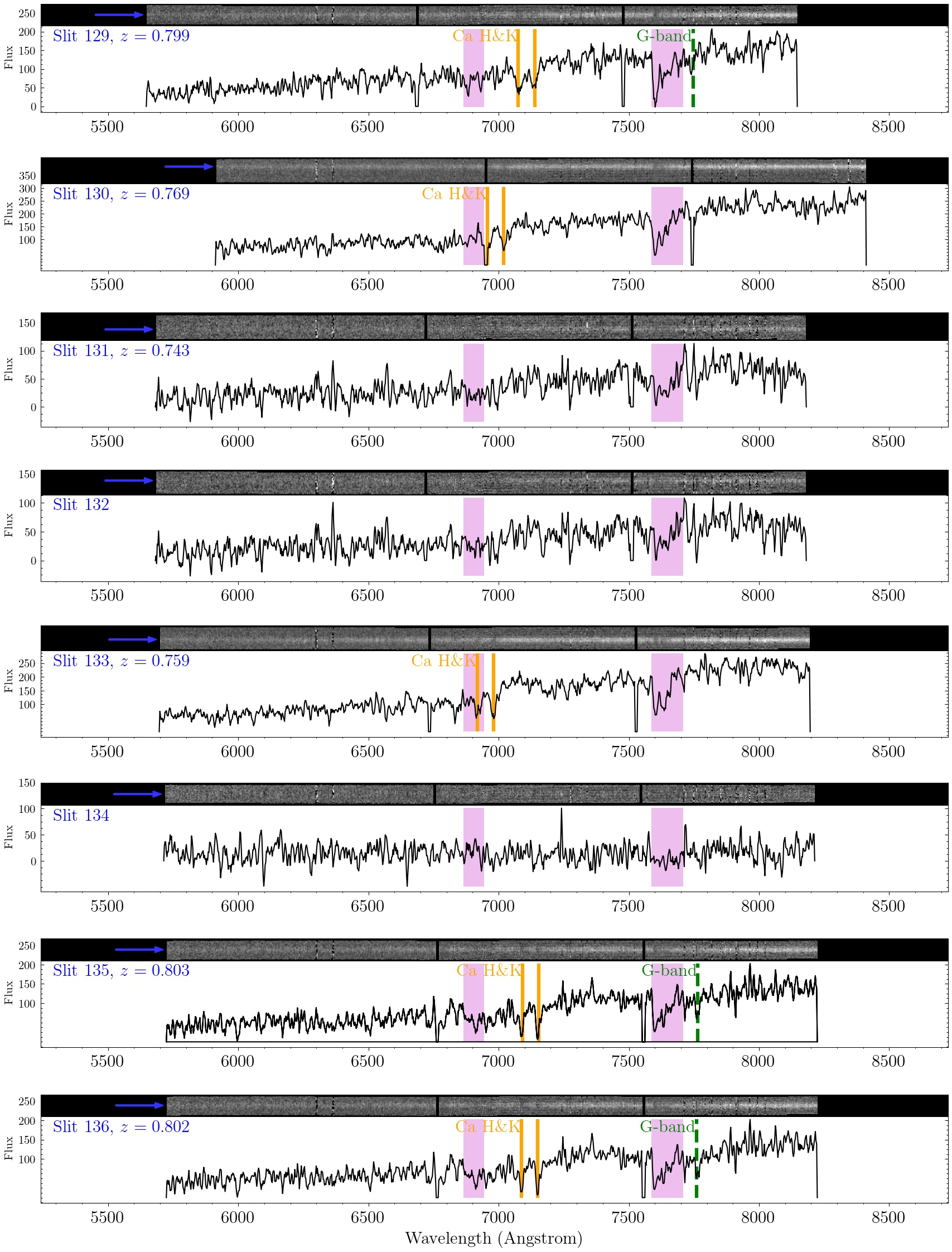}
    \caption{Galaxy Spectra of Slit~\#129--136.}
\end{figure*}

\clearpage
\begin{figure*}
    \centering     \includegraphics[width=\textwidth, height=0.96\textheight, keepaspectratio]{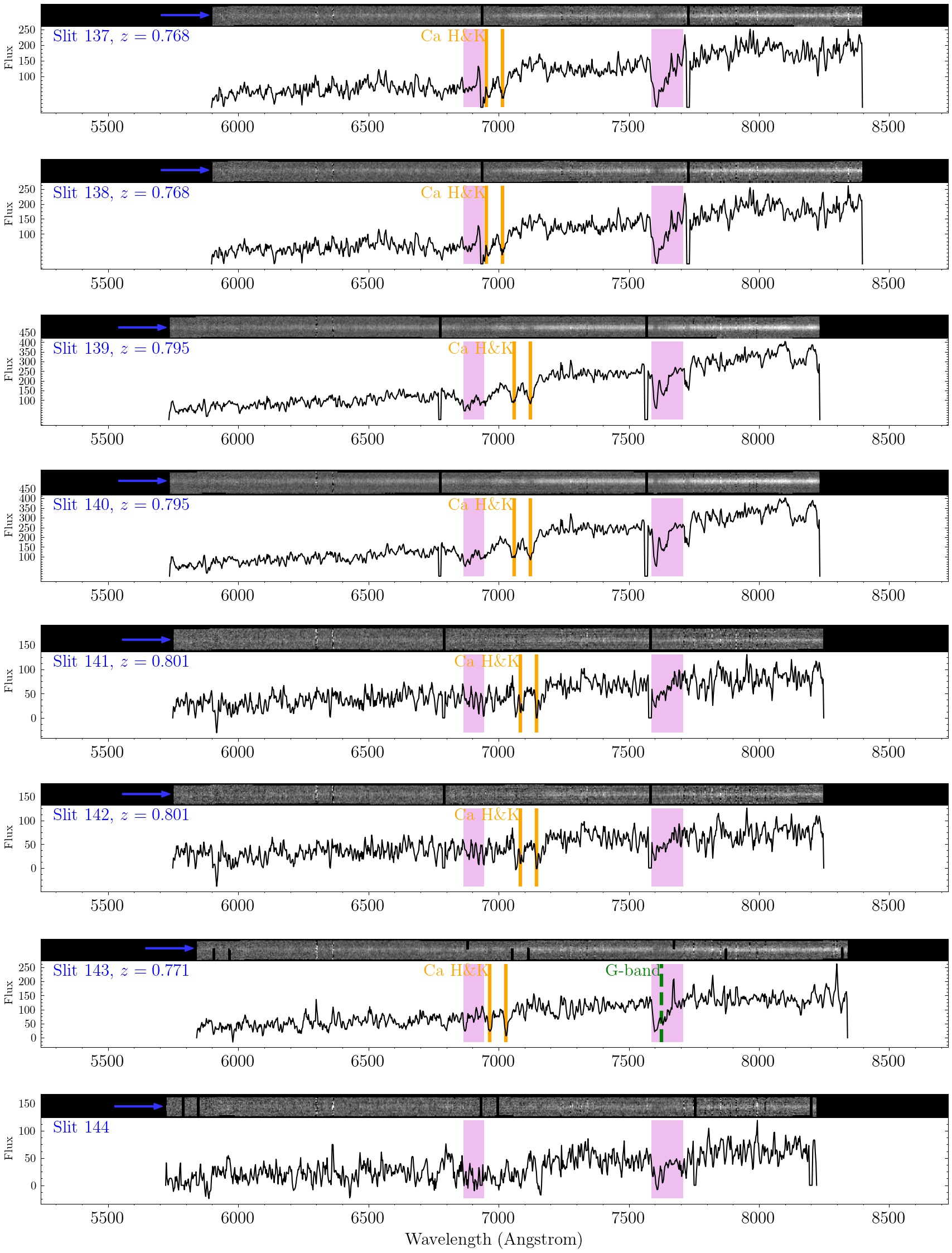}
    \caption{Galaxy Spectra of Slit~\#137--144.}
\end{figure*}

\clearpage
\begin{figure*}
    \centering     \includegraphics[width=\textwidth, height=0.96\textheight, keepaspectratio]{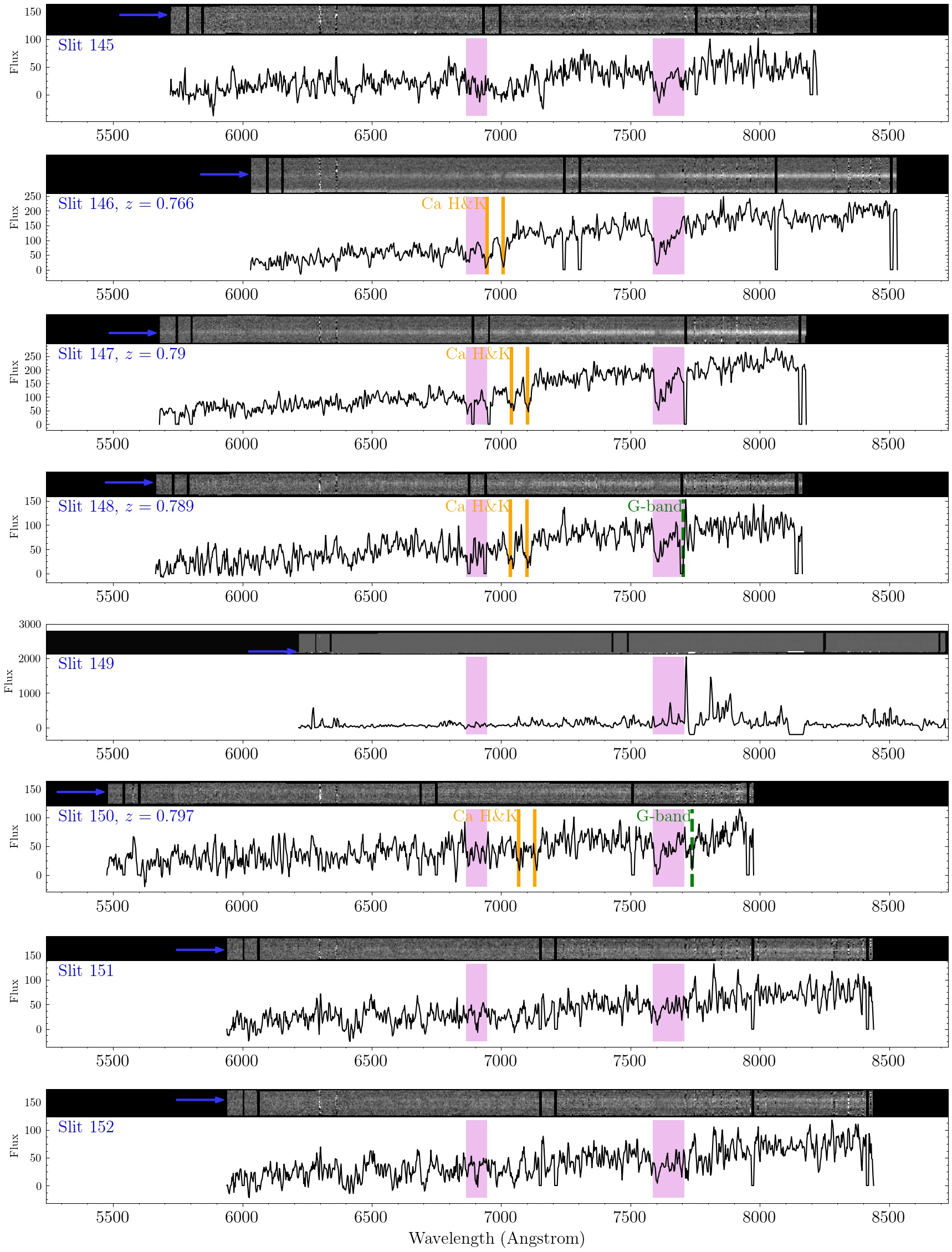}
    \caption{Galaxy Spectra of Slit~\#145--152.}
\end{figure*}

\clearpage
\begin{figure*}
    \centering     \includegraphics[width=\textwidth, height=0.96\textheight, keepaspectratio]{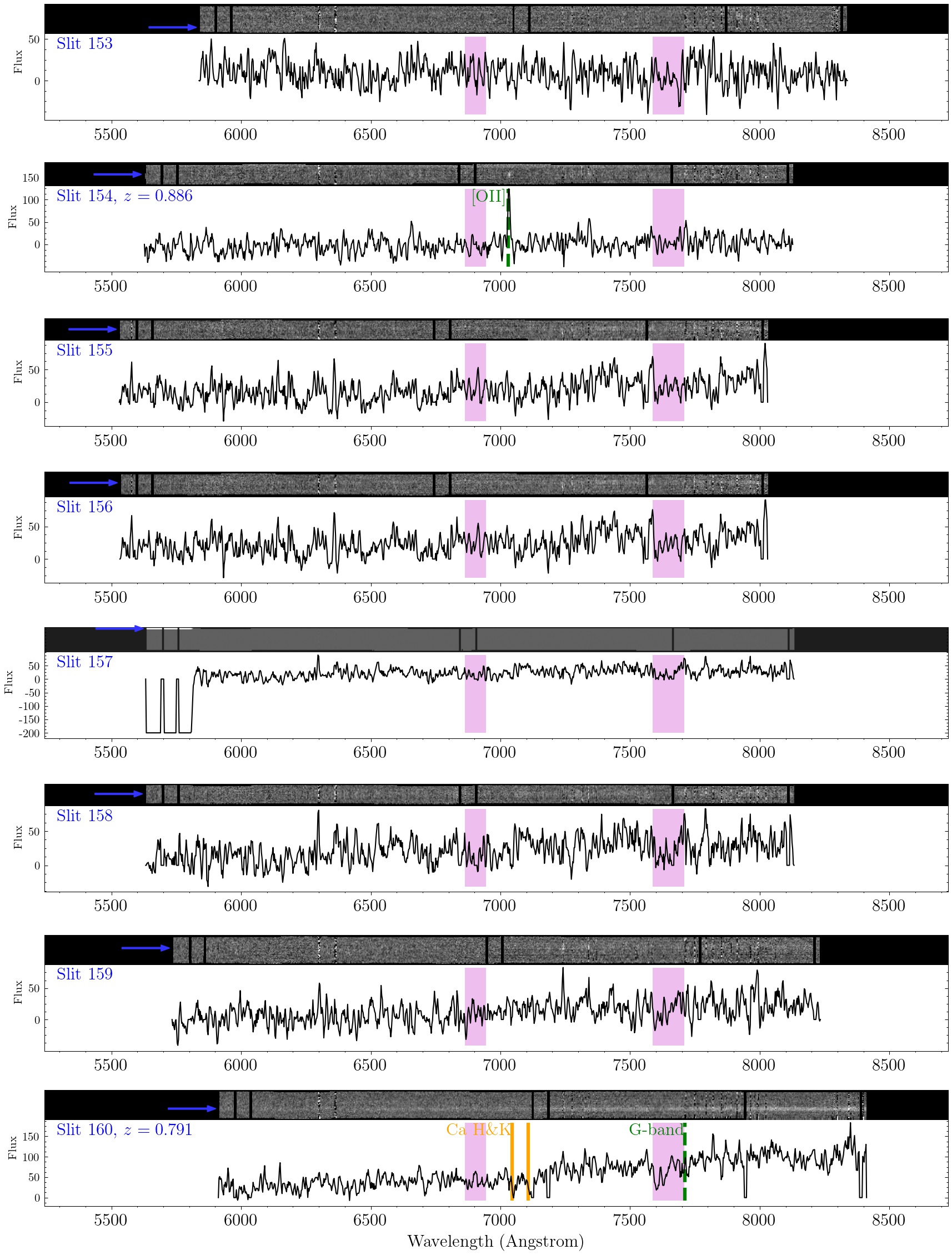}
    \caption{Galaxy Spectra of Slit~\#153--160.}
\end{figure*}

\clearpage
\begin{figure*}
    \centering     \includegraphics[width=\textwidth, height=0.96\textheight, keepaspectratio]{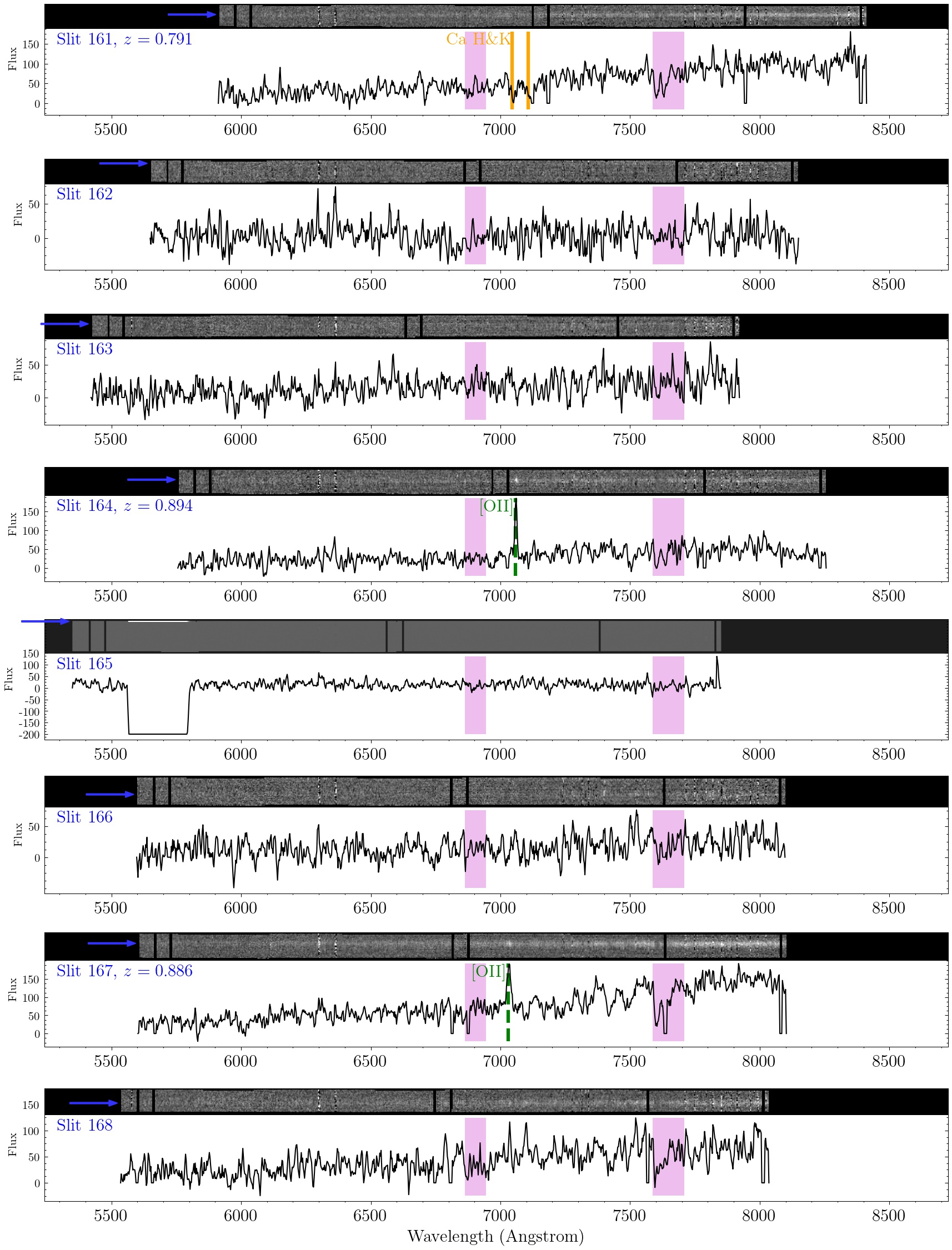}
    \caption{Galaxy Spectra of Slit~\#161--168.}
\end{figure*}

\clearpage
\begin{figure*}
    \centering     \includegraphics[width=\textwidth, height=0.96\textheight, keepaspectratio]{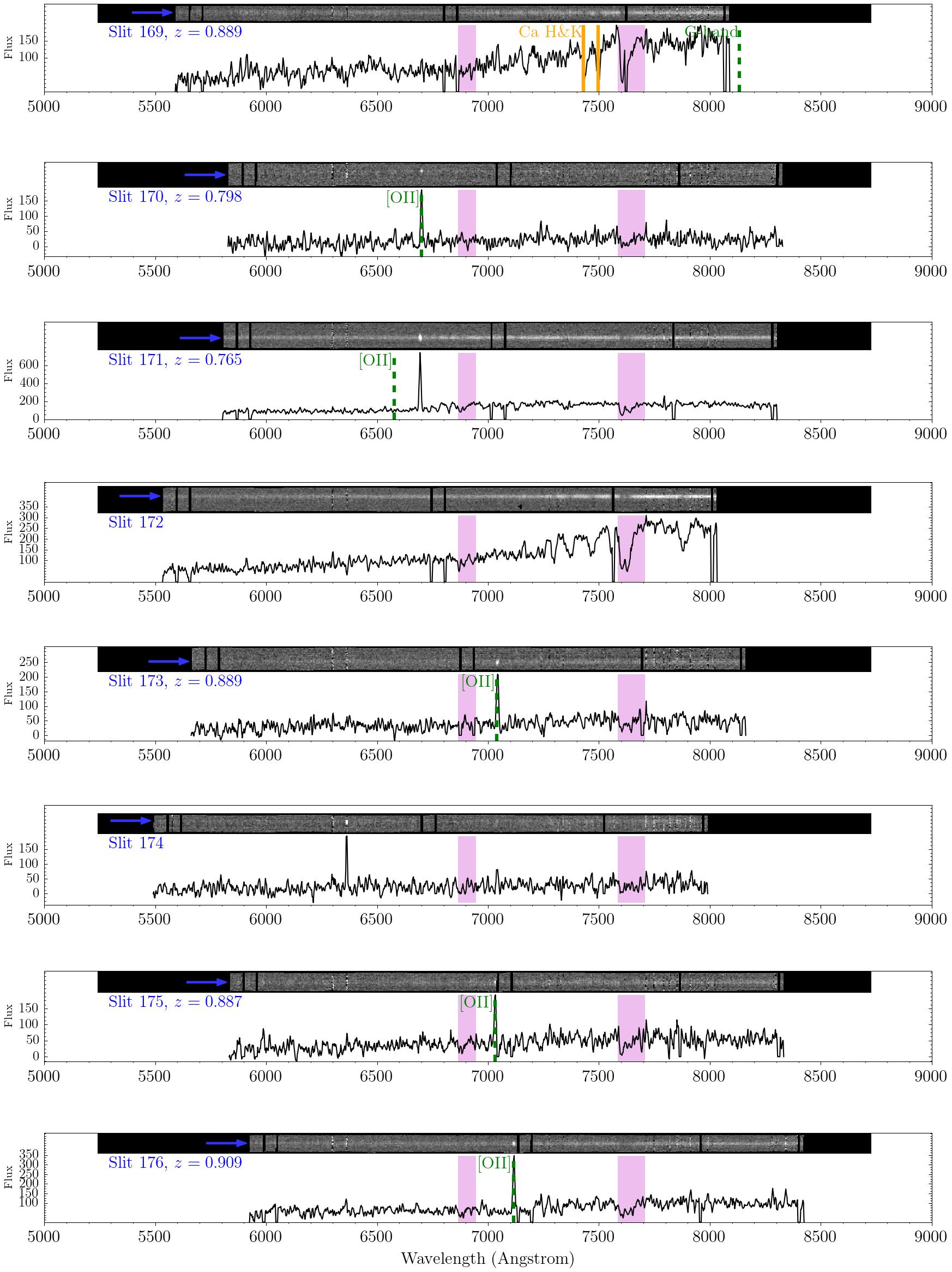}
    \caption{Galaxy Spectra of Slit~\#169--176.}
\end{figure*}

\clearpage
\begin{figure*}
    \centering     \includegraphics[width=\textwidth, height=0.96\textheight, keepaspectratio]{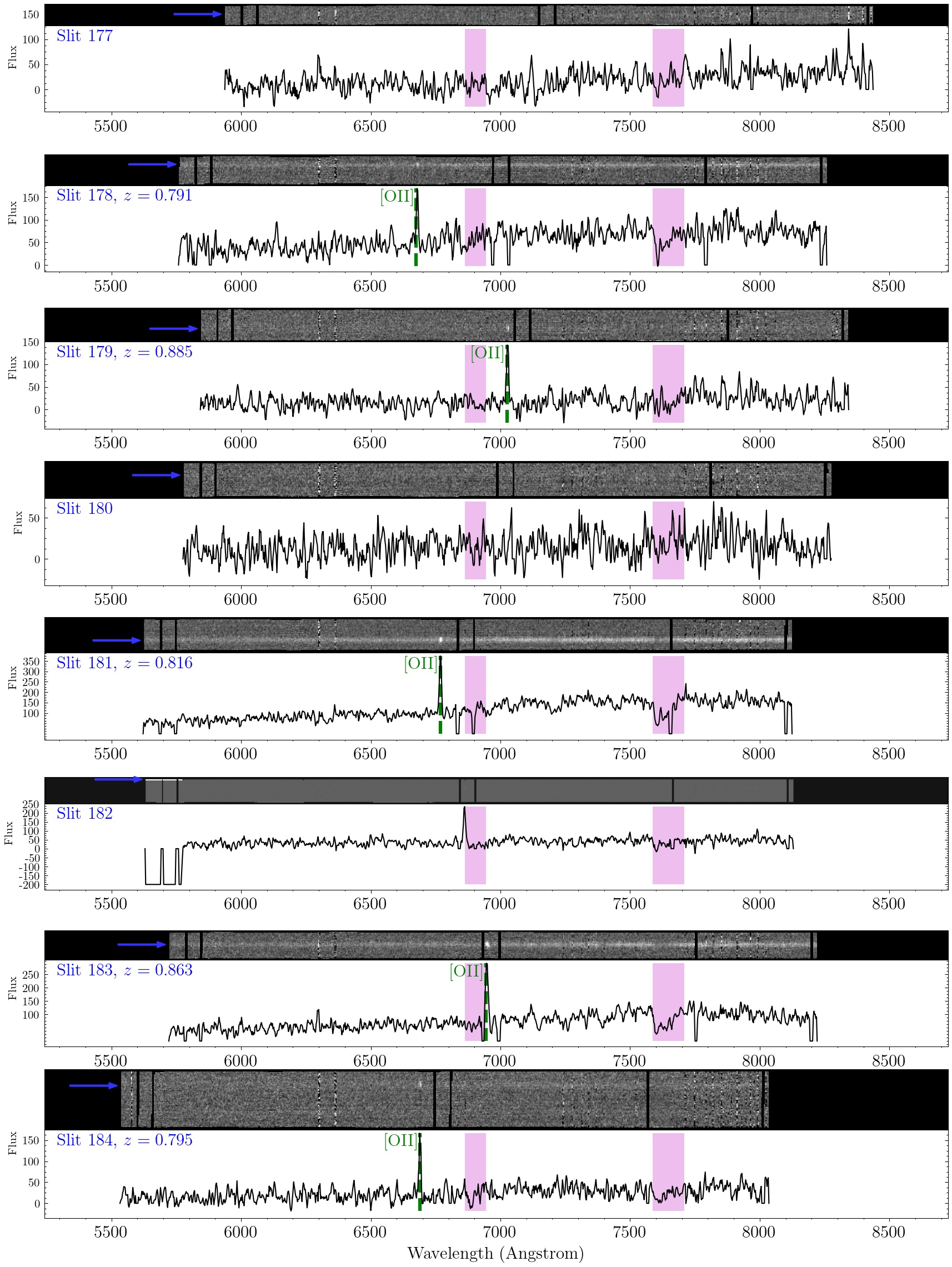}
    \caption{Galaxy Spectra of Slit~\#177--184.}
\end{figure*}

\clearpage
\begin{figure*}
    \centering     \includegraphics[width=\textwidth, height=0.96\textheight, keepaspectratio]{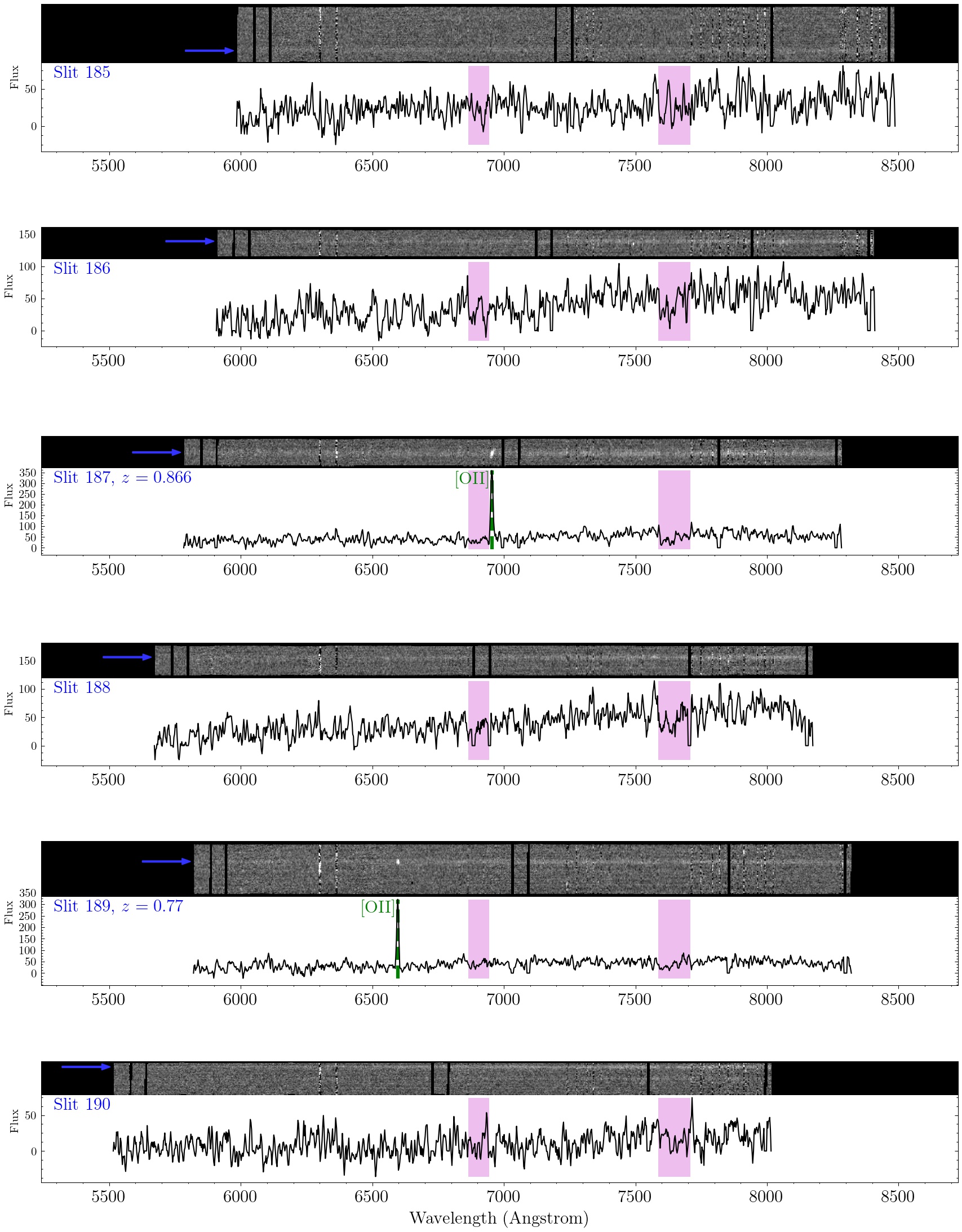}
    \caption{Galaxy Spectra of Slit~\#185--190.}
\end{figure*}

\clearpage
\begin{figure*}
    \centering     \includegraphics[width=\textwidth, height=0.96\textheight, keepaspectratio]{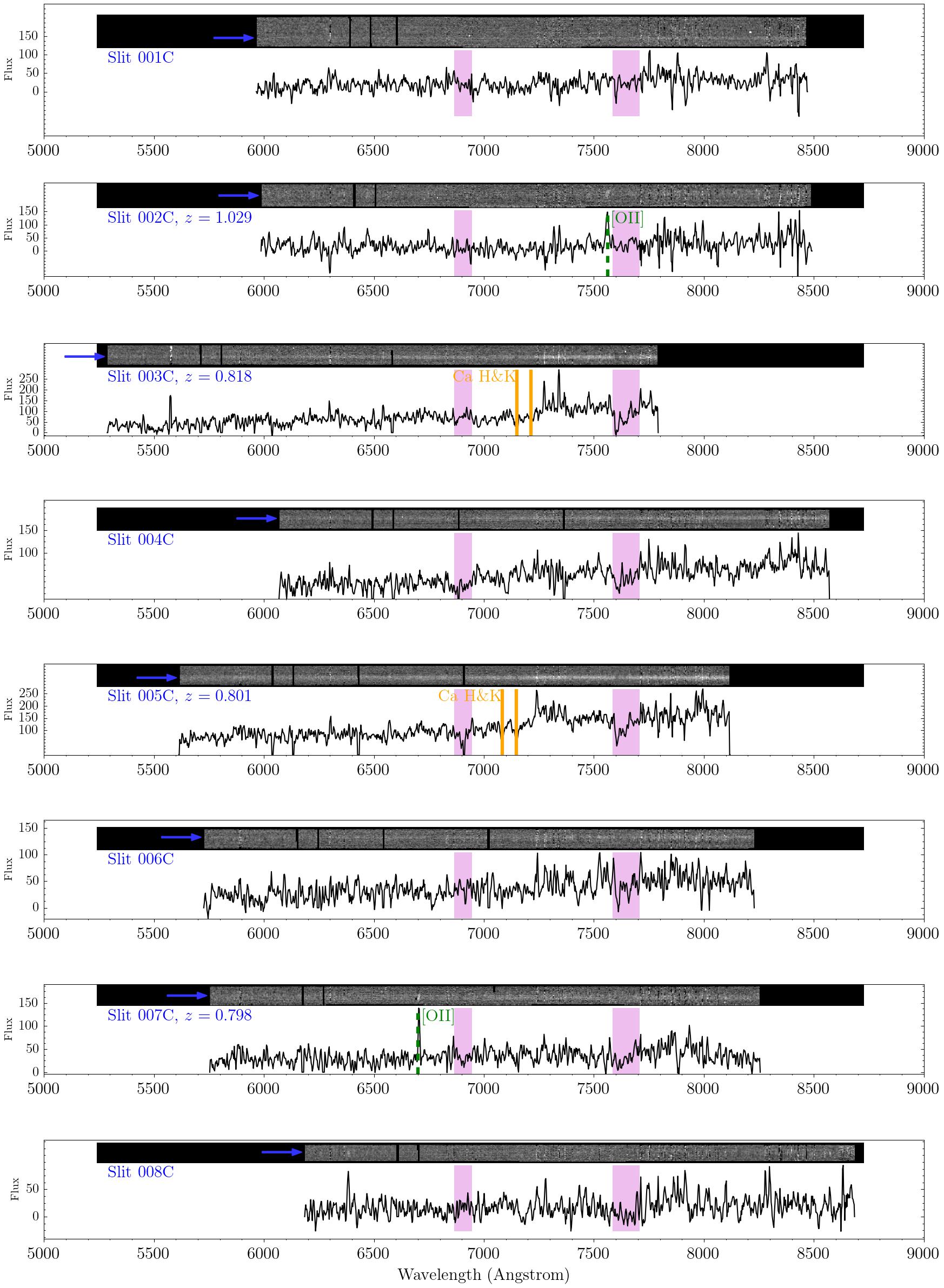}
    \caption{Galaxy Spectra of Slit~\#1C--8C.}
\end{figure*}

\clearpage
\begin{figure*}
    \centering     \includegraphics[width=\textwidth, height=0.96\textheight, keepaspectratio]{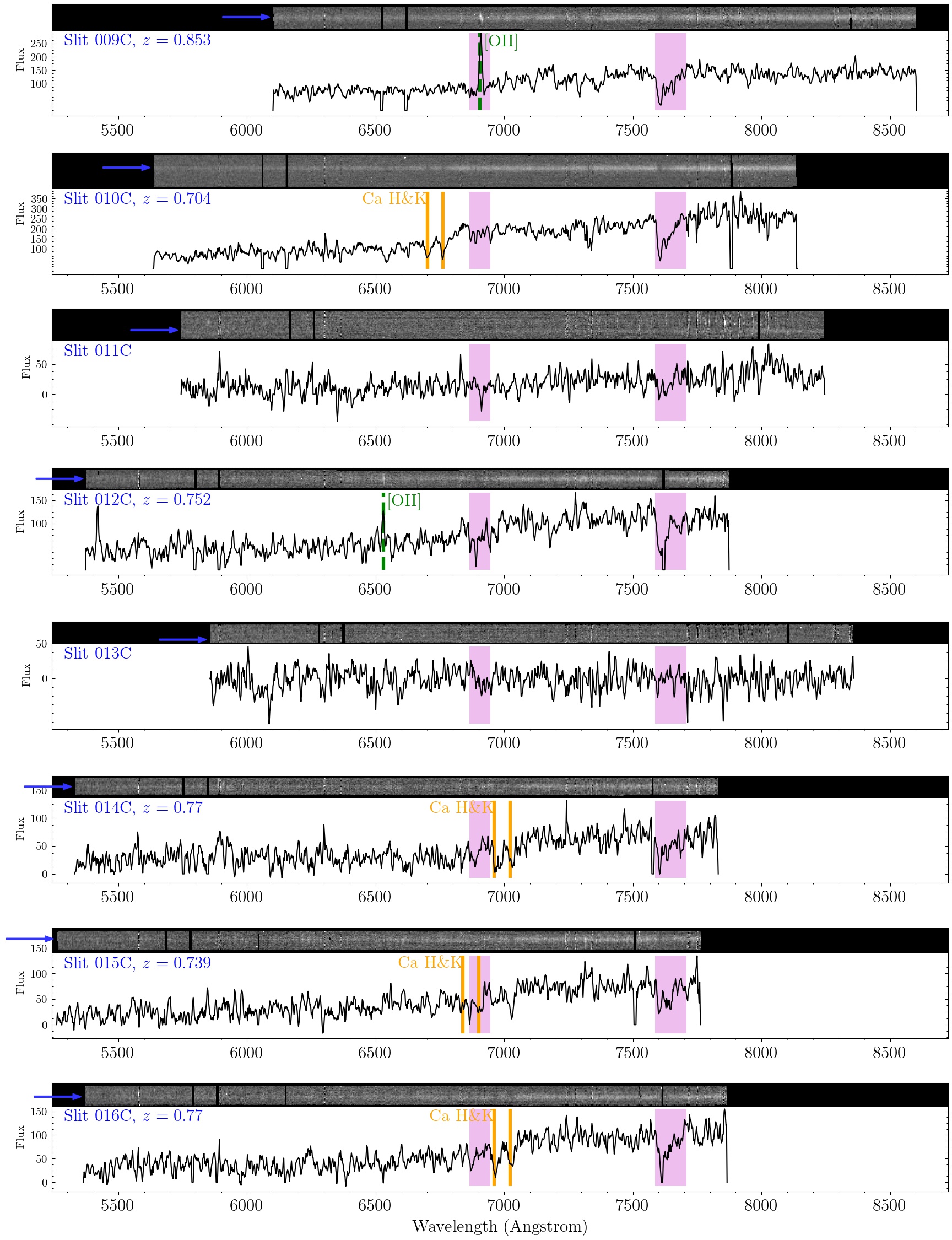}
    \caption{Galaxy Spectra of Slit~\#9C--16C.}
\end{figure*}

\clearpage
\begin{figure*}
    \centering     \includegraphics[width=\textwidth, height=0.96\textheight, keepaspectratio]{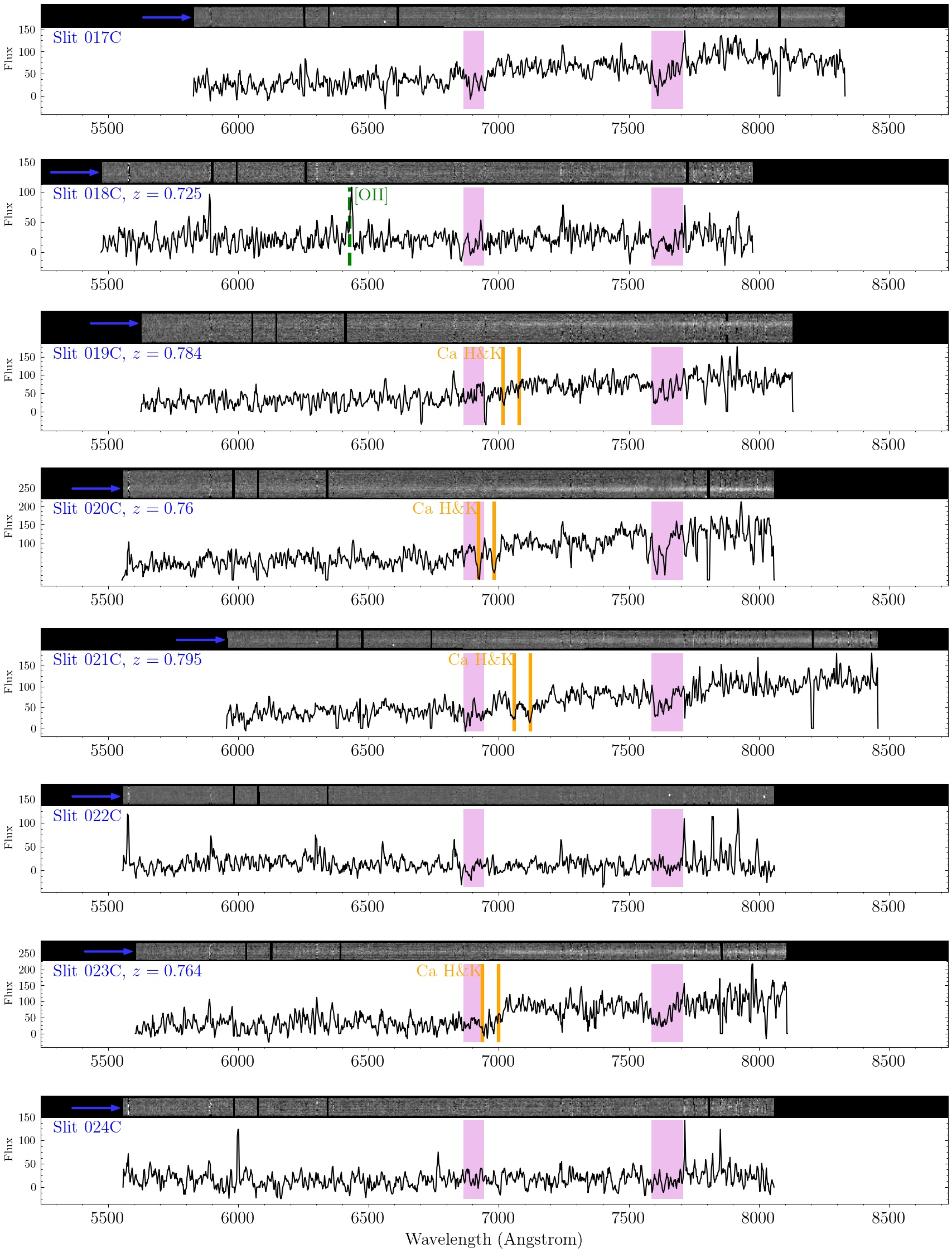}
    \caption{Galaxy Spectra of Slit~\#17C--24C.}
\end{figure*}

\clearpage
\begin{figure*}
    \centering     \includegraphics[width=\textwidth, height=0.96\textheight, keepaspectratio]{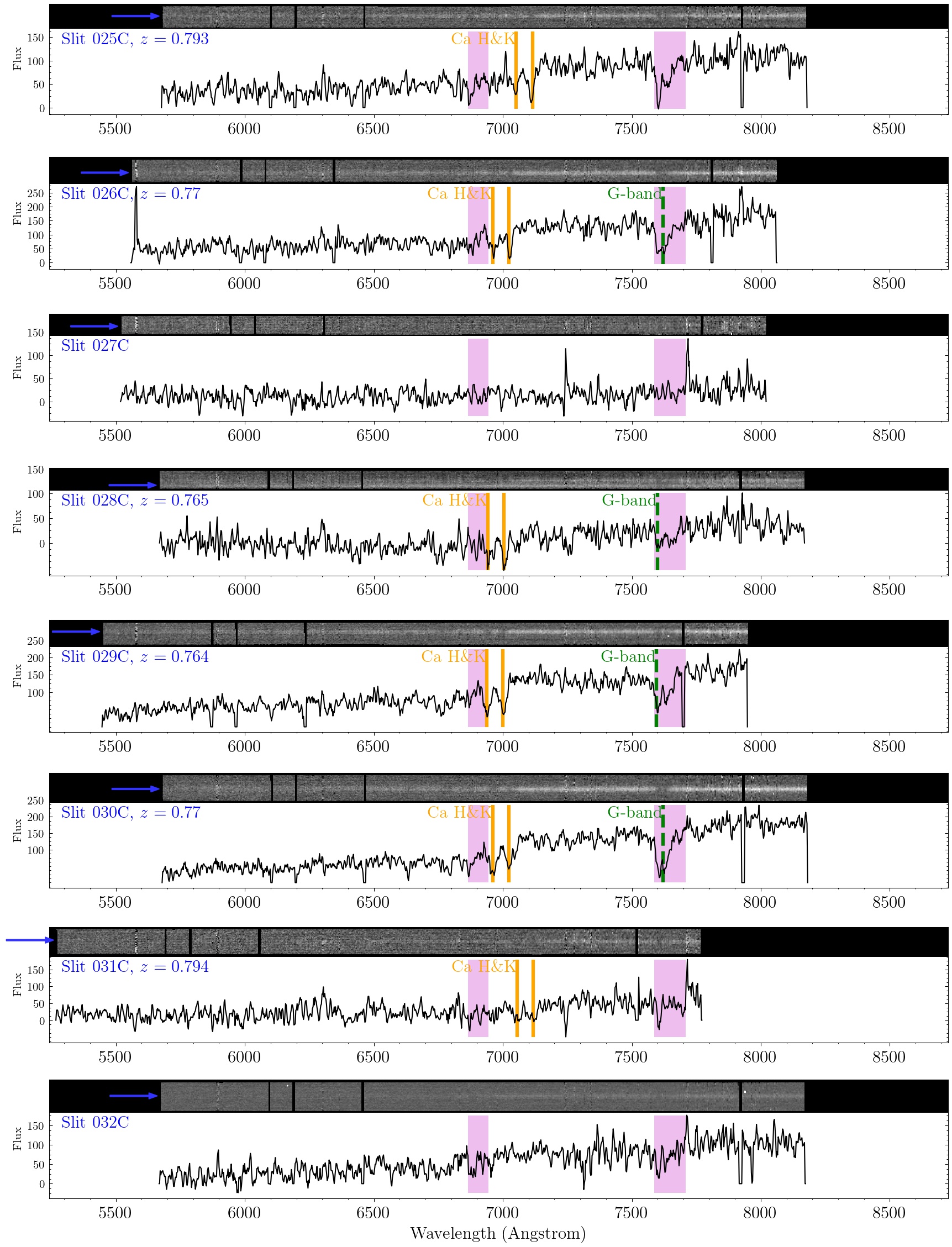}
    \caption{Galaxy Spectra of Slit~\#25C--32C.}
\end{figure*}

\clearpage
\begin{figure*}
    \centering     \includegraphics[width=\textwidth, height=0.96\textheight, keepaspectratio]{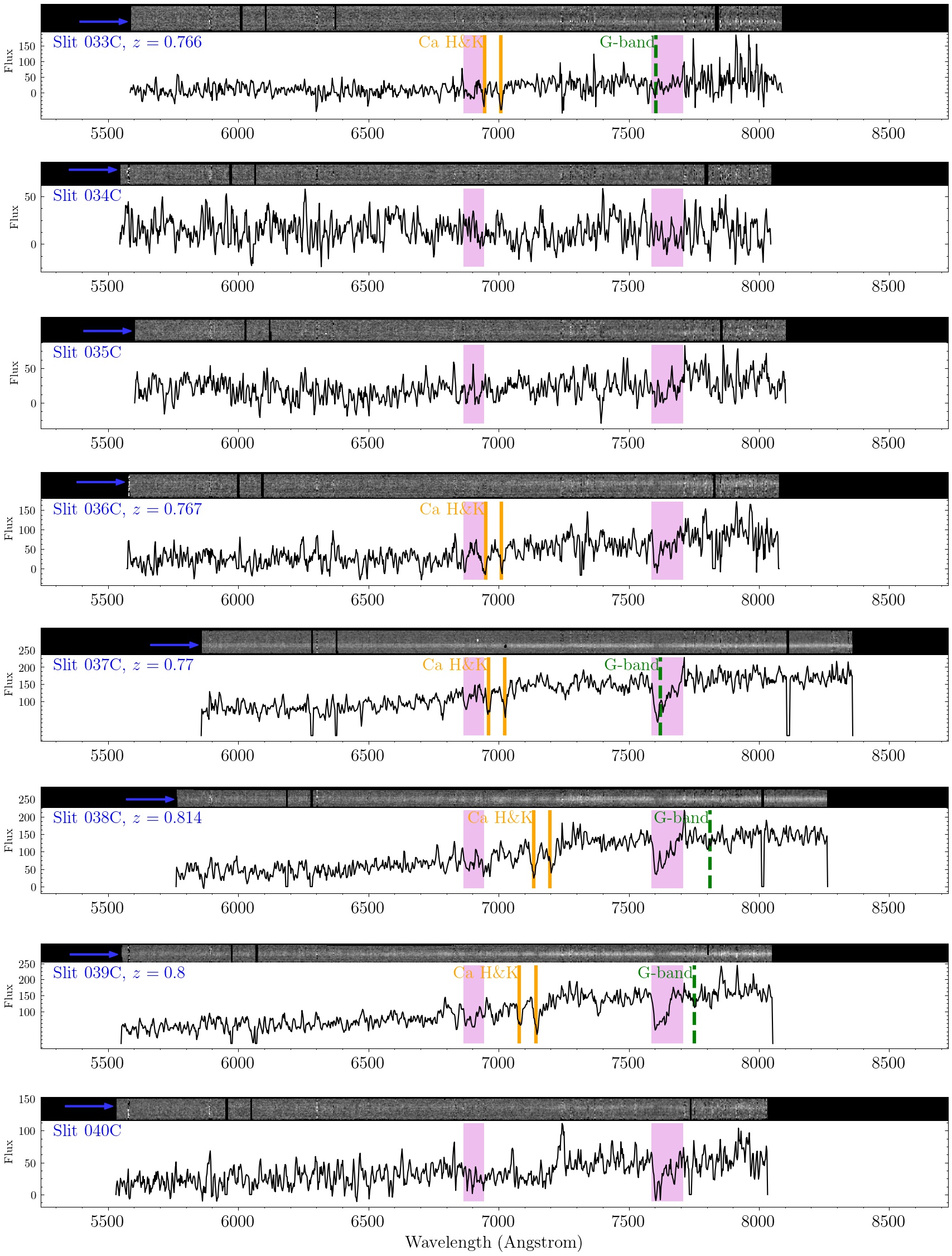}
    \caption{Galaxy Spectra of Slit~\#33C--40C.}
\end{figure*}

\clearpage
\begin{figure*}
    \centering     \includegraphics[width=\textwidth, height=0.96\textheight, keepaspectratio]{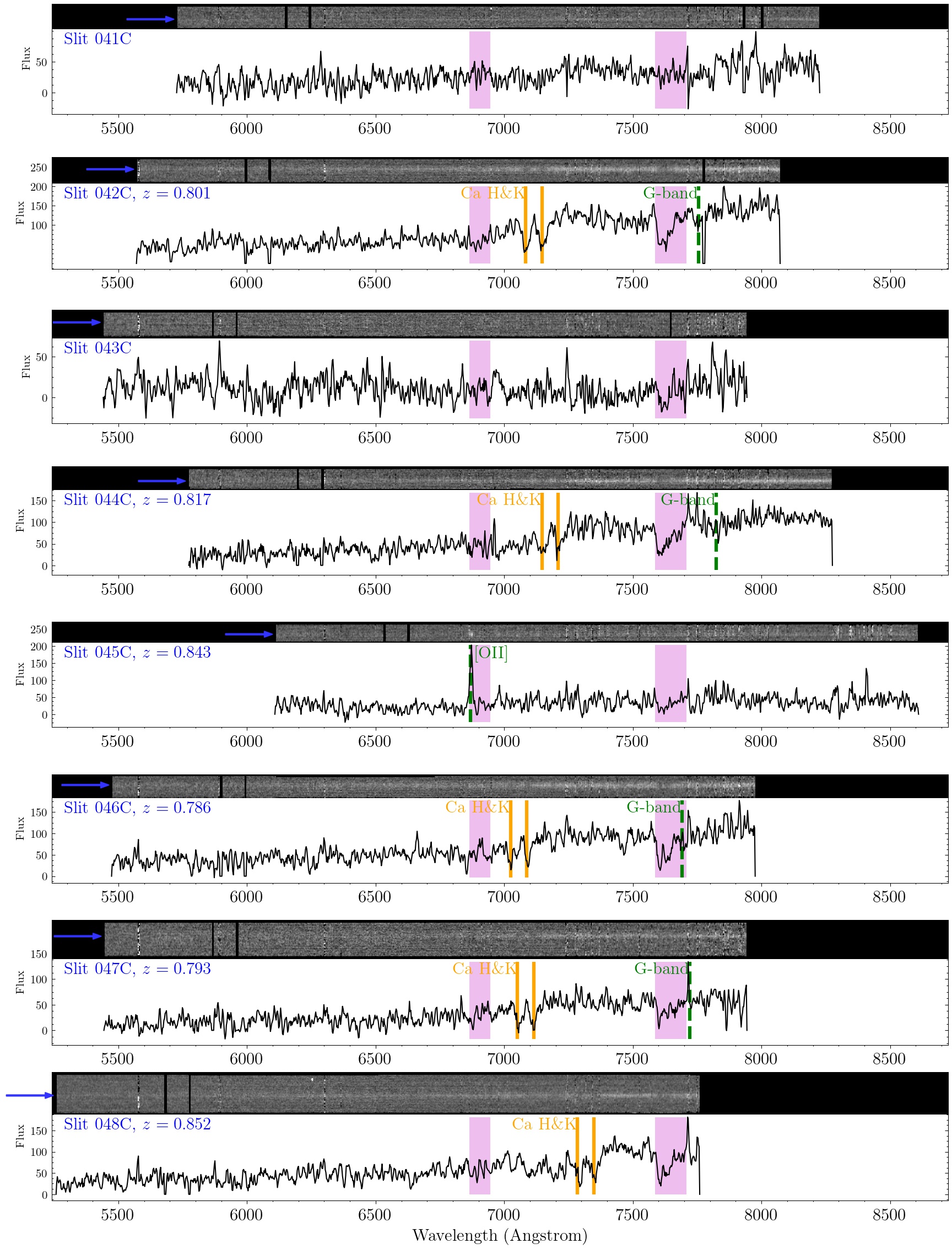}
    \caption{Galaxy Spectra of Slit~\#41C--48C.}
\end{figure*}

\clearpage
\begin{figure*}
    \centering     \includegraphics[width=\textwidth, height=0.96\textheight, keepaspectratio]{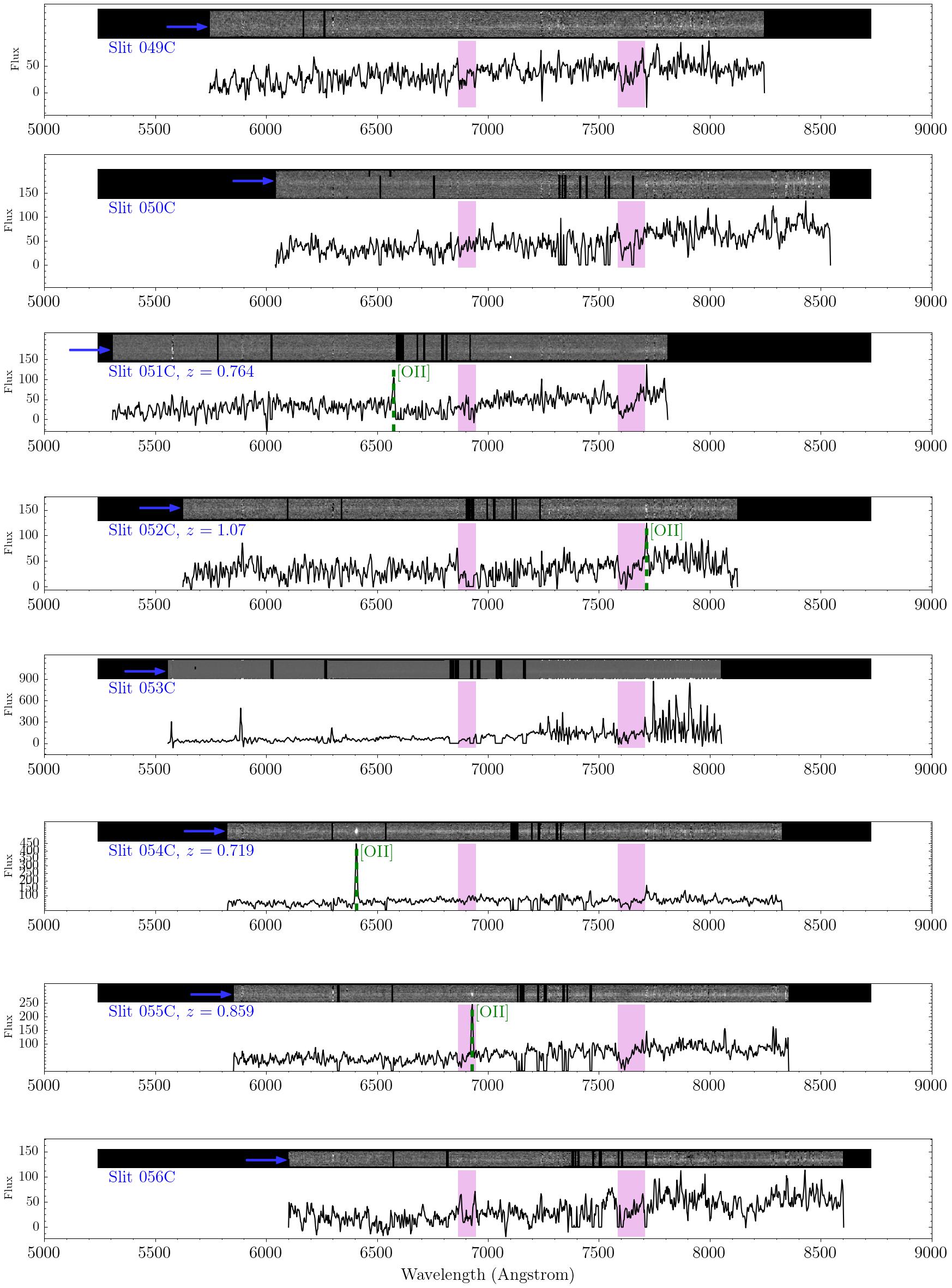}
    \caption{Galaxy Spectra of Slit~\#49C--56C.}
\end{figure*}

\clearpage
\begin{figure*}
    \centering     \includegraphics[width=\textwidth, height=0.96\textheight, keepaspectratio]{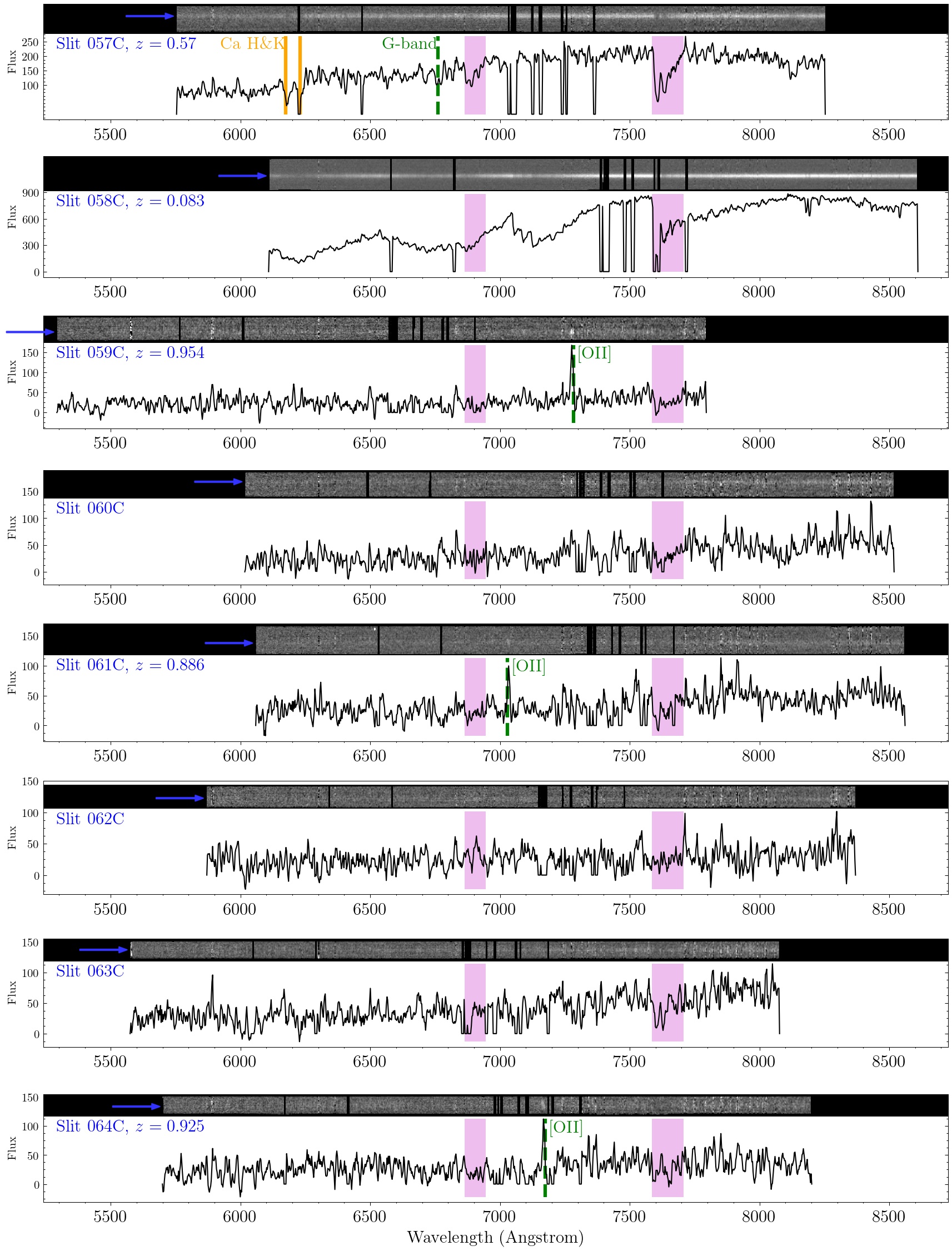}
    \caption{Galaxy Spectra of Slit~\#57C--64C.}
\end{figure*}

\clearpage
\begin{figure*}
    \centering     \includegraphics[width=\textwidth, height=0.96\textheight, keepaspectratio]{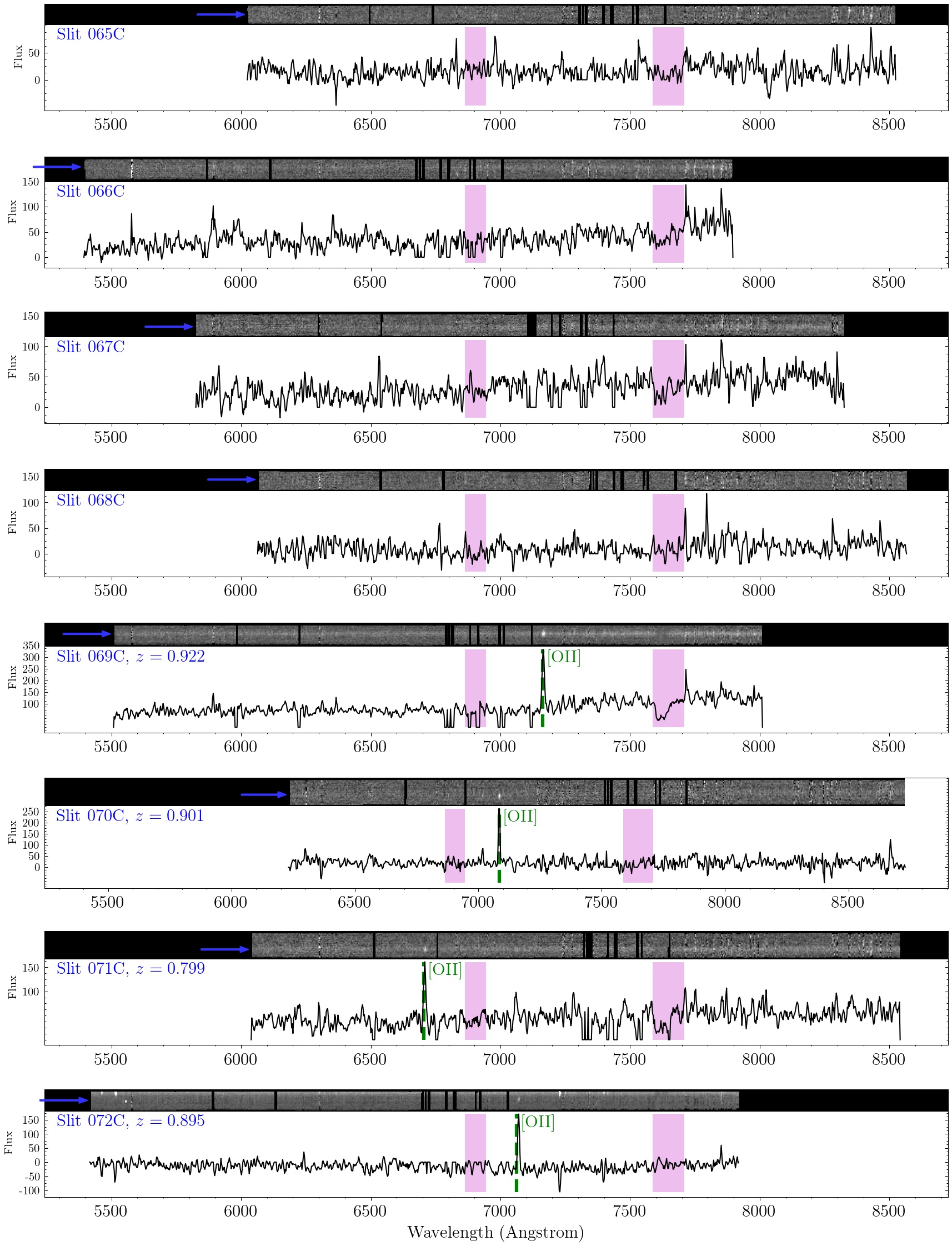}
    \caption{Galaxy Spectra of Slit~\#65C--72C.}
\end{figure*}

\clearpage
\begin{figure*}
    \centering     \includegraphics[width=\textwidth, height=0.96\textheight, keepaspectratio]{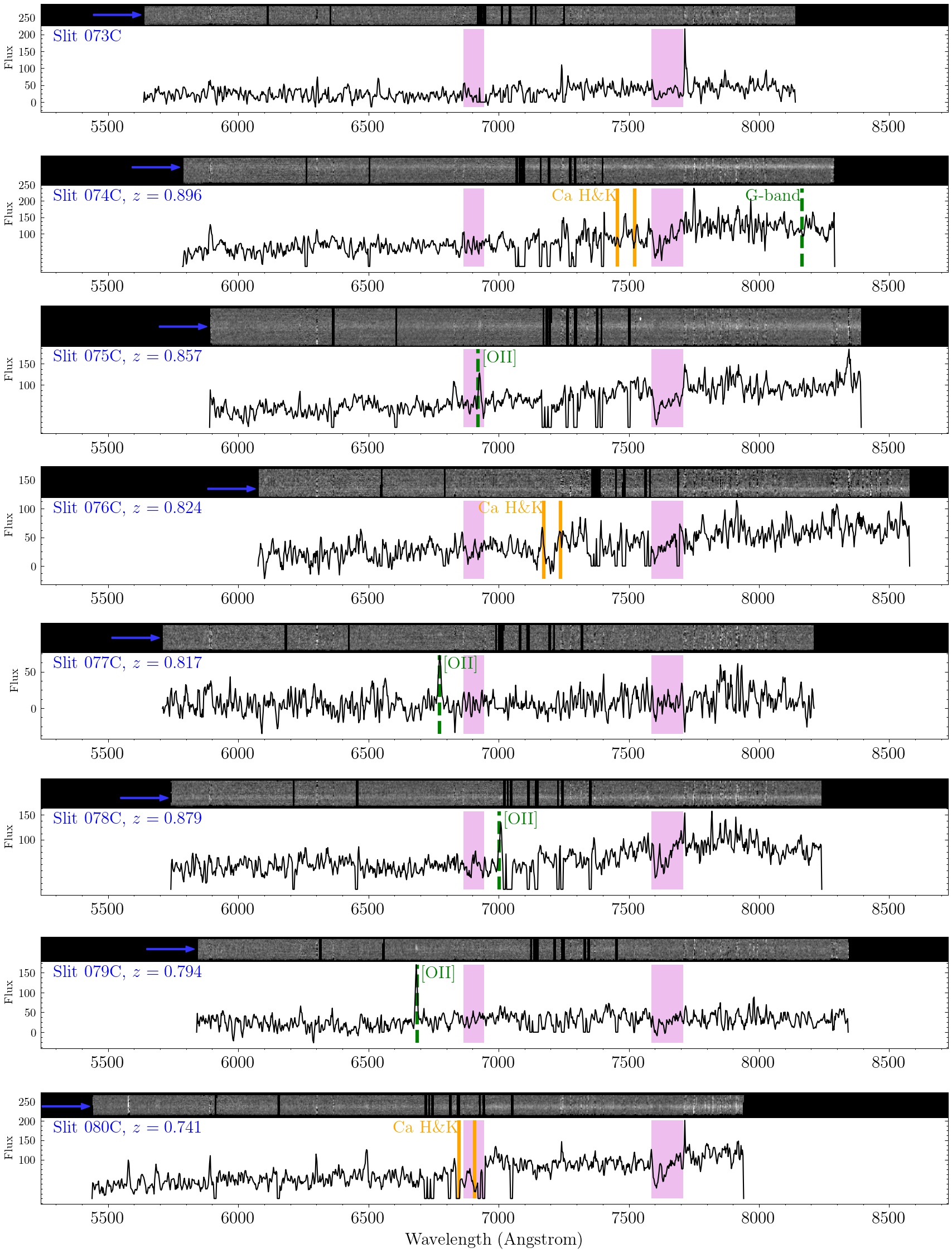}
    \caption{Galaxy Spectra of Slit~\#73C--80C.}
\end{figure*}

\clearpage
\begin{figure*}
    \centering     \includegraphics[width=\textwidth, height=0.96\textheight, keepaspectratio]{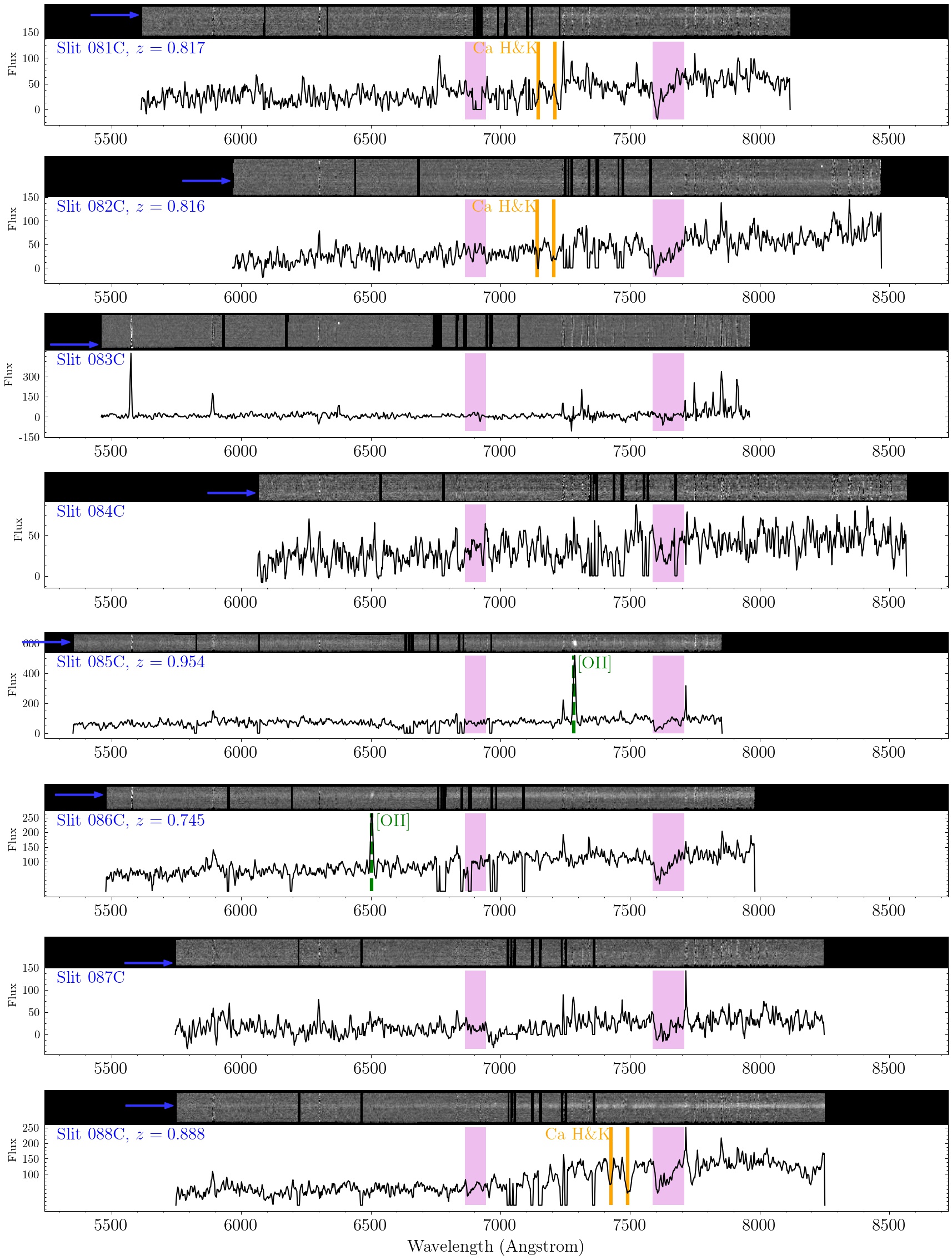}
    \caption{Galaxy Spectra of Slit~\#81C--88C.}
\end{figure*}

\clearpage
\begin{figure*}
    \centering     \includegraphics[width=\textwidth, height=0.96\textheight, keepaspectratio]{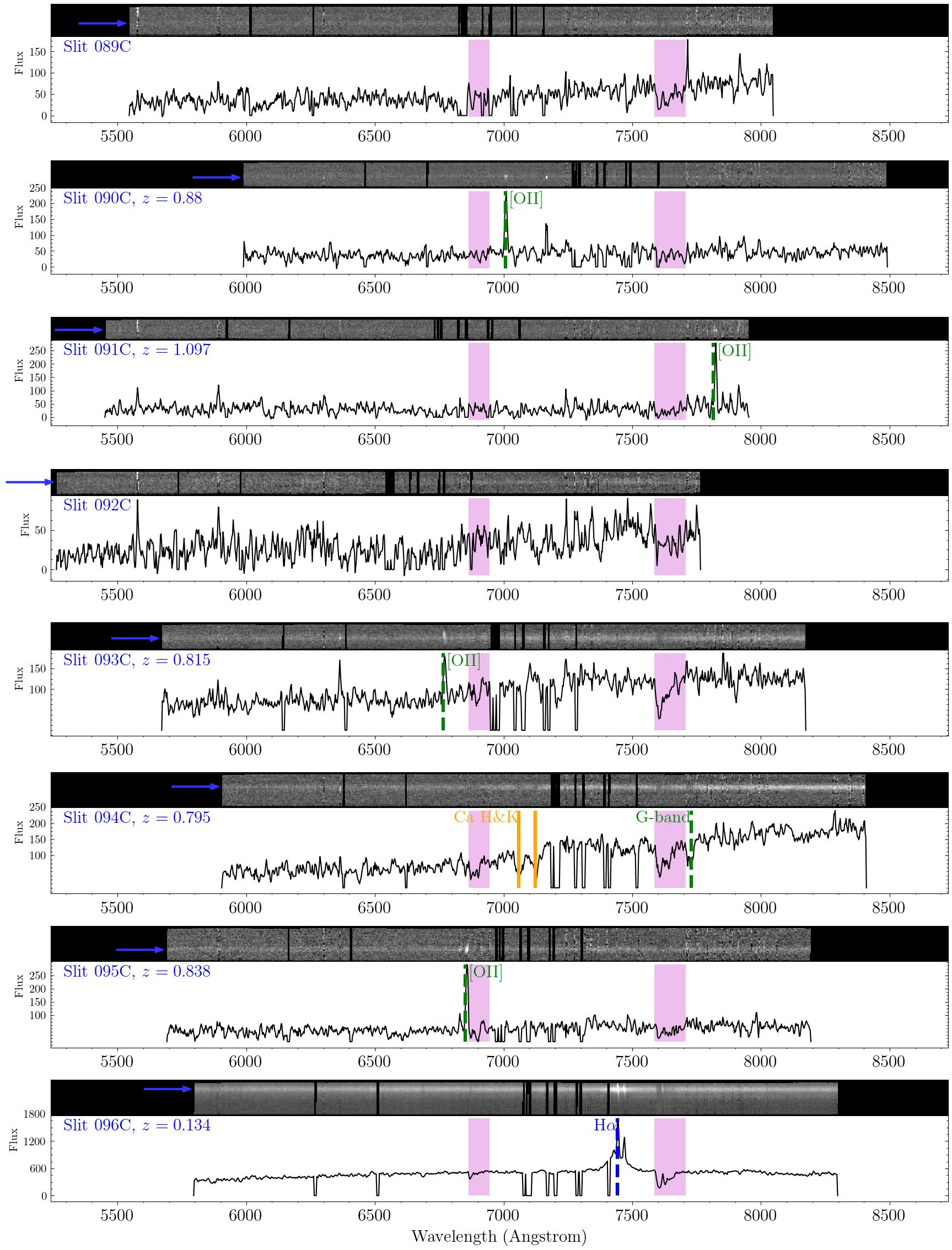}
    \caption{Galaxy Spectra of Slit~\#89C--96C.}
\end{figure*}

\clearpage
\begin{figure*}
    \centering     \includegraphics[width=\textwidth, height=0.96\textheight, keepaspectratio]{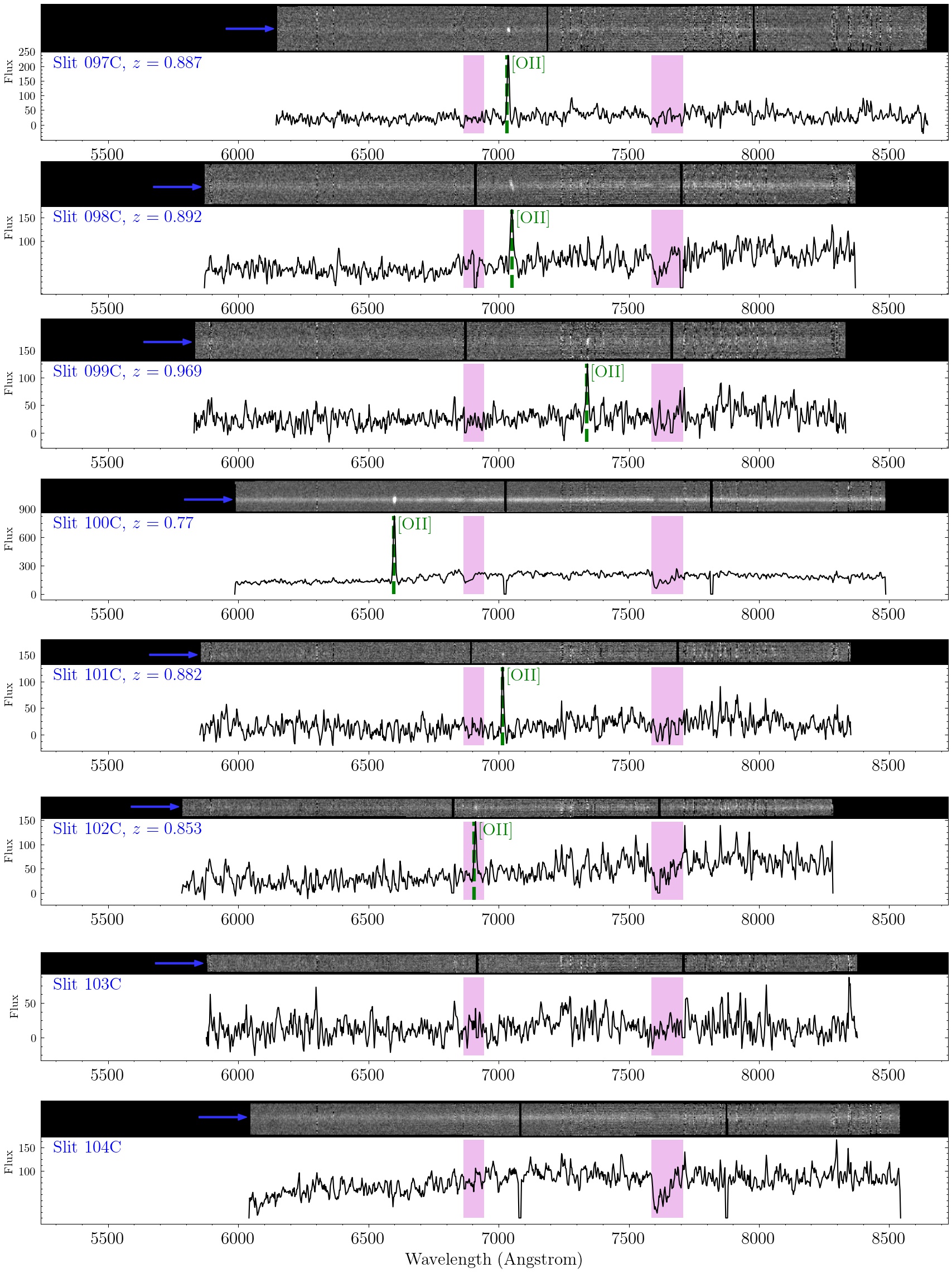}
    \caption{Galaxy Spectra of Slit~\#97C--104C.}
\end{figure*}

\clearpage
\begin{figure*}
    \centering     \includegraphics[width=\textwidth, height=0.96\textheight, keepaspectratio]{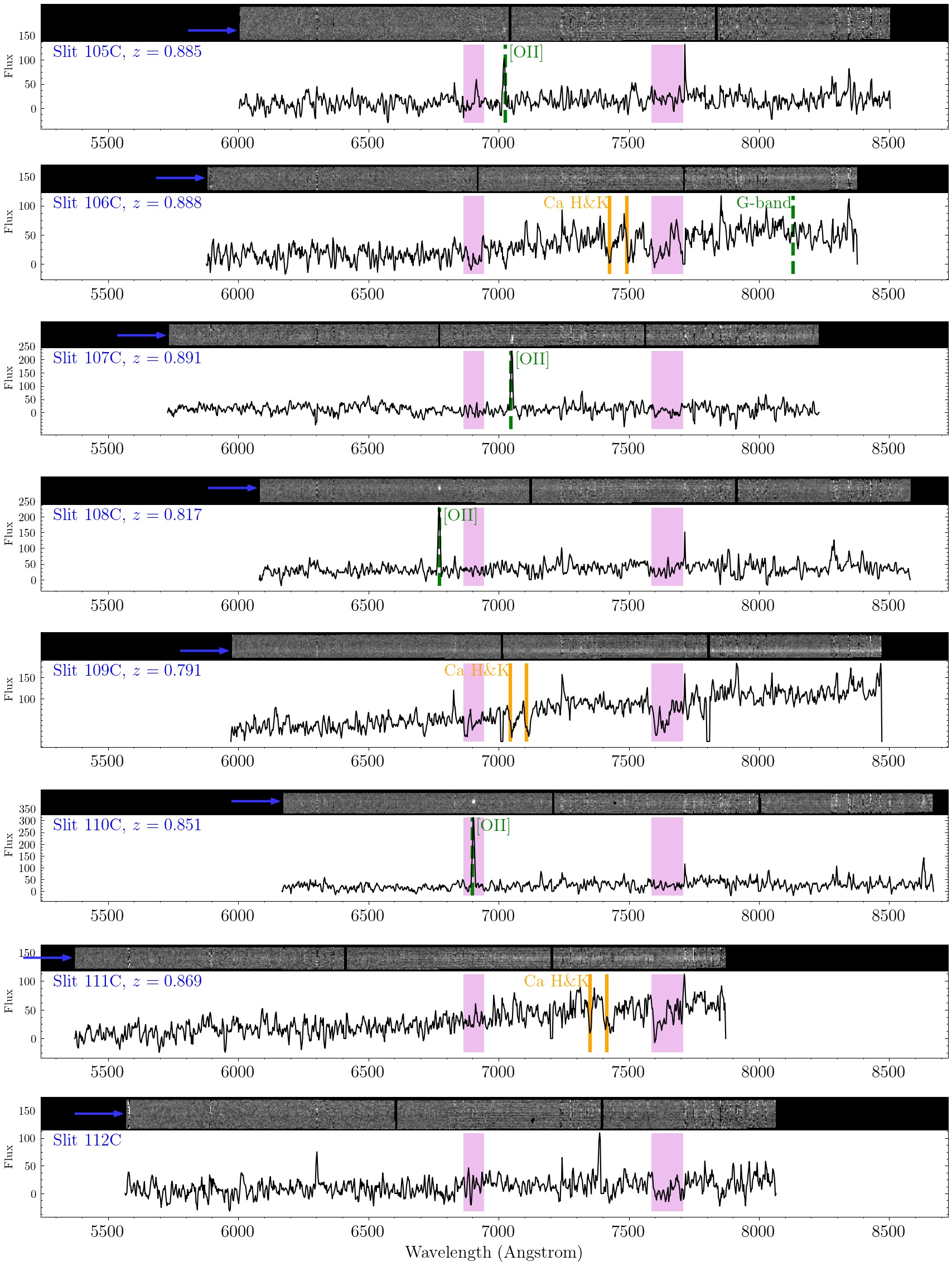}
    \caption{Galaxy Spectra of Slit~\#105C--112C.}
\end{figure*}

\clearpage
\begin{figure*}
    \centering     \includegraphics[width=\textwidth, height=0.96\textheight, keepaspectratio]{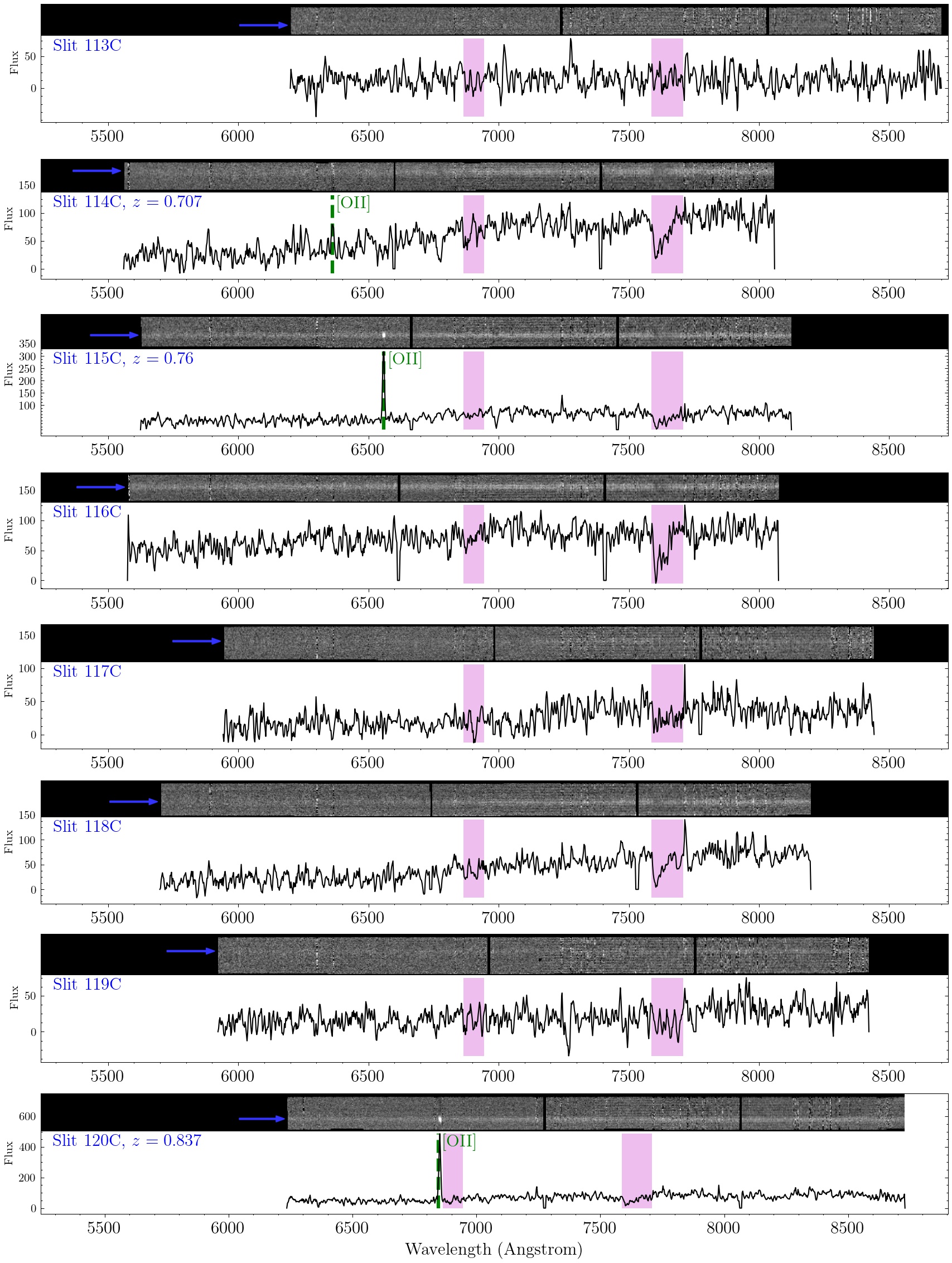}
    \caption{Galaxy Spectra of Slit~\#113C--120C.}
\end{figure*}

\clearpage
\begin{figure*}
    \centering     \includegraphics[width=\textwidth, height=0.96\textheight, keepaspectratio]{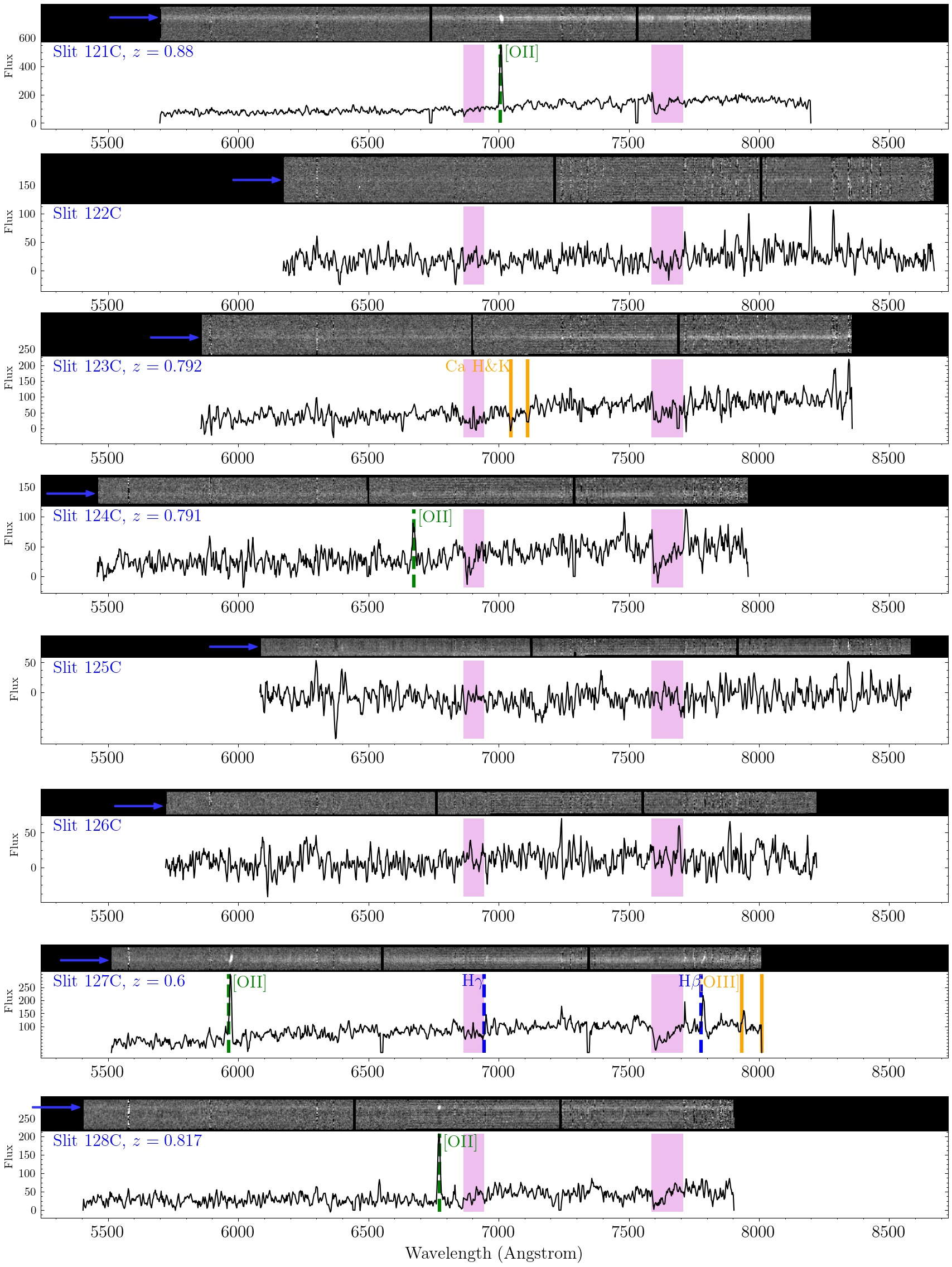}
    \caption{Galaxy Spectra of Slit~\#121C--128C.}
\end{figure*}

\clearpage
\begin{figure*}
    \centering     \includegraphics[width=\textwidth, height=0.96\textheight, keepaspectratio]{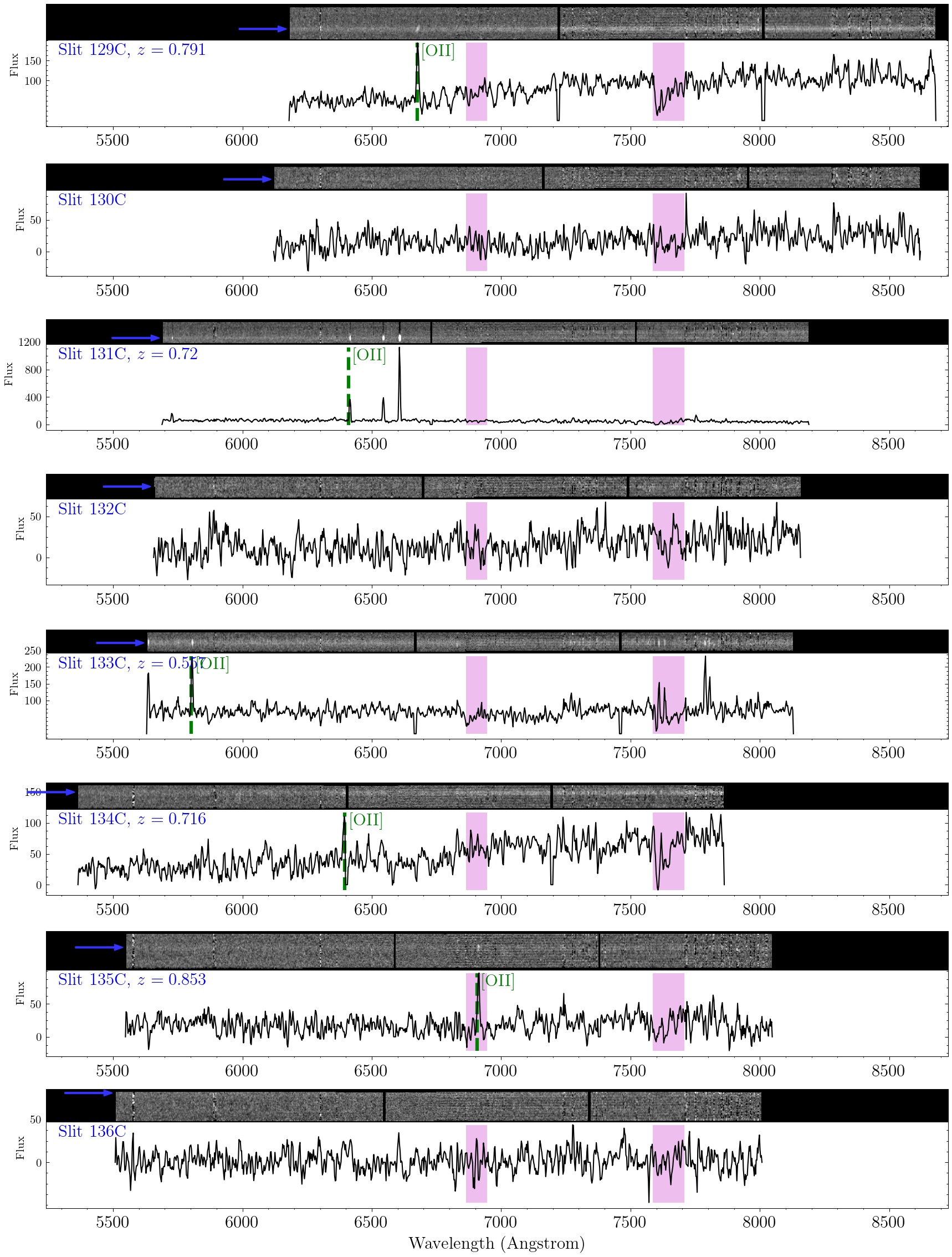}
    \caption{Galaxy Spectra of Slit~\#129C--136C.}
\end{figure*}

\clearpage
\begin{figure*}
    \centering     \includegraphics[width=\textwidth, height=0.96\textheight, keepaspectratio]{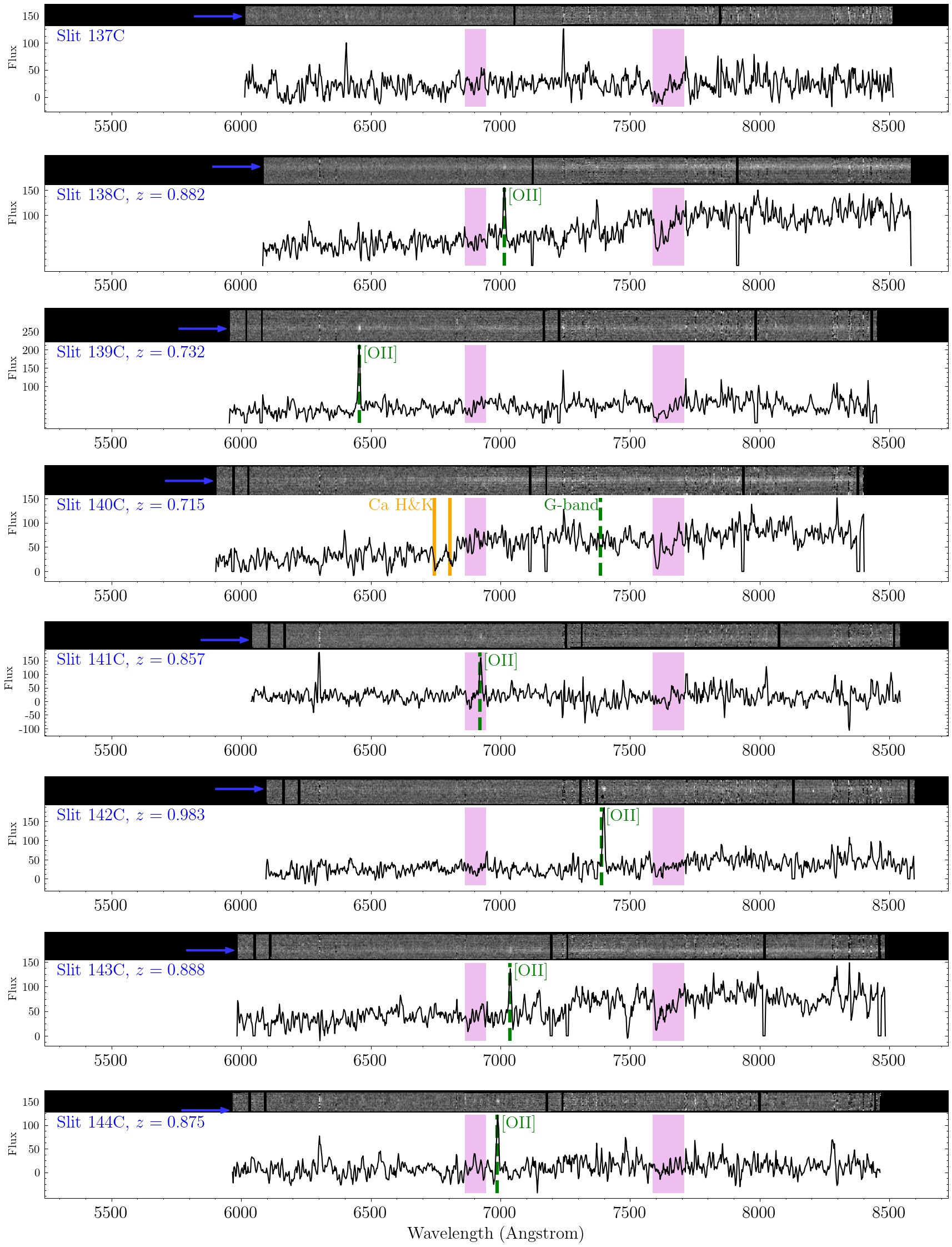}
    \caption{Galaxy Spectra of Slit~\#137C--144C.}
\end{figure*}

\clearpage
\begin{figure*}
    \centering     \includegraphics[width=\textwidth, height=0.96\textheight, keepaspectratio]{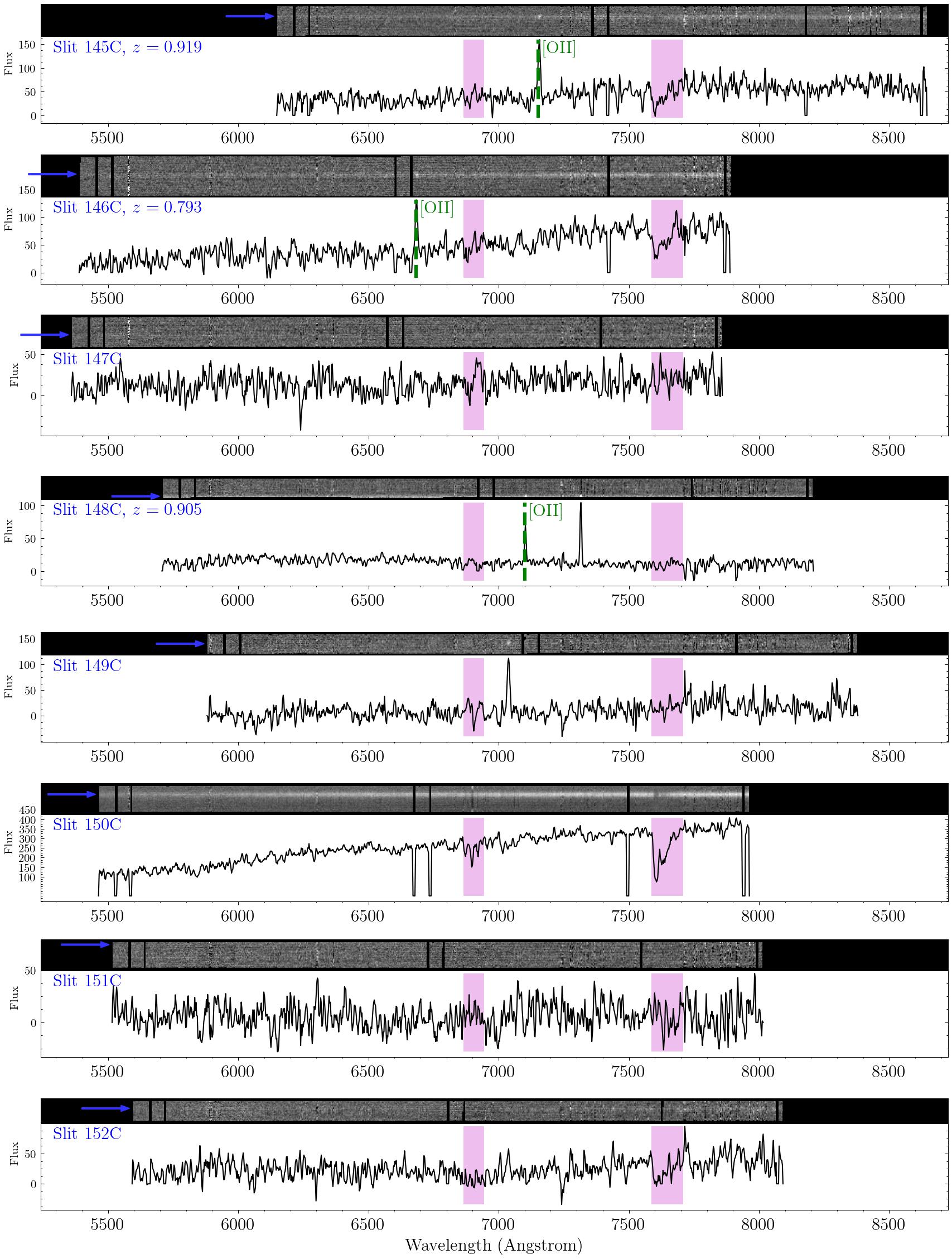}
    \caption{Galaxy Spectra of Slit~\#145C--152C.}
\end{figure*}

\clearpage
\begin{figure*}
    \centering     \includegraphics[width=\textwidth, height=0.96\textheight, keepaspectratio]{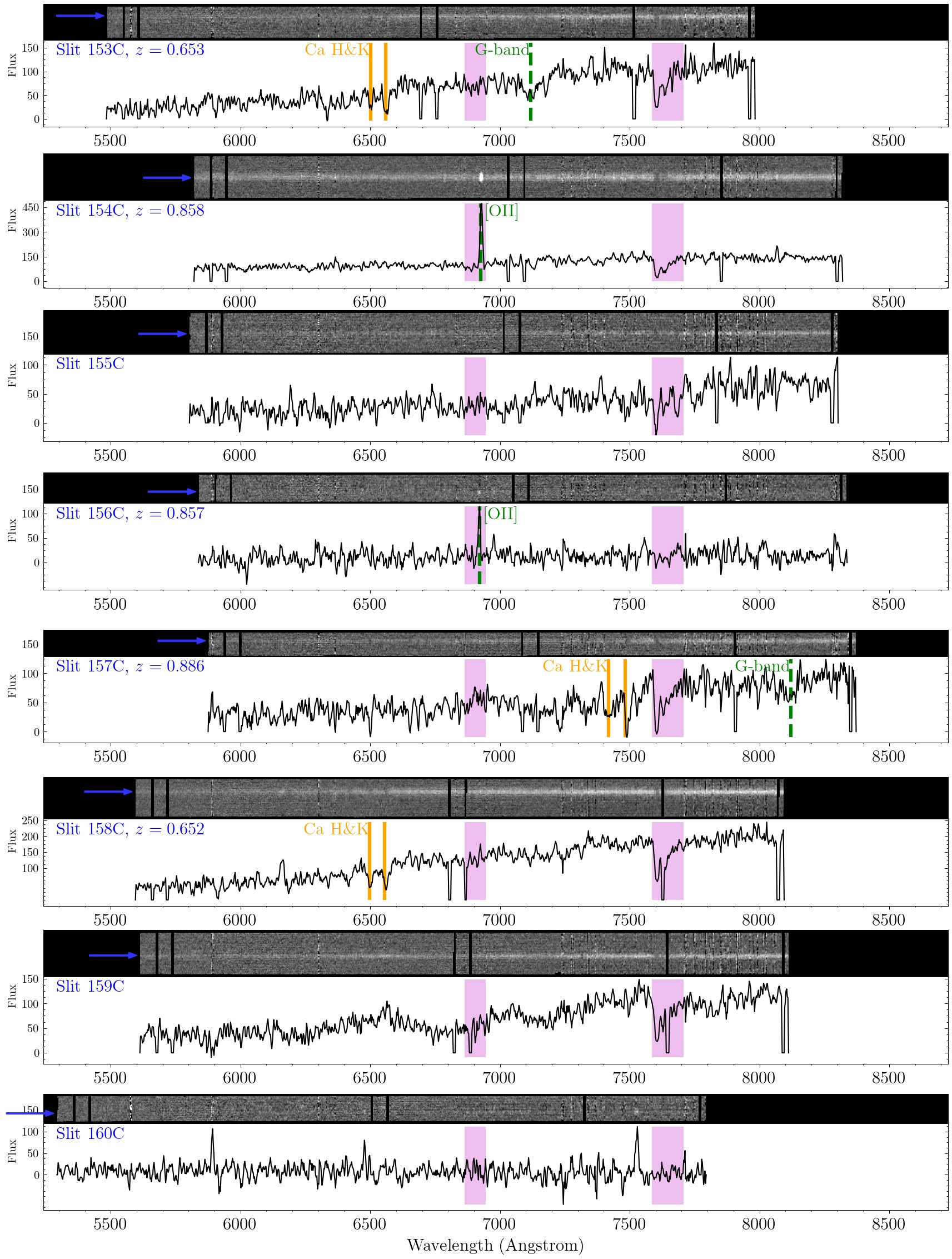}
    \caption{Galaxy Spectra of Slit~\#153C--160C.}
\end{figure*}

\clearpage
\begin{figure*}
    \centering     \includegraphics[width=\textwidth, height=0.96\textheight, keepaspectratio]{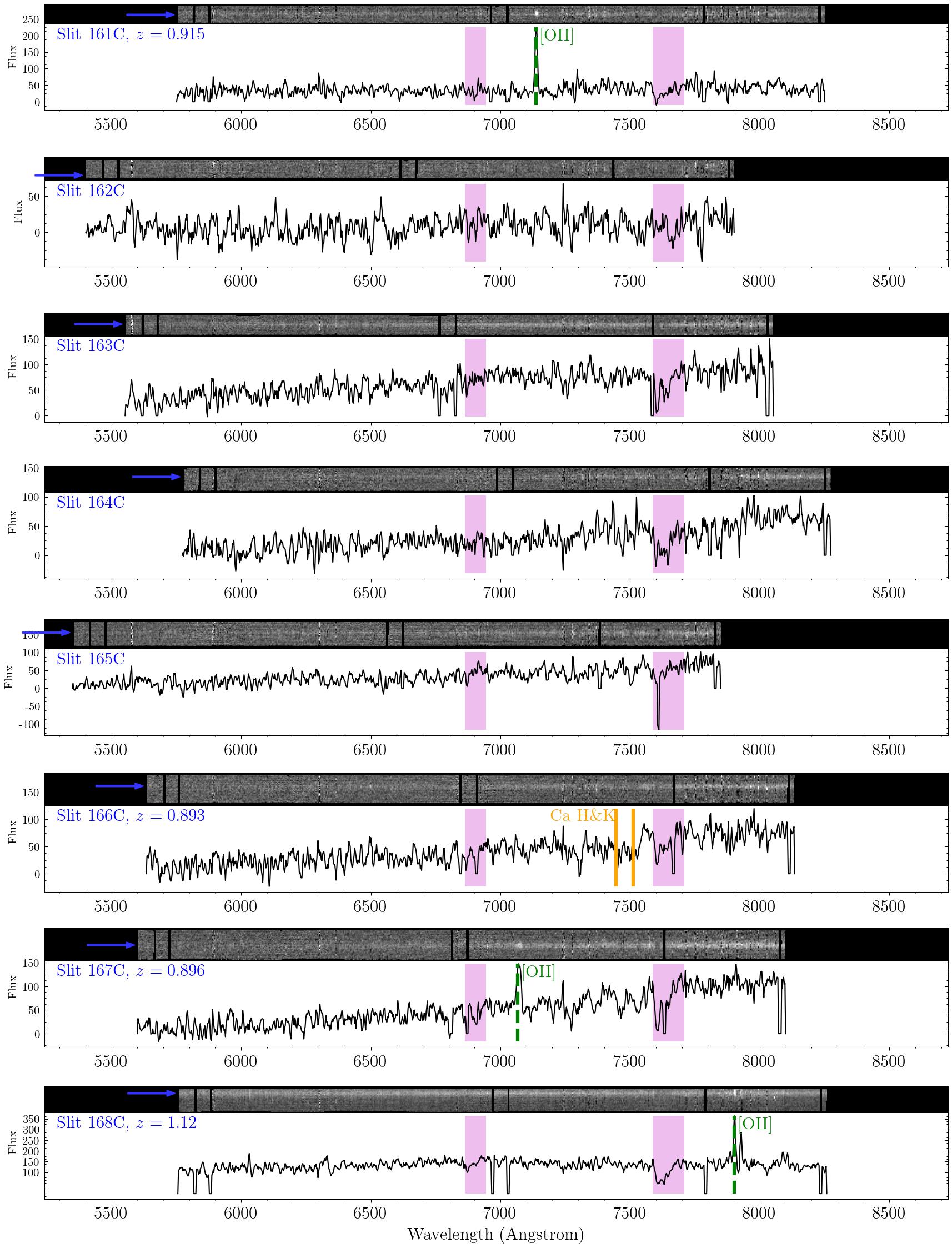}
    \caption{Galaxy Spectra of Slit~\#161C--168C.}
\end{figure*}

\clearpage
\begin{figure*}
    \centering     \includegraphics[width=\textwidth, height=0.96\textheight, keepaspectratio]{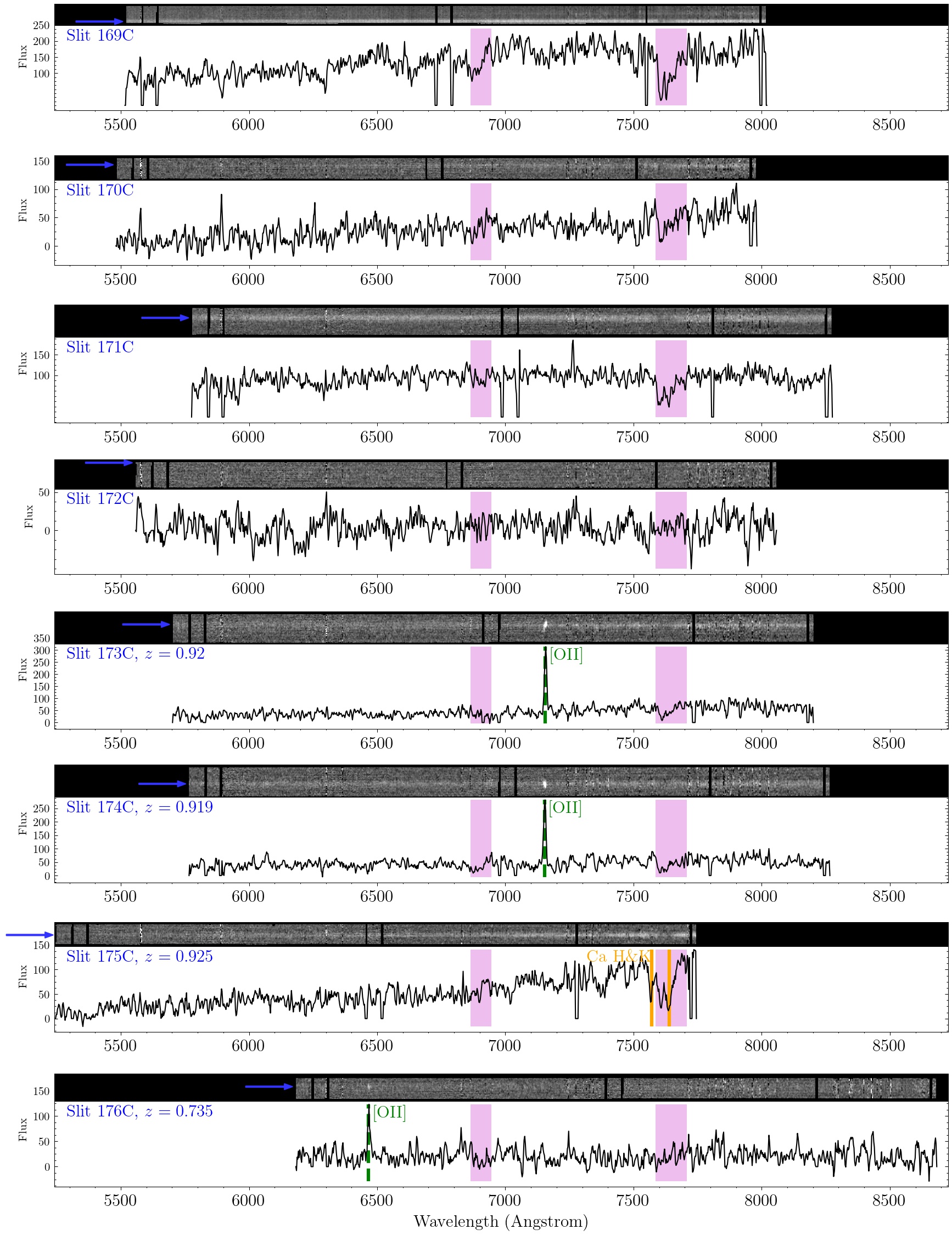}
    \caption{Galaxy Spectra of Slit~\#169C--176C.}
\end{figure*}

\clearpage
\begin{figure*}
    \centering     \includegraphics[width=\textwidth, height=0.96\textheight, keepaspectratio]{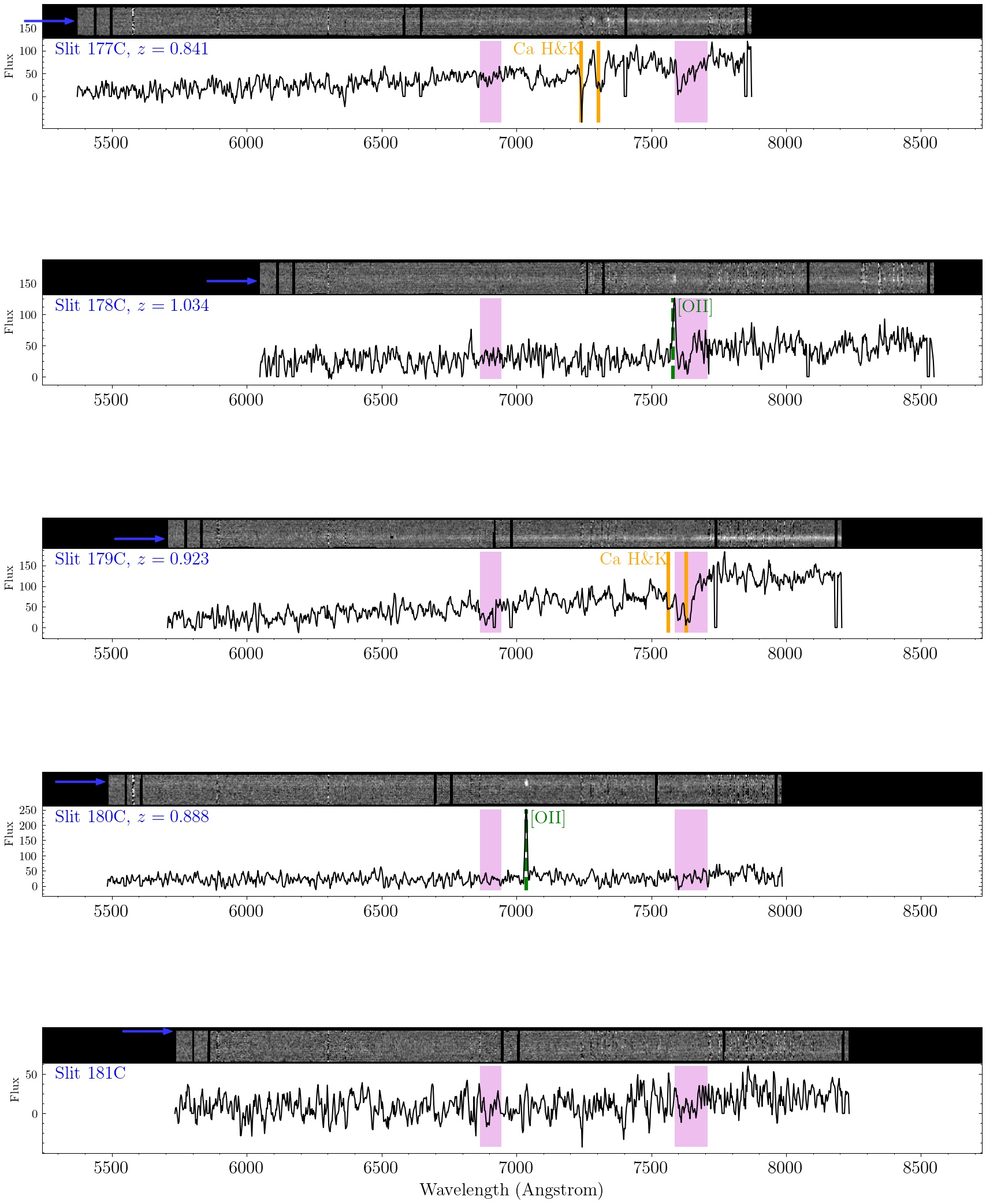}
    \caption{Galaxy Spectra of Slit~\#177C--181C.}
\end{figure*}

\clearpage